\documentclass{article}
\usepackage[a4paper, total={6in, 9in}]{geometry}
\usepackage{amssymb,amsmath}
\usepackage{bm}
\usepackage{cite}
\usepackage{graphicx,bbm,pstricks}
\usepackage{subcaption}
\usepackage{dsfont}
\usepackage{cancel}
\usepackage{bbm,multirow}
\usepackage{authblk}
\usepackage{algorithm,algorithmic}

\usepackage{amsthm}

\renewcommand{\b}[1]{\mathbf{#1}}
\newcommand{\btheta}{\boldsymbol{\theta}}
\newcommand{\bs}[1]{\boldsymbol{#1}}
\newcommand{\cO}{{\mathcal{O}}}
\newcommand{\GP}{{\mathcal{GP}}}
\newcommand{\D}{\mathrm{d}}
\numberwithin{equation}{section}

\DeclareMathOperator{\logit}{logit}
\begin{document}
\title{Hierarchical Bayesian data selection}
\author{Simon L. Cotter\thanks{simon.cotter@manchester.ac.uk}}
\affil{Department of Mathematics, University of Manchester, M13 9PL, UK}
\date{}
\maketitle
\begin{abstract} There are many issues that can cause problems when attempting to infer model parameters from data. Data and models are both imperfect, and as such there are multiple scenarios in which standard methods of inference will lead to misleading conclusions; corrupted data, models which are only representative of subsets of the data, or multiple regions in which the model is best fit using different parameters. Methods exist for the exclusion of some anomalous types of data, but in practice, \emph{data cleaning} is often undertaken by hand before attempting to fit models to data. In this work, we will employ hierarchical Bayesian data selection; the simultaneous inference of both model parameters, and parameters which represent our belief that each observation within the data should be included in the inference. The aim, within a Bayesian setting, is to find the regions of observation space for which the model can well-represent the data, and to find the corresponding model parameters for those regions. A number of approaches will be explored, and applied to test problems in linear regression, and to the problem of fitting an ODE model, approximated by a finite difference method. The approaches are simple to implement, can aid mixing of Markov chains designed to sample from the arising densities, and are broadly applicable to many inferential problems.
\end{abstract}

\begin{center}
\end{center}

\section{Introduction} 
The problem of fitting models to data is a ubiquitous one. In almost all applications across science, engineering and statistics, we are interested in understanding the world around us through observation, and through the construction of models which can approximately replicate the observed data. Through these models we can understand the underlying system they are designed to replicate, in some cases forming the basis for predictions or forecasts. Statistics provides us with a plethora of tools that enable us to combine models and data in order to arrive at estimates of the model parameters which might best fit the given data. In particular, the Bayesian statistical framework has been particularly useful in incorporating complex models, data, and prior information, in order to arrive at posterior densities that represent our updated belief in the value of the model parameters, conditioned on the data~\cite{stuart2010inverse}. However, there are scenarios where standard methods of inference may lead us to incorrect conclusions.

Data is invariably imperfect, and this is a key part of statistical inference, since it is often assumed that data is noisy, and that the discrepancy between model and data can be represented by random variables. These variables may have an assumed form, or hierarchical methods can be employed in which hyperparameters of the observational noise can be simultaneously inferred alongside model parameters~\cite{good1980some}. However, even in this situation there are often assumptions that the observational noise model is the same across all observations. Data which has become corrupted or has anomalous entries may in turn cause a corrupted inference unless the affected data are removed or their effects mitigated. Removal of problematic data can be conducted by hand, with some elements of automation, so called ``data cleaning"~\cite{rahm2000data}, but this is a laborious and subjective process which can potentially have a big impact on the results of the inference. One aspect that is well studied since as early as the 19th Century~\cite{edgeworth1887observations} is that of anomaly detection~\cite{chandola2009anomaly}, where methods have been developed to identify outliers in data. In some cases, the aim is to identify and remove outlier data from the inference. In other applications the artefacts or anomalies are themselves the objects of interest, leading to the development of application-specific methods, from credit card fraud detection~\cite{aleskerov1997cardwatch}, to detection of cancer from MRI imaging~\cite{spence2001detection}. Due to the prevalence of these problems across many subject areas, there are many different communities working on these problems, including those working in statistics and machine learning~\cite{hodge2004survey}, but also in computer science\cite{patcha2007overview} and data security~\cite{spalka2005comprehensive}.

Models are also imperfect. They are often simplified representations, not necessarily designed to exactly mimic reality, but incorporate important features that are key to understanding the system of interest. There are often a range of possible models that one could use to fit to a particular set of data, and that choice can have a big impact on the success of the inference. For example, suppose we are interested in modelling a system of chemical reactions of which we have time series data of the fluctuations of concentrations of chemical species. We could choose to fit our data to an ordinary differential equation, an It\^{o} stochastic differential equation, or a continuous-time Markov process, all of which would be valid choices depending on the type of data you wish to assimilate. The choice of model may not be an obvious decision, and as such model selection may be undertaken~\cite{ando2010bayesian,ding2018model}. In particular there are several popular approaches to Bayesian model selection, in which the aim is to use the data to infer which of the models best describes the data. For example, the Bayes' factor can be computed, which is the ratio of the marginal likelihoods of a pair of posterior distributions arising from the same data with different models~\cite{berger1996intrinsic}. The Bayesian information criterion~\cite{schwarz1978estimating} is another popular method for model selection, due to the ease with which the criterion can be computed and then compared between models. However, what these methods do not consider, is the idea that a given model may be an excellent representation, not of the whole data, but a substantial subset. Moreover, different considerations may be needed if a model can be representative of all of the data, but with different parameter estimates in different regions of observation space; for example the case of evolving model parameters of Covid-19 during the pandemic~\cite{nikhra2021evolving}.

Another approach to problems with discrepancies between models and data is to account for or infer model error alongside model parameters. This is a particularly prevalent approach in data assimilation and meteorology~\cite{houtekamer2009model,tremolet2007model,danforth2007estimating}, where learning the differences between already very large and complex models and reality is key to the accuracy of forecasting. The model error can, for example, be represented by a function that is then inferred as part of the overall inference. Naturally this can have significant computational complexity ramifications, but is an important methodology in a range of applications, including economics~\cite{jacquier2000bayesian} and engineering~\cite{skelton1989model}. In computer model calibration, Gaussian process priors are often placed on the function describing the model error, and then inferred alongside the unknown parameters~\cite{kennedy2001bayesian}.

In this paper we will explore and expand on the approach first presented in \cite{forsyth2022unlabelled} on hierarchical Bayesian data selection; the simultaneous inference of model parameters, and parameters which dictate the sensitivity of the posterior to a given observation. In \cite{forsyth2022unlabelled} the authors were aiming to solve an unlabelled landmark matching problem where there are two point clouds that represent the cell centres of an embryo taken from two images. The first image represents the final frame of a real time imaging (RTI) experiment where the aim is to track each cell and its microenvironment, e.g. its position in the embryo and the number of neighbouring cells. The embryo is then fixed, to prevent further development, and imaged a second time using immunostaining methods in order to ascertain the fate of each cell. The aim is then to identify which cells in each image correspond with each other, so that the two data sets can be combined and analysis carried out to analyse the hypothesis that historical cell microenvironment is a significant indicator of cell fate. One issue is that there may be cells in both images which do not have a corresponding match in the other image, due to mitosis or apoptosis occurring in the intervening time between the final frame of the RTI and the embryo being fixed, or due to issues with cells being out of focus or being incorrectly segmented. In this case, standard methods for inference can lead to the incorrect matching of cells across the images. Only by identifying which cells do not have a match using data selection can one arrive at a robust method for matching the unlabelled point clouds. The method that was developed for this application, will be explored in more detail in Section \ref{sec:NPF}.

The methods that we will present are not only straightforward to implement, but have other benefits including the smoothing of posterior distributions, enabling faster mixing of sampling methodologies such as Markov chain Monte Carlo (MCMC)~\cite{brooks2011handbook}. The aim of the approach is to infer, alongside the model parameters, the regions in observation space for which the chosen model is a good representation of the observations made. This gives us a wealth of information beyond that of a standard inference, and furthermore makes the inference more robust to data corruption and model errors. Bayesian data selection has also been considered independently from a different perspective in \cite{weinstein2021bayesian}, in which the authors propose a score for performing both data selection and model selection, the ``Stein volume criterion", which is a generalisation of the marginal likelihood, which is often used within model selection problems. They consider the case of i.i.d. data, where they are interested in finding subspaces (the foreground) of observation space, such that the data projected onto those subspaces can be well fit to the proposed mode. Although this approach is divergent from our own, and concerns a less general setting, it highlights the need for methods that can efficiently and robustly perform data selection in an objective manner.

In Section \ref{sec:pModel} we introduce Bayesian data selection where we introduce a probability $p_i$ that the $i$th observation should be included in the inference, when the standard likelihood can be written in product form, as in the case of i.i.d. noise. This approach is computationally intractable for all but the simplest inverse problem with a small number of observations, and so in Section \ref{sec:fid} we consider an alternative where data fidelity parameters are inferred alongside the model parameters, whose value dictates the sensitivity of the posterior to each observation. We introduce a logit-GP prior for the underlying data fidelity field in Section \ref{sec:learn}, which allows us to identify regions where there are neighbourhoods of observations with good matches between model and data. In Section \ref{sec:NPF} we consider a version of the method which can be applied even when the likelihood is not product-form. In Section \ref{sec:interpret} we discuss how the posteriors arising from these approaches can be interpreted and used to improve inference. In Section \ref{sec:lin} we apply Bayesian data selection to linear regression in a range of problematics settings; corrupted data,  model error (i.e. non-linear behaviour of the data in some regions), and where the model is a piecewise fit to the data. In Section \ref{sec:ODE} we apply Bayesian data selection to an inverse problem for the initial condition of an initial value problem, where we are using a numerical approximation of the solution to the ODE whose error grows in time. We conclude in Section \ref{sec:discussion} with a discussion.

\section{The $p$ model} \label{sec:pModel}
Before introducing Bayesian data selection, we summarise the standard Bayesian inference for a set of unknowns
$\btheta \in \Theta$ given observations $\b{y}$, and introduce the notation. We assume that we have observed our
system with a set of $N$ independent observations
$y_i \in \mathbb{R}^d$, $i\in \{1,\ldots, N\}$. Due to independence,
we can write down the likelihood of each observation, given by:
\begin{equation}
  L_i(y_i|\btheta) \propto \exp( -\phi_i(\btheta;y_i)),
\end{equation}
where $\phi_i(\btheta;y_i)$ is the negative log-likelihood of the $i$th observation.
Therefore the likelihood can be written in product form,
\begin{eqnarray}
  L(\b{y}|\btheta) &\propto & \prod_{i=1}^N \exp \left (-\phi_i(\btheta;y_i) \right),\\
  &=& \exp \left ( - \sum_{i=1}^N \phi_i(\btheta;y_i) \right ), \\
  &=:& \exp \left ( -\Phi(\btheta; \b{y}) \right ),
\end{eqnarray}
where $\b{y} = [y_1, \ldots, y_N] \in \mathbb{R}^{d\times N}$.
Assuming a prior distribution with density $\pi_0(\btheta)$, then by
Bayes' rule, the posterior distribution has density
\begin{equation}
\pi(\btheta) = \frac{1}{Z} L(\b{y}|\btheta) \pi_0(\btheta),
\end{equation}
where \[Z = \int_\Omega  L(\b{y}|\btheta) \pi_0(\btheta) \, d\btheta,\]
is a normalisation constant.

We now consider the problem where we are not sure which observations
should be included in the inference. This uncertainty can arise for
several reasons, for example the model we are using to compute the
likelihood may not be a good representation of reality in some regions
of the observation space, or some of the observations were corrupted
by outside factors. To incorporate this additional uncertainty we
introduce a set of independent parameters $p_i \in [0,1]$ for $i\in \{1, \ldots,
N\}$ which represents the probability that observation $y_i$ should be
included in the likelihood evaluation. We refer to this approach to Bayesian data selection as the $p$ model.

Given $N \in \mathbb{N}$ observations, there are $2^N$ possible
likelihoods incorporating subsets of those observations. Each of these
$2^N$ possibilities can be described by a vector $\bs{\iota} =
[\iota_1, \ldots, \iota_N] \in
\{0,1\}^N$, so that
\begin{eqnarray}
L_{\bs{\iota}} &=&  \prod_{i=1}^N  \exp \left
                   (-\iota_i\phi_i(\btheta;y_i) \right),\\
  &=:&  \exp \left ( -\Phi_{\bs{\iota}}(\btheta; \b{y}) \right ).
\end{eqnarray}

Each of these likelihoods gives rise to a corresponding posterior
distribution with density
\begin{equation}
  \pi_{\bs{\iota}}(\btheta) = \frac{1}{Z_{\bs{\iota}}} L_{\bs{\iota}}(\b{y}|\btheta) \pi_0(\btheta),
\end{equation}
where $Z_{\bs{\iota}} \in \mathbb{R}^+$ is a normalisation constant.

Noting that the probability that our posterior is given by
$\pi_{\bs{\iota}}(\btheta)$ is equal to
\begin{equation} \b{p}^{\bs{\iota}}(\mathbbm{1} - \b{p})^{(\mathbbm{1} - \bs{\iota})}
  := \prod_{i=1}^N p_i^{\iota_i}(1- p_i)^{(1-\iota_i)},\end{equation}
we arrive at the density on $\btheta$ conditioned on $\b{p}$,
given by
\begin{eqnarray}
\pi(\btheta|\b{p}) = \sum_{\bs{\iota} \in \{0,1\}^N} \b{p}^{\bs{\iota}}(
\mathbbm{1} - \b{p})^{(\mathbbm{1} - \bs{\iota})} \pi_{\bs{\iota}}(\btheta).
\end{eqnarray}
Therefore the joint density on $\btheta$ and $\b{p}$ is given by
\begin{equation}\pi(\btheta,\b{p}) = \pi_0(\b{p}) \sum_{\bs{\iota} \in \{0,1\}^N} \b{p}^{\bs{\iota}}(
  \mathbbm{1} -  \b{p})^{(\mathbbm{1} - \bs{\iota})}
  \pi_{\bs{\iota}}(\btheta)
\end{equation}
\subsection{Marginal distribution on $\btheta$} \label{sec:marge}
If we pick $\pi_0(\b{p})$ to be conjugate to this joint density, then
we will be able to integrate the joint density with respect to the $p_i$ and arrive at a
marginal density on $\btheta$. Each term in the sum in the joint
density is proportional to a product of independent Bernoulli
densities, and as such we pick $\pi_0$ to be an independent product of
Beta distributions on each of the $p_i$, with hyperparameters $\alpha,
\beta >0$. Given this prior, we arrive at the joint density:
\begin{eqnarray}\pi(\btheta,\b{p}) &=& \frac{1}{B(\alpha,\beta)^N}
    \sum_{\bs{\iota} \in \{0,1\}^N}\b{p}^{\alpha - 1}(\mathbbm{1} - \b{p})^{\beta - 1}\b{p}^{\bs{\iota}}(
  \mathbbm{1}- \b{p})^{(\mathbbm{1} - \bs{\iota})}
                                           \pi_{\bs{\iota}}(\btheta), \\
  &=& \frac{1}{B(\alpha,\beta)^N}
    \sum_{\bs{\iota} \in \{0,1\}^N} \pi_{\bs{\iota}}(\btheta) \prod_{i=1}^N
      p_i^{\alpha - 1 + \iota_i}(1 - p_i)^{\beta - \iota_i}\\
\end{eqnarray}
where $\b{p}^x = [b_1^x,\ldots,b_N^x]$ denotes elementwise
exponents, and $B(\alpha,\beta) =
\frac{\Gamma(\alpha)\Gamma(\beta)}{\Gamma(\alpha + \beta)}$ is the
Beta function.

Noting that
\begin{equation} \int_0^1 p^{\alpha - 1 + \iota}(1-p)^{\beta - \iota} dp =
  B(\alpha + \iota, \beta + 1- \iota),\end{equation}
it is now trivial to find the marginal distribution on $\btheta$,
by integrating with respect to $p_1, \ldots , p_N$, giving us
\begin{eqnarray}\pi(\btheta) &=& \int_{[0,1]^N}
                                     \pi(\btheta,\b{p}) \, \D p_1
                                     \ldots \D p_N,\\
                                 &=&  \frac{1}{B(\alpha,\beta)^N}
    \sum_{\bs{\iota} \in \{0,1\}^N} \pi_{\bs{\iota}}(\btheta)
                                     \prod_{i=1}^N  B(\alpha +
                                     \iota_i, \beta + 1- \iota_i),\\
                                     &=&  \frac{1}{B(\alpha,\beta)^N}
    \sum_{\bs{\iota} \in \{0,1\}^N} \pi_{\bs{\iota}}(\btheta)
                                     B(\alpha +
                                     1, \beta)^{n_{\bs{\iota}}}
                                         B(\alpha, \beta + 1)^{N -
                                         n_{\bs{\iota}}},\\
                                         &=:& 
    \sum_{\bs{\iota} \in \{0,1\}^N} C_{\bs{\iota}}\pi_{\bs{\iota}}(\btheta),                                    
\end{eqnarray}
where $n_{\bs{\iota}} := \sum_{i=1}^N \iota_i$ is the number of observations
included in $L_\iota$. We note that this density is a mixture
distribution of the possible posterior densities $\pi_{\bs{\iota}}$ with coefficients
\begin{equation}
C_{\bs{\iota}} = \frac{B(\alpha +
                                     1, \beta)^{n_{\bs{\iota}}}
                                         B(\alpha, \beta + 1)^{N -
                                         n_{\bs{\iota}}}}{B(\alpha,\beta)^N}.
  \end{equation}

This gives us a method by which we can sample the density on
$\btheta$ with data selection without explicitly computing the
data selection. The main obstructions to the practical use of this
approach are:
\begin{enumerate}
  \item The number of terms in this sum grows exponentially with the
    number of observations,
\item We are required to know $Z_\iota$, the normalisation
  constants for each of the $2^N$ possible posterior densities.
\end{enumerate}

It is also worth noting that we must be careful when picking
priors for the parameters $\btheta$, since the likelihood is not
proper for $\iota = [0,\ldots,0]$. As such, we must pick proper priors
for all parameters. This method can be used directly for some simple but still
very commonly used cases, assuming the number of observations is small, as we shall see in Section \ref{sec:lin}, where we apply Bayesian data selection to linear regression problems. Although this method is not useful in a practical sense, it is a method with a strong theoretical justification, and exploring its use for simple examples and comparing with the results of other methods is instructive.

\section{Data/Model Fidelity}\label{sec:fid}
The $p$ model approach as presented in Section \ref{sec:pModel} is
problematic for all but the simplest cases, since it requires
knowledge of the normalisation constants of $2^N$ posterior
distributions, where $N$ is the number of observations made. This
simply makes it computationally intractable for almost any application
of interest. As such, we might wish to consider other methods for
characterising our belief in a particular observation. Our method is based on a hierarchical approach where parameters are introduced which, similarly to the $p_i$ in the $p$ method, control the sensitivity of the posterior to each observation. 

Consider the $p$ model, where we have set $p_i = 1$ for all $i \in
\{1,\ldots,N\}$ bar one value $p_k \in [0,1]$. The parameter $p_k \in [0,1]$ describes a parameterisation of a path on the space of probability distributions, from
\[  \pi(\btheta) \propto \pi_0(\btheta) \prod_{i=1}^N \exp \left (-\phi_i(\btheta; y_i)
  \right ),\]
at $p_k=1$ to the density
\[ \pi_{\scalebox{0.75}[1.0]{-} k}(\btheta) \propto \pi_0(\btheta) \prod_{i=1, i\neq k}^N \exp \left (-\phi_i(\btheta;y_i)
  \right ),\]
  at $p_k=0$. The path is a Euclidean geodesic in the space of probability densities, and can be simply represented by 
\begin{equation}
    \pi(p_k) = p_k \pi + (1-p_k)\pi_{\scalebox{0.75}[1.0]{-} k}.
\end{equation}
   We now consider an alternative procedure that maintains many of the features of the $p$-model, but which has comparable computational burden to standard Bayesian inference.

Consider the $i$th observation, with a tempering parameter, or what we will refer to as a \emph{fidelity} parameter, $\tau_i \in
[0,1]$,
so that
\begin{equation}
L_i(\btheta) = \frac{1}{Z(\tau_i)} \exp \left (-\tau_i \phi_i(\btheta;y_i) \right ),
\end{equation}
where $Z(\tau_i)$ is a normalisation constant such that
\begin{equation}
Z(\tau_i) = \int_{\mathbb{R}^{d}} \exp \left (-\tau_i \phi_i(\btheta;y_i) \right ) dy_i.\label{eq:norm}
\end{equation}
Since the likelihood is usually a standard distribution, such as a Gaussian PDF, these can be
found explicitly in many cases. Note that in practice we only need to know the normalisation constant's functional dependence on the $\tau_i$.

\begin{figure}[htbp]
    \centering
    \begin{subfigure}[b]{0.48\textwidth}
         \begin{center}
    \includegraphics[width=\textwidth]{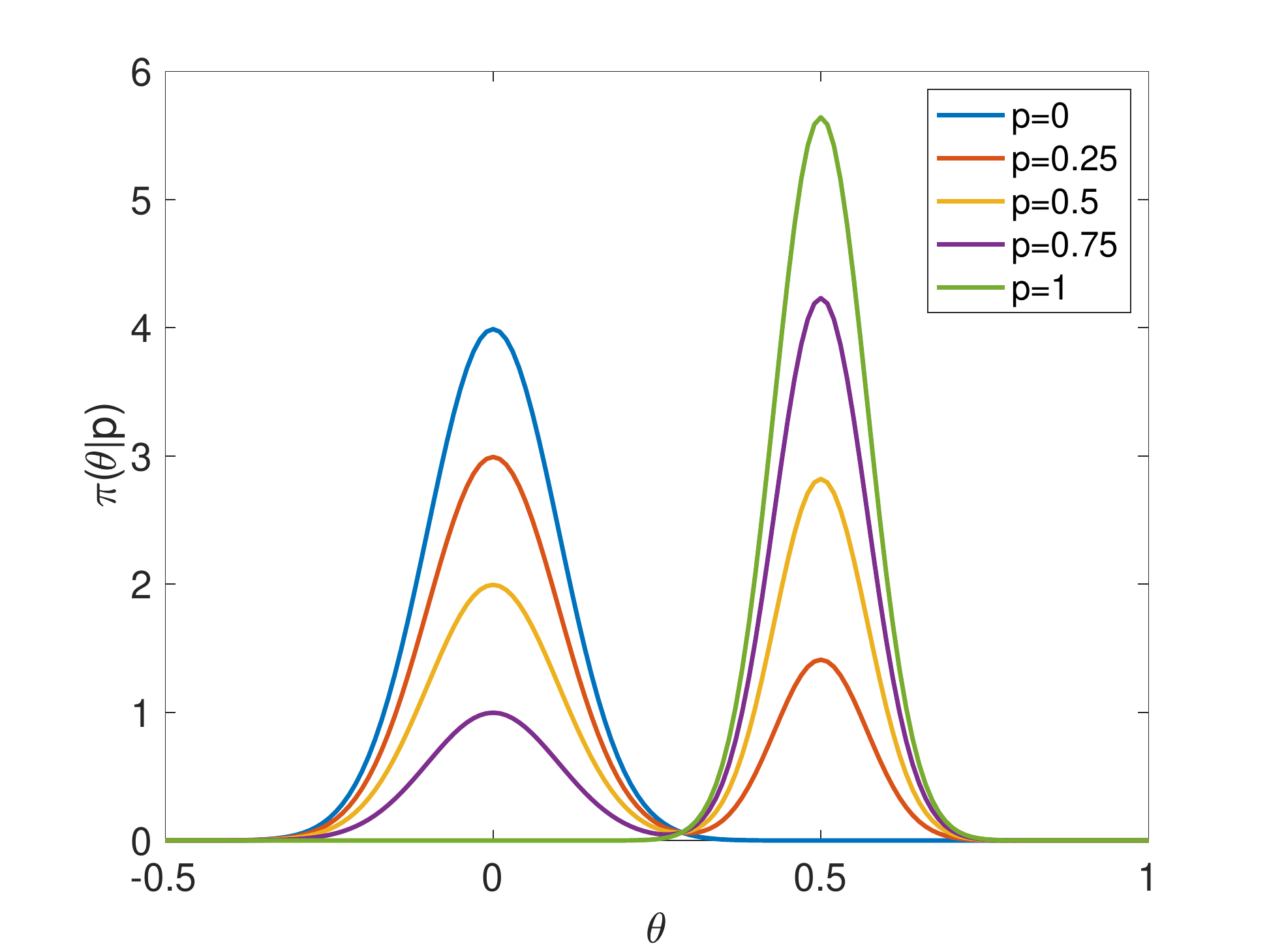}\caption{$p$ method path.}
    \end{center}
    \end{subfigure}
    \begin{subfigure}[b]{0.48\textwidth}
         \begin{center}
    \includegraphics[width=\textwidth]{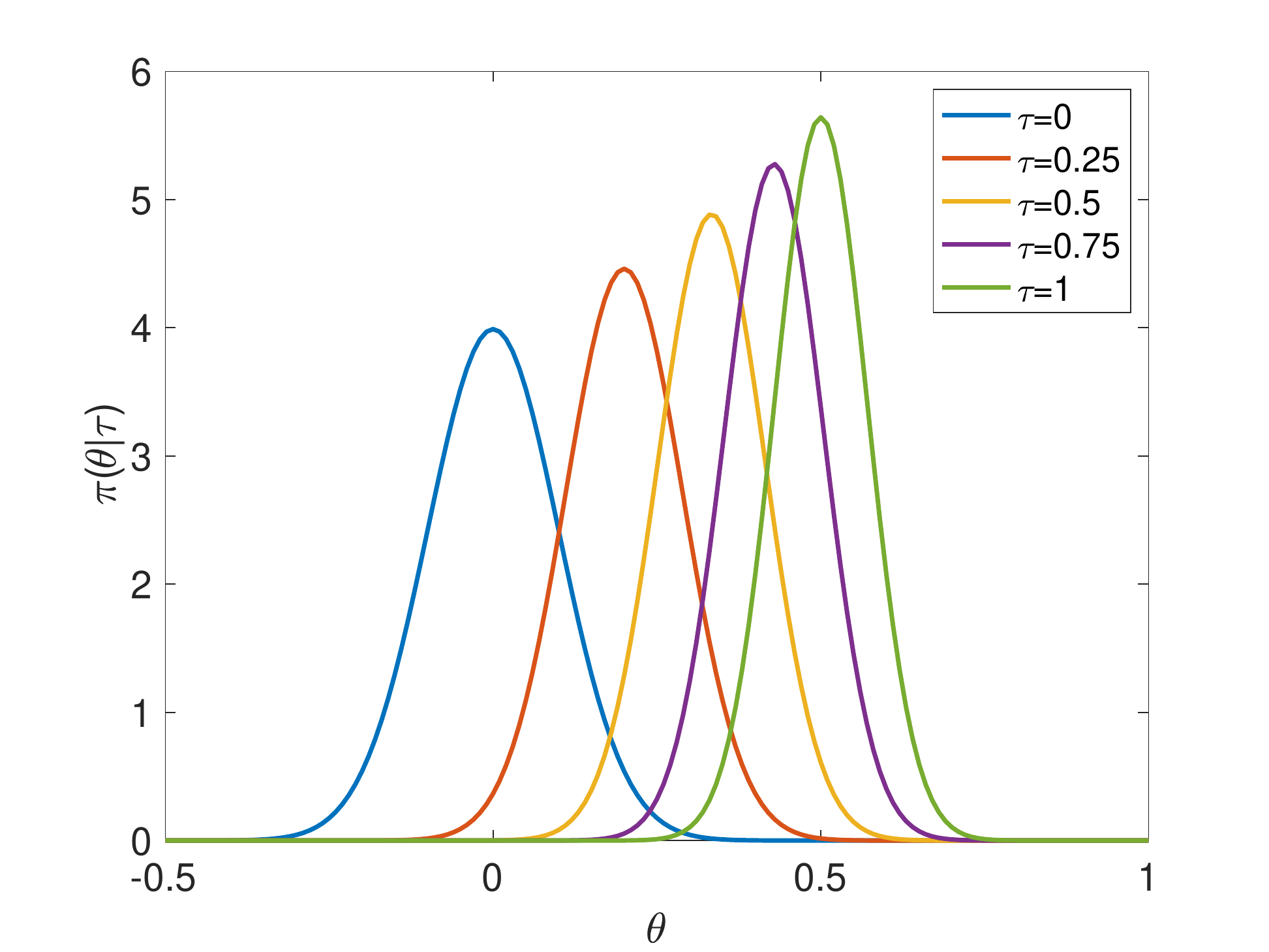}\caption{Data fidelity path.}
    \end{center}
    \end{subfigure}
    \caption{Plots to show the paths on the space of probability distributions between two Gaussians using (a) the $p$ method, (b) data fidelity.}
    \label{fig:P}
\end{figure}

Changing the parameter $\tau_k$ from $\tau_k = 1$ to $\tau_k = 0$, as with the $p$-model, describes a path between $\pi$ and $\pi_{\scalebox{0.75}[1.0]{-} k}$. However, the behaviour is different, since tuning $p_k$ simply changes the weights between two distributions, moving along the Euclidean geodesic between the two distributions. In contrast, changing $\tau_i$ transports the probability density smoothly from $\pi$ to the mode of $\pi_{\scalebox{0.75}[1.0]{-} k}$, similarly to that seen in optimal transport.
Both of these parameters represent our belief in the
fidelity of this observation. Figure \ref{fig:P} shows example paths between two Gaussians using each of these methods.

This method is similar to that presented in \cite{wang2017robust}. In this work, the authors suggest a method for mitigation of model-data discrepancies involving the introduction of weights $w_i$ for each observation which are inferred alongside the model parameters. The likelihood of each observation is raised to the power $w_i$, such that the new posterior is given by:
\begin{equation}
\pi(\btheta, {\bs w}) \propto \pi_0(\btheta)\pi_0({\bs w}) \prod_{i=1}^N L_i(\btheta)^{w_i}.
\end{equation}
What differs from our approach, is that they do not view the change in $w_i$ as a change in the statistical model for the likelihood of $i$th observation, which is reflected in the fact that they lose the dependence of the normalisation constant on these weights, as described by \eqref{eq:norm}.

In \cite{wang2017robust}, they explore the use of some interesting priors on these weights, including a product of Beta distributions, a product of Gamma distributions, and scaled Dirichlet priors. In this work we will look at some cases where conjugate priors can be picked which enable analytical marginalisation of the $\tau_i$. This has significant benefits, since using Bayesian data selection increases the dimensionality of the state space by $N$, the number of observations. Since there are likely to be strong correlations between the model parameters and the $\tau_i$, or even multiple modes (as we will see in Section \ref{sec:MFPV}), this could lead to slow mixing of MCMC chains.

Additionally, in Section \ref{sec:learn}, we will consider an alternative prior for the fidelity parameters, derived from a logit-Gaussian process, which will enable us to learn the regions which have neighbourhoods of observation points where the model and data are consistent.

\subsection{Gaussian likelihood case}\label{sec:GLC}
We consider the Gaussian likelihood case, but similar derivations can easily be found for other standard distributions. We will demonstrate how
$Z(\tau_i)$ can be easily computed, and then we will pick a conjugate
prior for ${\bs{\tau}} = [\tau_1, \ldots, \tau_N]^\top \in [0,1]^N$ which will allow us to find the marginal distribution
on $\btheta$. 

Firstly we define our observations in this case, in the language of \cite{stuart2010inverse}, to be of the form:
\begin{equation}
    {y_i} = \mathcal{G}_i(\btheta) + \eta_i, \qquad \eta_i \sim \mathcal{N}(0,\Sigma),
\end{equation}
where the $\eta_i$ are i.i.d. for all $i=1,\ldots,N$, $\mathcal{G}_i:\btheta \to \mathbb{R}^d$ is the observation operator of the $i$th observation, and $\Sigma \in \mathbb{R}^{d \times d}$ is a positive definite symmetric covariance matrix. In this case, the data fidelity likelihood of the $i$th
observation is given by:
\begin{eqnarray}
L_i(y_i|\btheta, \tau_i) & = & \frac{1}{Z(\tau_i)} \exp \left (
  -\frac{\tau_i}{2} (\mathcal{G}_i(\btheta) - y_i)^\top \Sigma^{-1}
                              (\mathcal{G}_i(\btheta) - y_i)
                              \right ).
                              \end{eqnarray}
Since this is simply a Gaussian density for $y_i$, we can easily compute the normalisation constant
\begin{eqnarray}
  Z(\tau_i) & = &  \sqrt{{\rm det} \left ( 2\pi \tau_i^{-1}
                              \Sigma \right )} \\
                              & \propto & \tau_i^{d/2}
\end{eqnarray}
Note that the fidelity parameter essentially becomes absorbed into the
noise covariance, and so appears in the normalisation constant as a
factor of $\tau_i^{d/2}$. Therefore the full likelihood is given by:
\begin{equation}
L(\b{y}|\btheta, \bs{\tau}) \propto \prod_{i=1}^N     \tau_i^{d/2} \exp \left (
  -\frac{\tau_i}{2} (\mathcal{G}_i(\btheta) - y_i) ^\top \Sigma^{-1} (\mathcal{G}_i(\btheta) - y_i) \right ).
  \end{equation}
  Invoking Bayes' rule, we arrive at the joint posterior density
  \begin{equation}
\pi(\btheta, \bs{\tau}) \propto \pi_0(\btheta) \prod_{i=1}^N
\pi_0(\tau_i)   \tau_i^{d/2} \exp \left (
  -\frac{\tau_i}{2} (\mathcal{G}_i(\btheta) - y_i) ^\top \Sigma^{-1} (\mathcal{G}_i(\btheta) - y_i) \right ).
\end{equation}
As before, we can pick a conjugate prior for each $\tau_i$ that
allows us to find the marginal distribution on $\btheta$.
We pick a truncated gamma distribution as the prior distribution
for each $\tau_i$, with hyperparameters $\alpha, \beta>0$, i.e.
\begin{equation} \pi({\tau}) = \frac{\beta^\alpha \tau^{\alpha - 1}\exp(-\beta
  \tau)}{\gamma(\alpha,\beta)} \mathbbm{1}_{[0,1]}(\tau).\end{equation}
Substituting this into the expression for the joint density gives us:
  \begin{equation}
\pi(\btheta, \bs{\tau}) \propto \pi_0(\btheta) \mathbbm{1}_{[0,1]^N}(\bs{\tau}) \prod_{i=1}^N
  \tau_i^{\alpha - 1 + d/2} \exp \left ( -\tau_i \left (\beta
  +\frac{1}{2} (\mathcal{G}_i(\btheta) - y_i) ^\top \Sigma^{-1} (\mathcal{G}_i(\btheta) - y_i) \right ) \right ).
\end{equation}
This is a product of truncated gamma densities, with parameters
$\hat{\alpha}_i = \alpha + d/2$ and $\hat{\beta}_i = \beta
+\frac{1}{2} (\mathcal{G}_i(\btheta) - y_i) ^\top \Sigma^{-1} (\mathcal{G}_i(\btheta) - y_i) $. Since
\begin{eqnarray}
\int_0^1 \tau_i^{\hat{\alpha}_i - 1}
                      \exp(-\hat{\beta}_i \tau_i) d \tau_i &=&
                                                               \frac{1}{\hat{\beta_i}^{\hat{\alpha}_i}}
                                                               \gamma(\hat{\alpha}_i,
                                                               \hat{\beta}_i),
\end{eqnarray}
the marginal density on $\btheta$ is given by
\begin{eqnarray}\label{eq:fid}
\pi(\btheta) &\propto& \pi_0(\btheta) \prod_{i=1}^N
  \left (   \beta
+\frac{1}{2} (\mathcal{G}_i(\btheta) - y_i) ^\top \Sigma^{-1}
                           (\mathcal{G}_i(\btheta) - y_i)    \right
                           )^{- \alpha - d/2} \\ \nonumber && \hspace{.5cm}
                                                               .\gamma
                                                               \left ( \alpha + d/2,
                                                               \beta
+\frac{1}{2} (\mathcal{G}_i(\btheta) - y_i) ^\top \Sigma^{-1} (\mathcal{G}_i(\btheta) - y_i) \right ).
\end{eqnarray}

\begin{figure}[htbp]
    \centering
    \begin{subfigure}[b]{0.48\textwidth}
         \begin{center}
    \includegraphics[width=\textwidth]{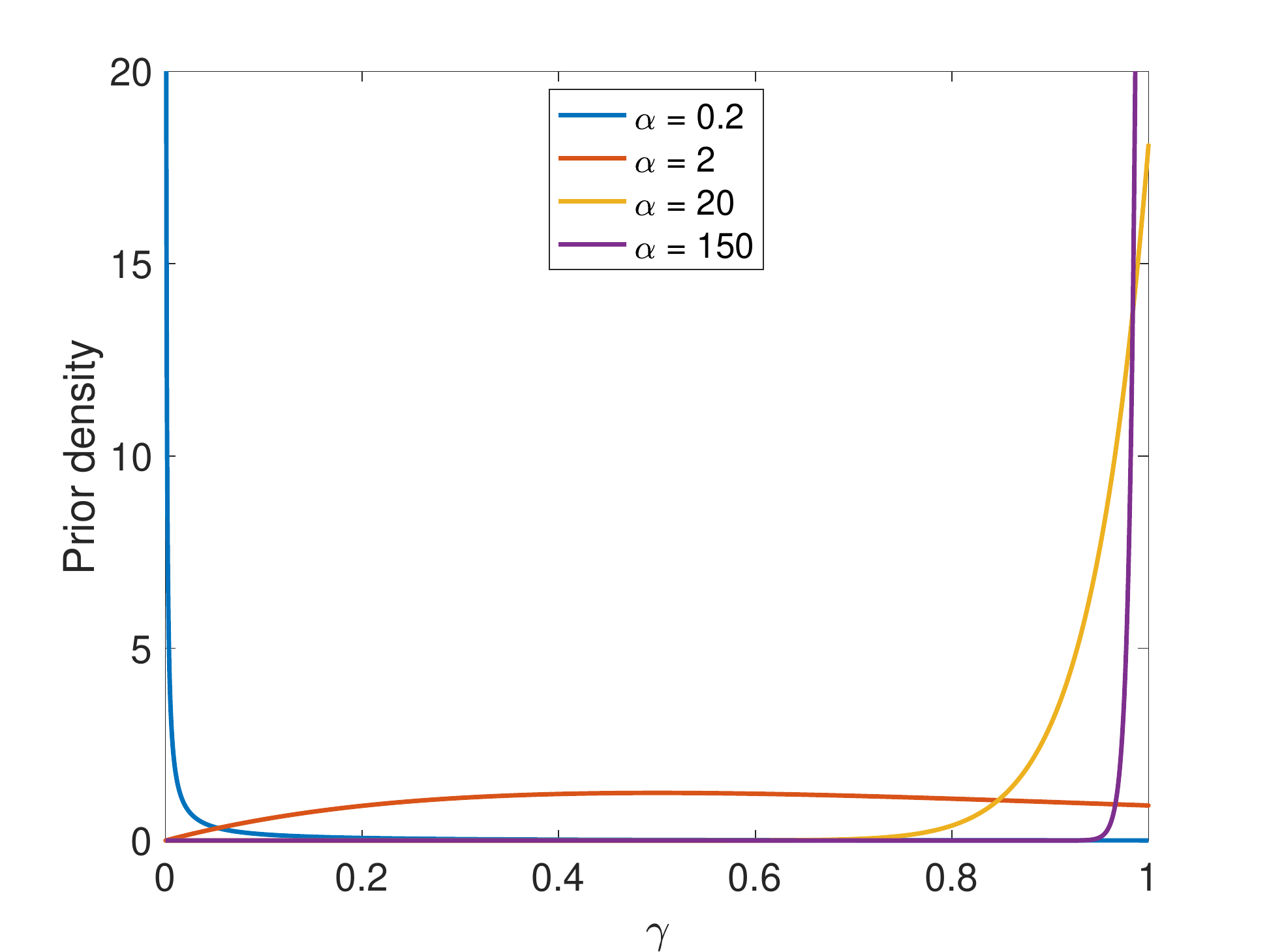}\caption{Examples of different fidelity priors. }\label{fig:FLa}
    \end{center}
    \end{subfigure}
    \begin{subfigure}[b]{0.48\textwidth}
         \begin{center}
    \includegraphics[width=\textwidth]{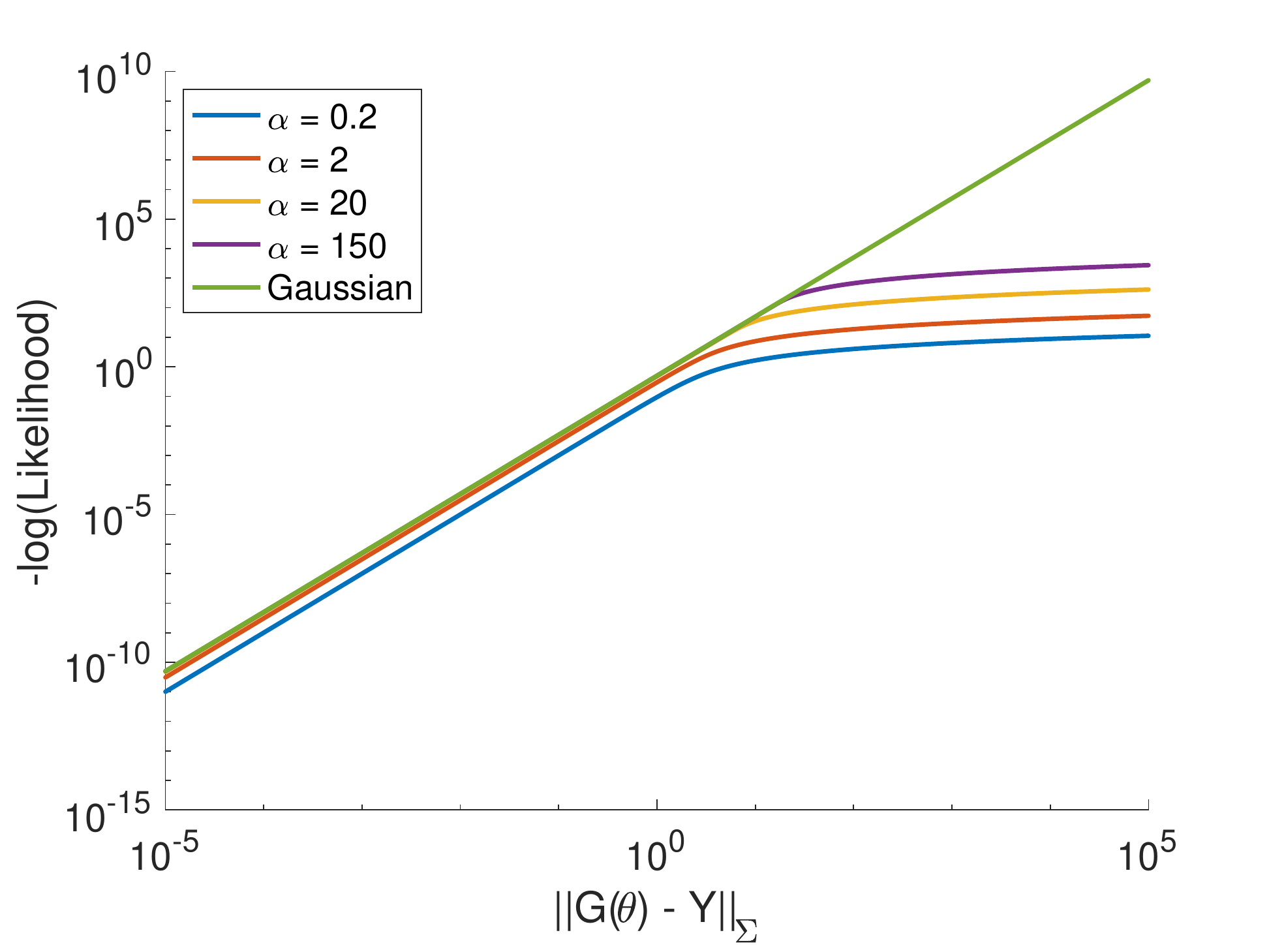}\caption{Standard and data selection likelihoods. }\label{fig:FLb}
    \end{center}
    \end{subfigure}
    \caption{Plots to show the effect of the prior specification of the fidelity parameters on the data selection likelihood function. (a) Plots of fidelity priors; truncated Gamma priors with \ensuremath{\beta = 2}, and a range of values of \ensuremath{\alpha}. (b) Plots of a univariate Gaussian likelihood, and corresponding data fidelity likelihoods with different fidelity priors, as functions of \ensuremath{\|\mathcal{G}(\btheta) - Y\|_\Sigma}.}
    \label{fig:FL}
\end{figure}

Figure \ref{fig:FL} shows how the resulting likelihoods differ from the original Gaussian likelihood for a single observation. All the likelihood functions are scaled such that the likelihood is equal to $1$ when $\|\mathcal{G}(\btheta) - Y\|_\Sigma = 0$. A range of hyperparameters are used with varying means of the fidelity priors, which are shown in Figure \ref{fig:FLa}. The resulting data fidelity likelihoods for a single observation are plotted in Figure \ref{fig:FLb}, as a function of $\|\mathcal{G}(\btheta) - Y\|_\Sigma$, alongside a Gaussian likelihood for comparison. The data fidelity likelihoods converge asymptotically to the Gaussian case as $\|\mathcal{G}(\btheta) - Y\|_\Sigma \to 0$. However, as the value of $\|\mathcal{G}(\btheta) - Y\|_\Sigma$ increases, the likelihood becomes less sensitive to changes in $\|\mathcal{G}(\btheta) - Y\|_\Sigma$, with the gradients levelling off. The different fidelity priors lead to a shift in the value of $\|\mathcal{G}(\btheta) - Y\|_\Sigma$ at which this effect becomes more pronounced. The fidelity likelihood converges to the Gaussian case as $\alpha \to \infty$, which corresponds to convergence to a Dirac on $\tau = 1$.

This behaviour is consistent with our aims, since if an observation cannot be well-matched to the model, we wish the likelihood to be less sensitive to the data-model mismatch for this observation, focussing instead on reducing the mismatch for observations which are more consistent with the model.

\subsubsection{Independent sampling of the fidelity parameters}\label{sec:ind}
In this section, we have shown how the fidelity parameters can be marginalised to leave a posterior density defined only on the model parameters. This is very helpful in the context of using MCMC methods to explore these distributions, since there are strong correlations between the model parameters and the fidelity parameters. Marginalising these extra parameters means that we can explore the model parameter space much more easily. However, as we will discuss in Section \ref{sec:interpret}, we may wish to recover the marginal densities on the fidelity parameters. This can be done very easily since the density of fidelity parameter $\tau_i$ conditioned on the value of the model parameters is dependent only on the current data-model mismatch for the $i$th observation. This density is given by
  \begin{equation}\label{eq:cond}
 \pi(\tau_i | \btheta) \propto  \mathbbm{1}_{[0,1]^N}(\bs{\tau}) \prod_{i=1}^N
  \tau_i^{\alpha - 1 + d/2} \exp \left ( -\tau_i \left (\beta
  +\frac{1}{2} (\mathcal{G}_i(\btheta) - y_i) ^\top \Sigma^{-1} (\mathcal{G}_i(\btheta) - y_i) \right ) \right ).
\end{equation}
This is simply a product of gamma distributions truncated to the unit interval with shape $\hat{\alpha}_i = \alpha + d/2$ and rate $ \hat{\beta}_i = \beta
  +\frac{1}{2} (\mathcal{G}_i(\btheta) - y_i) ^\top \Sigma^{-1} (\mathcal{G}_i(\btheta) - y_i)$. Sampling directly from these distributions is straightforward. Therefore we can implement a Metropolis-within-Gibbs (MwG) method which targets the marginal density on the model parameters given by \eqref{eq:fid}, and at each iteration  samples of the fidelity parameters are drawn independently from the conditional distribution \eqref{eq:cond}. This gives us a method that samples effectively from the joint density without any issues of strong correlations leading to slow mixing.

  \subsection{Learning the regions of high model fidelity with Gaussian process priors} \label{sec:learn}
So far we have considered the fidelity parameters $\tau_i$ to be independent random variables. However, in many contexts, as we will explore in Section \ref{sec:ODE}, we are aware that our model will only be a valid representation of reality for a subset of the observation space. Our aim in this scenario is twofold, in that we wish to identify that region, and then conduct inference using that subset of the data. This calls for a different approach, which leads us to consider the idea that there is an underlying scalar field on the observation space $\cO$, which can describe our belief that our model is able to represent our observations. This will naturally lead to priors on the $\tau_i$ which incorporate spatial correlations.

A natural choice for this is to consider a Gaussian process prior for the fidelity field $F:\cO \subset \mathbb{R}^D \to [0,1]$. In particular, we will deploy a logit-GP prior, where the logit function is applied element-wise:
\begin{equation}
\logit(F) := \begin{pmatrix} \logit(F_1) \\ \vdots  \\ \logit(F_D) \end{pmatrix} \sim \GP(m(\cdot), K(\cdot, \cdot)).
\end{equation}
Many different choices can be made regarding the mean function and kernel, which could be informed by information we may have about the limitations of the chosen model, but comparing the effects of these choices is beyond the scope of this work. Therefore we will keep things simple, choosing a constant function for $m(x) = m \in \mathbb{R}$, and a squared exponential kernel.

Since our likelihood is only affected by the values of this fidelity field at the positions in observation space that we have made observations, this prior simplifies to a multivariate logit-normal, where again the logit function is applied element-wise:
\begin{equation}
    \logit(\bs{\tau}) \sim \mathcal{N}\left ( m \mathbbm{1}_N, \Sigma^\tau \right ),
\end{equation}
where 
\[
{\Sigma^\tau}_{i,j} = K(x_i, x_j) = \sigma_\tau^2 \exp\left (-\frac{1}{2l^2}\|x_i - x_j\|_2^2 \right ),
\]
and where $\sigma_\tau, l>0$ are parameters to be chosen, and $\mathbbm{1}_N \in \mathbb{R}^N$ denotes a vector of ones. Since $l$ is meant to be a typical lengthscale, it makes sense to set it to be given by $l = \min_{i,j} \|x_i - x_j \|_2$.

This choice of the kernel ensures that the underlying field is smooth. This means that only those points in observation space where the model and data are a good fit in a neighbourhood of lengthscale $l$ are likely to have a high posterior mean fidelity. In particular, this prevents single observations where the model and data are coincidentally similar from affecting the inference (see Section \ref{sec:ODE} for an example of this). This leads to a more robust identification of the good regions of observation space, and in turn a higher quality inference on the model parameters.

\section{Bayesian data selection for non-product form densities}\label{sec:NPF}
In Sections \ref{sec:pModel} and \ref{sec:fid} we consider Bayesian data selection for
product form likelihoods. However, there are many scenarios in which
the likelihood is not in this form, but for which we might still wish
to conduct Bayesian data selection. We consider an example where we
have a Bayesian inverse problem with data $y_i \in \mathbb{R}^d$ of the form
 \[y_i = \mathcal{G}(\btheta) + \eta_i, \qquad \mathcal{N}(0,\Sigma), \qquad i = 1, \ldots, N,\]
but where we do not know \emph{a priori} the value of the observational noise
covariance $\Sigma$. In this scenario, we can take a hierarchical
Bayes approach, letting $\Sigma$ itself be a random variable to be
inferred from the data. We can pick a conjugate prior for $\Sigma$,
for instance the inverse Wishart prior, with hyperparameters $\nu  \in
\mathbb{R}$ the degrees of freedom, and $\Psi \in \mathbb{R}^{d \times d}$
the scale matrix. For a Gaussian likelihood and using this prior, and
given a prior $\pi_0(\btheta)$ on the unknown parameters $\btheta$,
$\Sigma$ can be marginalised from the posterior to give the marginal density:
\[ \pi(\btheta|\bs{y}) \propto \pi_0(\btheta) |\Psi + X
  X^\top|^{-(\nu + N)/2},\]
where $X = [X_1 = \bs{y}_1 - \mathcal{G}_1(\btheta), \ldots, X_N =
\bs{y}_N -
\mathcal{G}_N(\btheta)] \in \mathbb{R}^{d \times N}$ is the matrix of
all of the data-model mismatches. In this case, the target density has
lost its product form, and as such we cannot conduct data selection
either with the $p$-model, or by inferring the value of parameters
which alter the noise covariance for individual observations, as in
the previous two sections.

Instead, we take a different approach, by introducing fidelity parameters $\gamma_i \in [0,1]$ for each observation,  that controls the relative contribution of that observation to the likelihood. The scaled
mismatch for the $i$th observation is given by
\[\tilde{X}_i = \gamma_i (\bs{y}_i - \mathcal{G}_i(\btheta)).\] In effect, these $\gamma_i$ are  inverse annealing temperatures for each observation, with small values of $\gamma_i$ resulting in a likelihood which is not sensitive to the data-model mismatch for this observation. The aim is then to infer, as in Sections \ref{sec:pModel} and \ref{sec:fid}, the values of these parameters alongside the model parameters. 

Using these scaled mismatches, we arrive at the new posterior predictive
conditioned on $\Gamma = {\rm diag} (\gamma_1, \ldots, \gamma_N) \in \mathbb{R}^{N
  \times N}$ given by:
\[
L(\bs{y}|\btheta,\Gamma) = \frac{1}{Z(\Gamma)} |\Psi + (X \Gamma)
  (X \Gamma)^\top|^{-(\nu + N)/2},
  \]
 where 
\[Z(\Gamma) = \int |\Psi + (X \Gamma)
  (X \Gamma)^\top|^{-(\nu + N)/2} d X\]
is the normalisation constant which depends on $\Gamma$. We can
compute this constant by considering the substitution $Y = (X
\Gamma)$, which is equivalent to
\[y_{ij} = \gamma_i x_{ij}, \quad \forall i,j.\]
Therefore the Jacobian of the transformation from $X$ to $Y$ is
diagonal of the form
\[D_J = \begin{pmatrix} \gamma_1 I_d & & & \\
    & \gamma_2 I_d & & \\
    & & \ddots & \\
    & & & \gamma_N I_d \end{pmatrix},\]
with determinant
\[|{\rm det}(D_J)| = \prod_{i=1}^N \gamma_i^d.\]
Therefore,
\begin{eqnarray}
Z(\Gamma) &=& \int_{\mathbb{R}^{d\times N}} |\Psi + (X \Gamma)
(X \Gamma)^\top|^{-(\nu + N)/2} d X\\
&=& \left ( \prod_{i=1}^N \gamma_i^{-d} \right )\int_{\mathbb{R}^{d\times N}} |\Psi + Y
  Y^\top|^{-(\nu + N)/2} d Y\\
  &=& \left (\prod_{i=1}^N \gamma_i^{-d}  \right )\frac {\pi^{dN/2} \Gamma_d\left
      (\frac{\nu}{2} \right )}{|\Psi|^{\nu/2}\Gamma_d \left (
      \frac{\nu + N}{2} \right )}
  \end{eqnarray}

  Therefore, substituting this into the expression for the likelihood
  and dropping constants, we arrive at
  \begin{equation}
L(\bs{y}|\btheta,\Gamma) \propto \left (\prod_{i=1}^N
  \gamma_i^{d}\right ) |\Psi + (X \Gamma)
  (X \Gamma)^\top|^{-(\nu + N)/2},
    \end{equation}

    Lastly, we can invoke Bayes' to arrive at the joint posterior
    density:
  \begin{equation}\label{eq:NPF}
\pi(\btheta,\Gamma|\b{y}) \propto \pi_{0}(\btheta) \pi_0(\Gamma)
\left (\prod_{i=1}^N
  \gamma_i^{d}\right ) |\Psi + (X \Gamma)
  (X \Gamma)^\top|^{-(\nu + N)/2}.
\end{equation}

This is the approach that we took in \cite{forsyth2022unlabelled} where the aim was to match two unlabelled point clouds which represented the cell centers of an embryo taken from two different images, possibly using different image modalities. In this context it is difficult to estimate the observational noise covariance, and so it is natural to adopt a hierarchical approach, applying a prior to $\Sigma$ as we have in this section.

\section{Interpretation of Bayesian data selection posterior distributions} \label{sec:interpret}

In this section we discuss how the posterior distributions arising from the data selection methods that we have presented can be interpreted. We can either view the posterior density marginals on the model parameters as our overall target for characterisation, or we can view the marginal distributions of the fidelity parameters as our primary concern. Both could be valid interpretations, but are likely best employed in different scenarios.

As we can see from Figure \ref{fig:FL}, the fidelity likelihood function is approximately proportional to the standard likelihood function until the data-model mismatch reaches a threshold. If the majority of the data-model mismatches w.r.t. the posterior density are below this threshold, then the likelihood function is largely unaffected. Similarly, if there are only a few larger mismatches, the fidelity likelihood damps the effect of these observations on the posterior, preventing corruption, and leading to posterior distributions concentrated closer to the ground truth.

However, the situation is different if there are large data-model mismatches in significant regions of the observation space. Observations on the boundary between regions for which the model is a good fit may also cause a shift in the posterior modes. However, the marginal distributions on the fidelity parameters provide significant information about the regions in observation space for which the model can be a good match to the data; which data we should use in our inference, and for what regions of observation space is the model a valid approximation of the truth. This idea is central to our philosopy; our aim is not to fit data to a model, which would not provide any insight of note, but to identify regions in which the model is a good fit, and to identify the model parameters which provide that fit in each region. There is no guarantee that the data selection methods in Sections \ref{sec:fid}-\ref{sec:NPF} sufficiently mitigate for the effect of large model-data discrepancies, but they are effective at identifying regions of strong agreement between model and data, especially when employing the logit-GP prior approach presented in Section \ref{sec:learn}.

In this scenario, one can set a threshold, for example, for the inclusion of data in a second stage of inference, using the marginal distributions on the fidelity parameters. In particular we might adopt this approach when using the logit-GP prior for the fidelity parameters, which aids us in identifying regions where the model and data are consistently in agreement. From the point of identification and trunctation of the data to this region, inference on the model parameters with or without data selection should yield a very similar marginal distribution on the model parameters. We will see this later in the example in Section \ref{sec:usingLGP}.

\section{Application to Bayesian linear regression} \label{sec:lin}
We consider a linear regression problem, where we wish to fit a linear
model to some noisily observed inputs $\{x_i\}_{i=1}^N$ and
corresponding response $\{y_i\}_{i=1}^N$. A linear model is the simplest possible model, but it is still a method which is used ubiquitously, in particular in many statistical applications. We choose to study it here since we are able to compute the data selection posteriors using the $p$ model for this scenario, as long as the number of observations does not grow too big. This enables us to compare the different approaches to Bayesian data selection that we have presented in Sections \ref{sec:pModel}-\ref{sec:NPF}. 

Our aim in a one-dimensional linear regression is to infer the values
of $a,b \in \mathbb{R}$ which define the linear model $y = ax + b$,
conditioned on the observations. We will start by considering this
problem without data selection.

\subsection{Without data selection}
For simplicity we assume that the
observational noise is mean zero i.i.d. Gaussian  with known covariance, such that
\begin{equation} y_i = \mathcal{G}_i(a,b) + \eta = ax_i + b + \eta_i,
  \quad \eta_i \sim \mathcal{N}(0,\sigma^2).\end{equation}
This can be rewritten in vector form as
\begin{equation} \b{y} = \mathcal{G}(a,b) + \bs{\eta}  = A \begin{pmatrix} a \\ b \end{pmatrix} + \bs{\eta}, \qquad
  \bs{\eta} \sim \mathcal{N}(0, \sigma^2 I),\end{equation}
where
\begin{equation}
A = \begin{pmatrix} x_1 & 1 \\ x_2 & 1 \\ \vdots & \vdots \\ x_N & 1 \end{pmatrix}.
\end{equation}
By linearity of the observation operator $\mathcal{G}$ in the
unknowns, the likelihood is Gaussian. Further, if we pick Gaussian
priors for $\btheta = [a,b]^\top$ such that $\btheta \sim
\mathcal{N}(\mu_0, \Sigma_0)$, and assuming all the $x_i$ are
distinct, then the posterior is Gaussian with
covariance:
\begin{equation}\Sigma_p = \left (\Sigma_0^{-1} + \sigma^{-2} (A^\top A) \right
  )^{-1} \end{equation}
and mean
\begin{equation}
\mu_p = \Sigma_p \left ( \Sigma_0^{-1}\mu_0 + \sigma^{-2} A^\top \b{y}
\right ).
\end{equation}

\subsection{The $p$ model}
We now follow the approach given in Section \ref{sec:pModel}. Firstly
we need to consider the $2^N$ possible likelihoods which arise from
the different subsets of the observations being considered. Consider
the likelihood $L_{\bs{\iota}}$ with $\bs{\iota} \in \{0,1\}^N$. The data
are in the same format, but we now can only include $y_i$ in the likelihood expression if $\iota_i
= 1$. We define
$\mathcal{I}^{\bs{\iota}} \in \mathbb{R}^{n_{\bs{\iota}}}$ where
$n_{\bs{\iota}} = \sum_{i=1}^N \iota_i$ by
\begin{equation}
\mathcal{I}^{\bs{\iota}}_i = \min \left \{j \left | \sum_{k=1}^j \right . \iota_k = i
\right \}.
  \end{equation}
That is, $\mathcal{I}^{\bs{\iota}} $ is the ordered vector containing
all indices $i$ for which $\iota_i = 1$.

Our
subset of the data can then be described by
\begin{equation}
\b{y}_{\bs{\iota}} = M_{\bs{\iota}} \b{y},
\end{equation}
where
\begin{equation}
  M_{\bs{\iota}} = \begin{pmatrix} e_{\mathcal{I}_1}^\top \\ 
    \\ \vdots \\ e_{\mathcal{I}_n}^\top \end{pmatrix},
  \end{equation}
  where $e_i \in \mathbb{R}^N$ is the $i$th canonical basis vector,
  and where we have dropped the dependencies on $\bs{\iota}$ for ease
  of presentation.

  Then our observation operator restricted to the observations with
  indices in $\mathcal{I}^{\bs{\iota}}$ is given by:
  \begin{equation}
\mathcal{G}_{\bs{\iota}} (a,b) = M_{\bs{\iota}} A \begin{pmatrix} a \\
  b \end{pmatrix}.
    \end{equation}
Our restricted observations are then given by
\begin{equation} \b{y}_{\bs{\iota}} = \mathcal{G}_{\bs{\iota}} (a,b) + \bs{\eta}_{\bs{\iota}}, \qquad
  \bs{\eta}_{\bs{\iota}} \sim \mathcal{N}(0, \sigma^2 I_{n _{\bs{\iota}}}).
\end{equation}
Following the same arguments as in the last section, this leads to a
Gaussian posterior with covariance matrix:
\begin{equation}\Sigma _{\bs{\iota}} = \left (\Sigma_0^{-1} +
      \sigma^{-2} (A^\top M _{\bs{\iota}}^\top M _{\bs{\iota}} A) \right
  )^{-1} \end{equation}
and mean
\begin{equation}
\mu _{\bs{\iota}} = \Sigma _{\bs{\iota}} \left ( \Sigma_0^{-1}\mu_0 + \sigma^{-2} A^\top M
  _{\bs{\iota}}^\top \b{y}
\right ).
\end{equation}

Finally, following the result in Section \ref{sec:marge}, we arrive at
the marginal density on $\btheta$, given by the following mixture
of multivariate Gaussian densities:
\begin{equation}\label{eq:LR:pmodel}\pi(\btheta)_p = 
    \sum_{\bs{\iota} \in \{0,1\}^N} C_{\bs{\iota}}\frac{1}{ \sqrt{\det\left (2 \pi
          \Sigma_{\bs{\iota}} \right )}}\exp \left(-\frac{1}{2} (\btheta
      -\mu_{\bs{\iota}})^\top \Sigma_{\bs{\iota}} ^{-1} (\btheta
      -\mu_{\bs{\iota}} ) \right ),
  \end{equation}
  where
\begin{equation}
C_{\bs{\iota}} = \frac{B(\alpha +
                                     1, \beta)^{n_{\bs{\iota}}}
                                         B(\alpha, \beta + 1)^{N -
                                         n_{\bs{\iota}}}}{B(\alpha,\beta)^N}.
  \end{equation}
  \subsection{Data selection with conjugacy of fidelity parameter
    priors}
 We now follow the approach of Section \ref{sec:fid}. We rewrite
 Equation \eqref{eq:fid} for this specific problem to arrive at:
\begin{eqnarray}\label{eq:LR:fid}
\pi(\btheta)_\tau \propto \pi_0(\btheta) \prod_{i=1}^N
  \left (   \beta
+\frac{1}{2 \sigma^2} (ax_i + b - y_i)^2    \right
                           )^{- \alpha - 1/2}\gamma
                                                               \left ( \alpha + 1/2,
                                                              \beta
+\frac{1}{2 \sigma^2} (ax_i + b - y_i)^2  \right ).
\end{eqnarray}

\subsection{Data selection for non-product form likelihoods}
We now follow the approach of Section \ref{sec:NPF}. Additionally to
the problems addressed in previous subsections, we no longer assume that we know the
value of $\sigma^2$. Since we are working with observations in only one dimension, the
inverse wishart prior of the observational noise covariance is
equivalent to an inverse gamma prior with $\alpha = \nu/2$ and
$\beta = \Psi/2$. We rewrite Equation \eqref{eq:NPF} for our specific
purposes here to arrive at:
  \begin{equation}\label{eq:NPF:fid}
\pi(\btheta,\Gamma|\b{y})_\gamma \propto \pi_{0}(\btheta) 
\left (\prod_{i=1}^N \pi_0(\gamma_i)
  \gamma_i^{d}\right )  \left (\Psi + \sum_{i=1}^N \gamma_i^2 (ax_i +
  b - y_i)^2 \right )^{-(\nu + N)/2}.
\end{equation}

\subsection{Numerical examples}
We will consider three different examples of one dimensional linear
regression using the three different approaches to Bayesian inference
with data selection
that we have described in Sections \ref{sec:pModel}-\ref{sec:NPF}, in comparison with standard Bayesian inference.
\subsubsection{Set up: priors and parameters}
In what follows, we will use the same
parameters and priors for each of the approaches. Namely, for all
approaches, we will assume a Gaussian prior on the vector of unknowns
for the linear regression $[a,b]^\top$ equal to
$\mathcal{N}([0,0]^\top, 100^2 I_2 )$. For the $p$-model approach with
marginal posterior given by \eqref{eq:LR:pmodel}, we
will use a prior on each $p_i$ equal to ${\rm Beta}(2, 50)$. For
the data fidelity approach with conjugate priors with marginal
posterior given by \eqref{eq:LR:fid}, we will use gamma priors
truncated to $[0,1]$ for the fidelity parameters $\tau_i$ with $\alpha
= 2$, $\beta = 2$. The same prior is chosen for the fidelity
parameters $\gamma_i$ for the approach for non-product form
likelihoods with joint posterior density given by
\eqref{eq:LR:fid}. Lastly, for the the non-product form likelihood, the chosen inverse wishart prior was given with hyperparameters $\nu = 100$, and $\Psi = 0.98$ (noting that this distribution is effectively an inverse gamma prior).

MCMC was not required to evaluate the posterior densities in
either the case of Bayesian linear regression without data selection,
or in the case of \eqref{eq:LR:pmodel}, since the posterior
distributions are given including normalisation constant. In the case
of the two data fidelity approaches, a standard random walk Metropolis
(RWM) algorithm was used for the unknown parameters $[a,b]^\top$, with
acceptance rates tuned close to the optimal 23.4\%~\cite{roberts2001optimal}. In the
case of the $\gamma_i \in [0,1]$, a RWM implementation is unwise since
the parameters only have support on the unit interval. Therefore for
the posterior given in \eqref{eq:NPF:fid} we implemented
MwG, using RWM for $[a,b]^\top$. For the
$\gamma_i$ we note that the function $T:[0,1] \to \mathbb{R}$
\[ T(\gamma_i)  = \log \left ( \frac{1}{\gamma_i} - 1 \right )\]
is bijective. We exploit this by producing proposals of the form
\[ \gamma_i' = T^{-1}(T(\gamma_i) + \beta \omega_i), \qquad \omega_i \sim
  \mathcal{N}(0,1),\]
which is essentially random walk on $T(\gamma_i)$ rather than $\gamma_i$.
The Metropolis-Hastings formula for sampling on the
$\gamma_i$ given this form of proposal leads to an acceptance probability given by
\begin{equation}a(\bs{\gamma}',\bs{\gamma}) = \min \left \{ 1,
    \frac{\pi(\bs{\gamma}') \prod_{i=1}^N (\gamma'_i -
      (\gamma'_i)^2)}{\pi(\bs{\gamma}) \prod_{i=1}^N (\gamma_i -
      \gamma_i^2)} \right \},\end{equation}
since
\begin{equation} \left|\frac{d}{dx} T^{-1}(x)\right | = \left |\frac{-\exp(x)}{(\exp(x) + 1)^2} \right| =
  T^{-1}(x) - (T^{-1}(x))^2,\end{equation}
  since $T^{-1}(x) \in [0,1]$.
Once again, the proposal variance was tuned to give acceptance rate averages reasonably close to
23.4\%. 

\subsubsection{Example 1: Model mismatch}\label{sec:ex1}
\begin{figure}[htp]
    \begin{center}
      \includegraphics[width=0.5\textwidth]{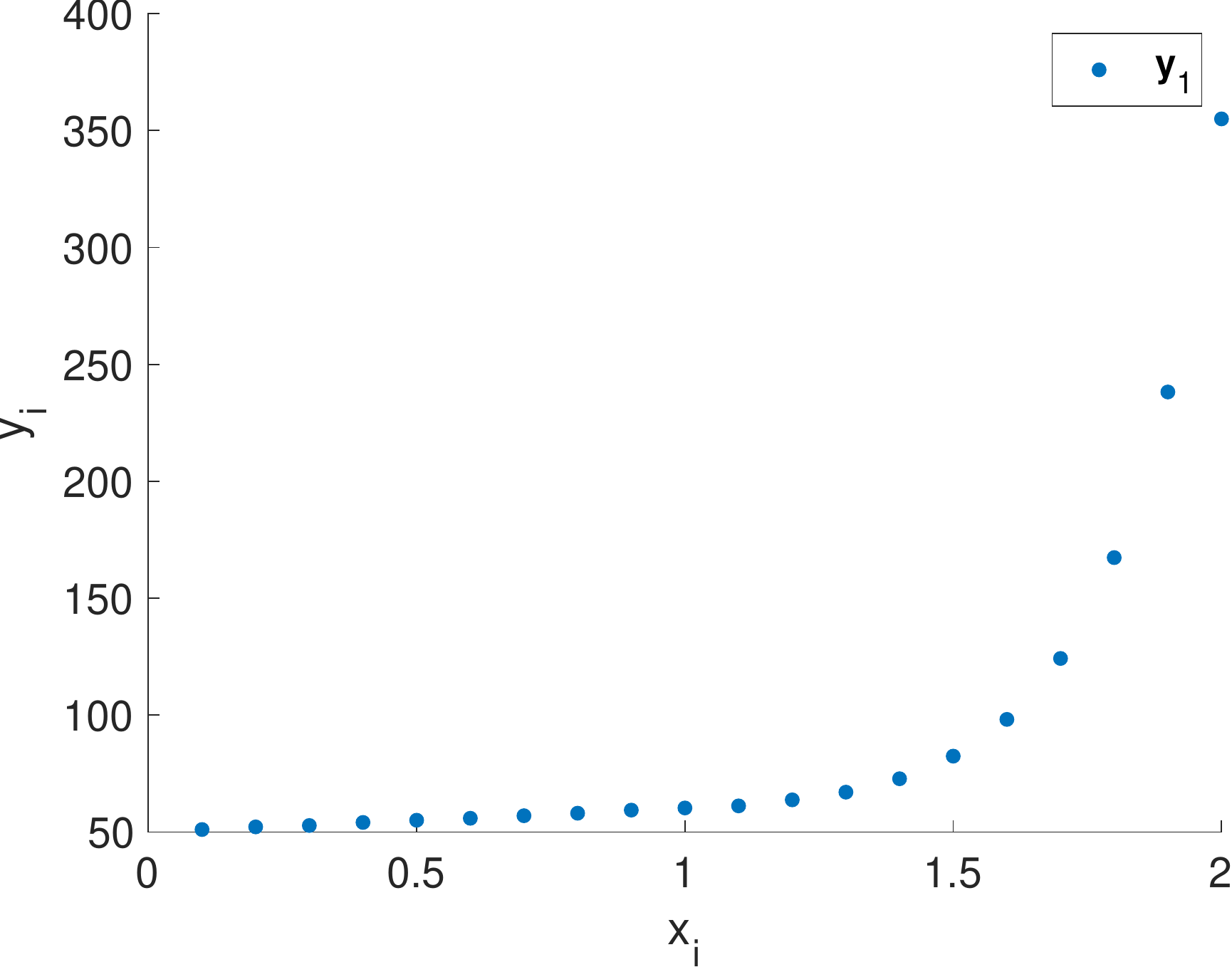}
      \caption{Plot to show data included in the vector $\b{y}_1$ on
        which we wish to conduct linear regression, displaying
        data-model mismatch. The parameters used were $a = 10$,
        $b=50$, $c=5$, $\sigma^2 = 0.01$.}\label{fig:y1_data}
    \end{center}
  \end{figure}

In the first example we consider the problem of conducting inference
when our model is only consistent with a portion of the data. We wish
to recover two main pieces of information: where is the model
consistent with data, and for which parameter values. We consider a
noisily observed function
\begin{equation}\label{eq:f1}
  f_1(x) = \begin{cases} ax + b & \qquad x \in [0,1)\\
    d \exp(cx) + e, & \qquad x \in [1,2].\end{cases}
  \end{equation}
  where, given $a, b, c \in \mathbb{R}$, we choose $d,e \in \mathbb{R}$ so that $f_1 \in C^1([0,2], \mathbb{R})$. We consider
  observations of this function given by $\b{y} = [y_1, \ldots
  y_{20}]^\top$ where
  \[ y_i = f(i/10) + \eta_i, \qquad \eta_i \sim \mathcal{N}(0,\sigma^2)
    \quad {\rm i.i.d.}, \qquad i = 1, \ldots, 20.\]
Figure \ref{fig:y1_data} shows a realisation of this data on which we will
then attempt to conduct Bayesian linear regression.
  
\begin{figure}[htp]
\begin{center}
    \begin{subfigure}[b]{0.48\textwidth}
         \begin{center}
         \includegraphics[width=\textwidth]{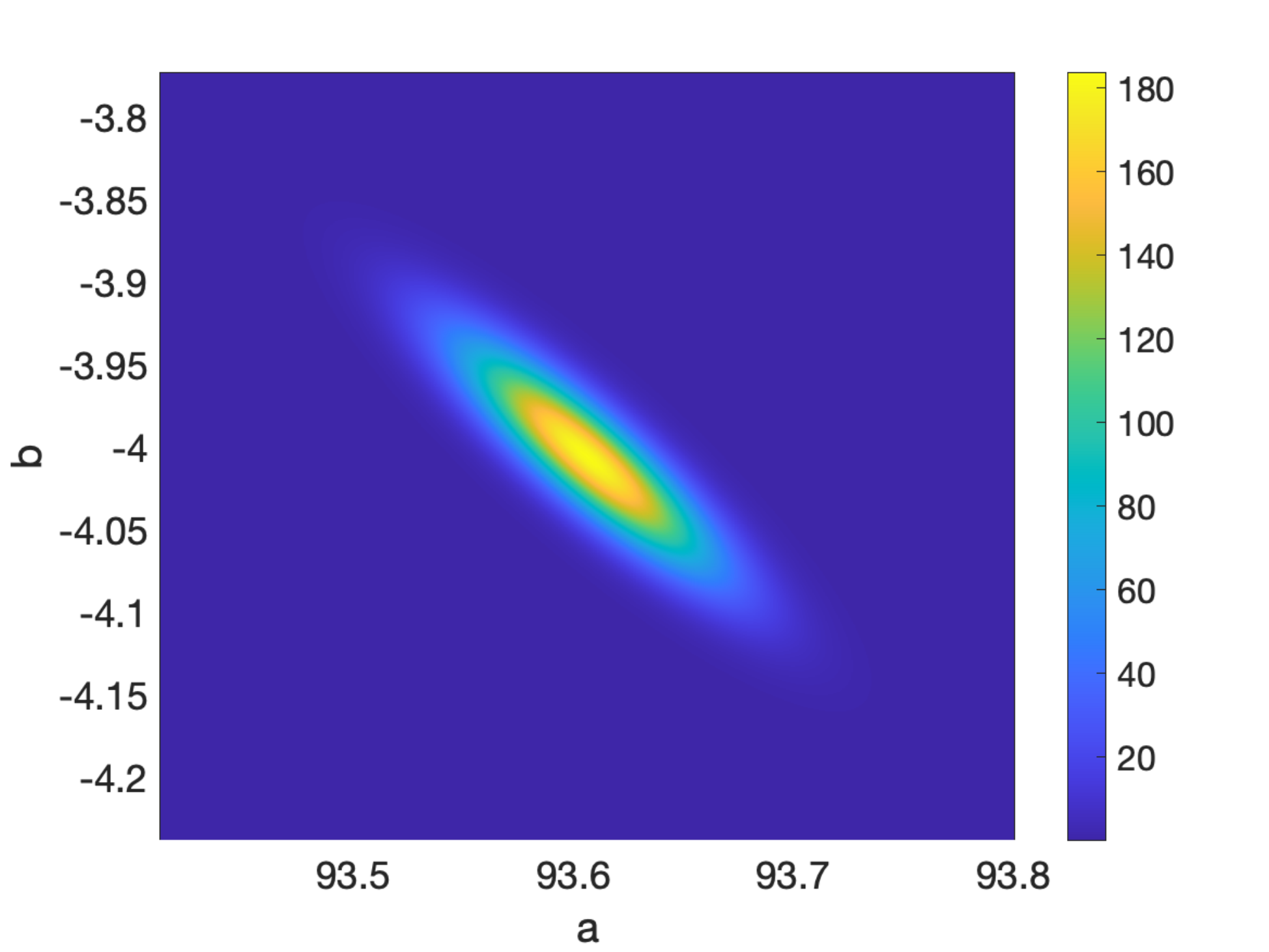}
         \caption{Standard Bayesian.}
         \label{fig:y1:SB}
         \end{center}
    \end{subfigure}
    \begin{subfigure}[b]{0.48\textwidth}
         \begin{center}
         \includegraphics[width=\textwidth]{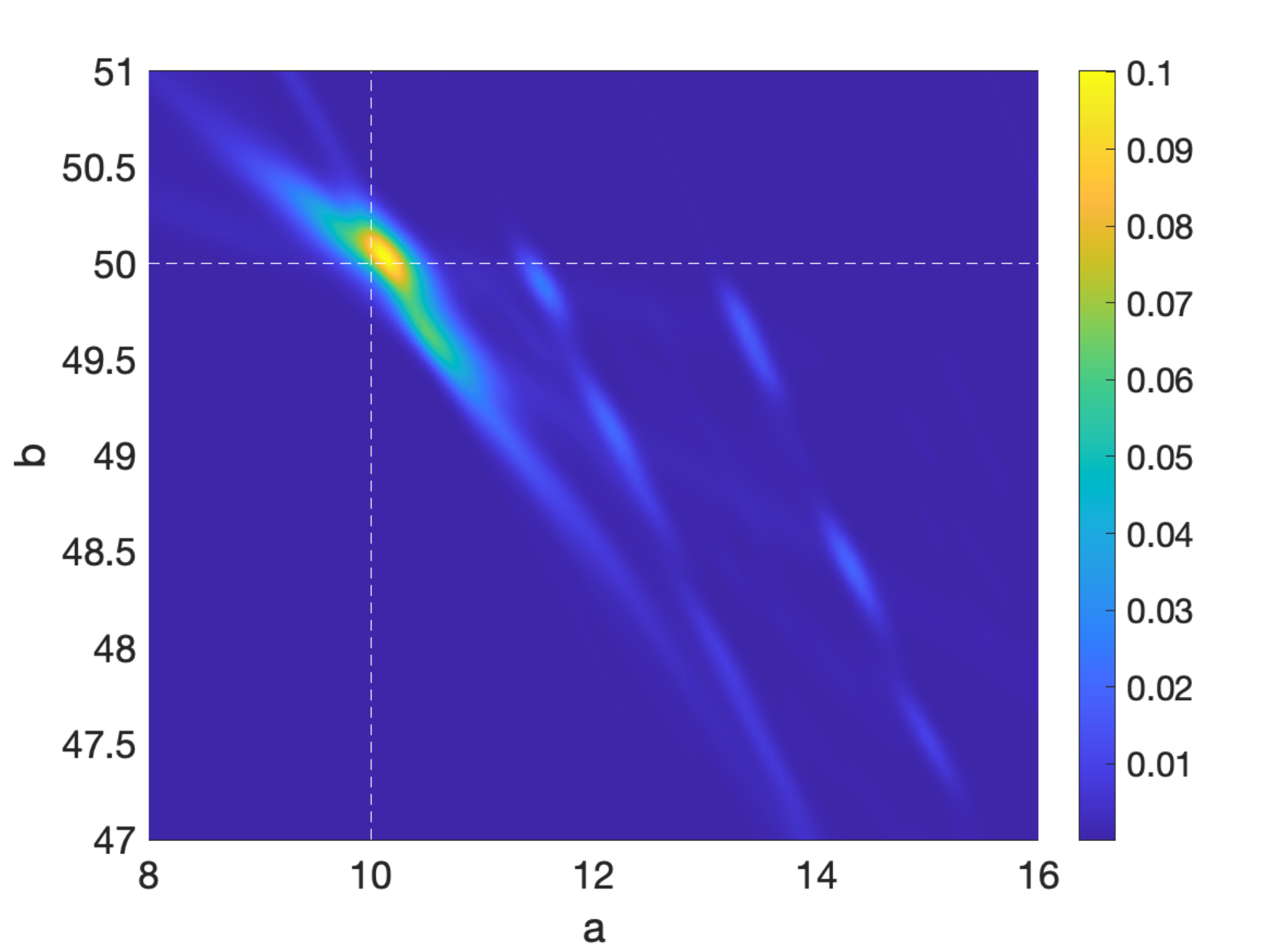}
         \caption{p-model.}
         \label{fig:y1:p}
         \end{center}
    \end{subfigure} \\
    \begin{subfigure}[b]{0.48\textwidth}
         \begin{center}
         \includegraphics[width=\textwidth]{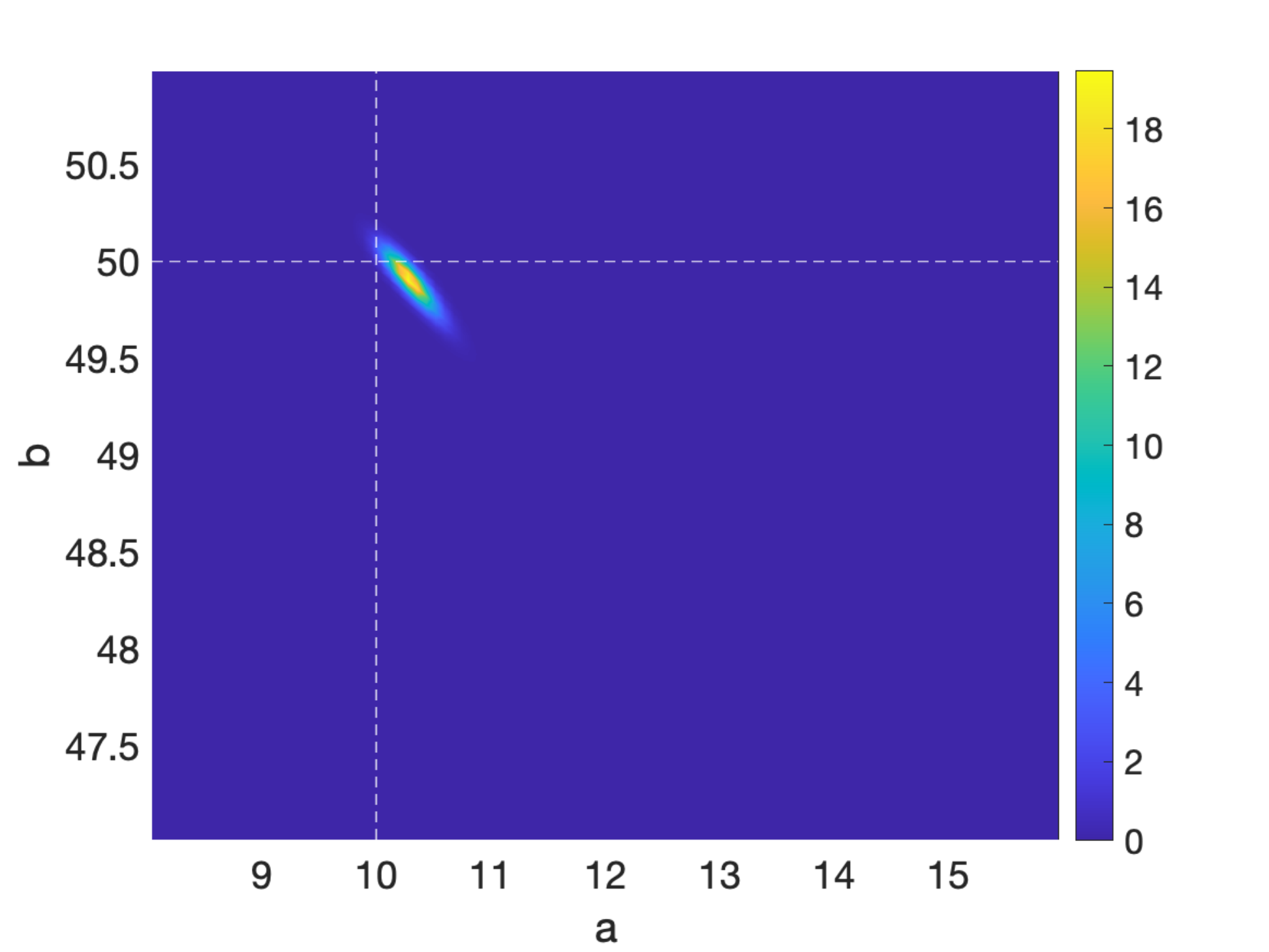}
         \caption{Data fidelity.}
         \label{fig:y1:DF}
         \end{center}
    \end{subfigure}
    \begin{subfigure}[b]{0.48\textwidth}
         \begin{center}
         \includegraphics[width=\textwidth]{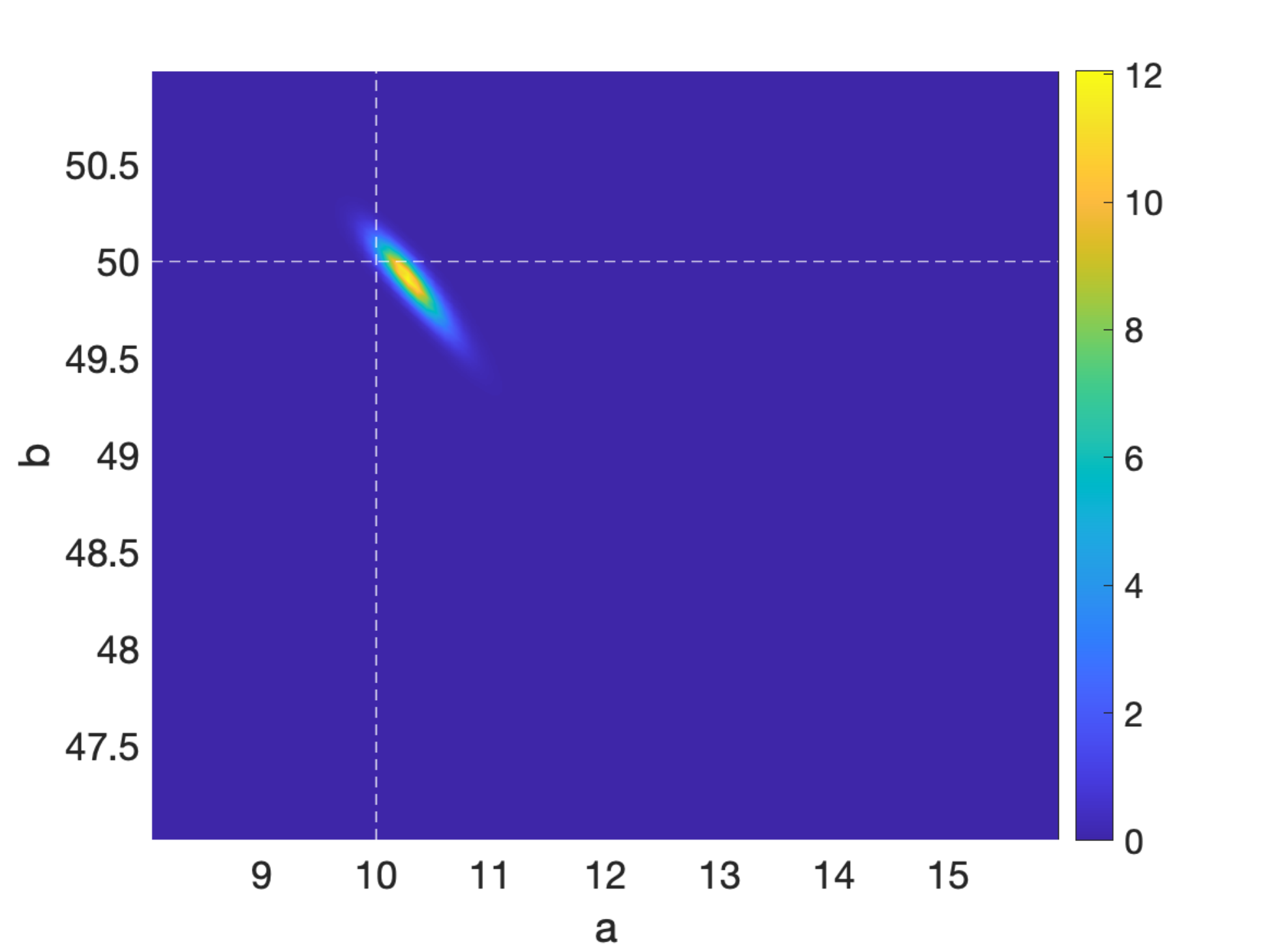}
         \caption{Non-product form.}
         \label{fig:y1:NPF}
         \end{center}
    \end{subfigure}
    \caption{The posterior densities arising from the data $\b{y}_1$, as given in Figure \ref{fig:y1_data}, using standard Bayesian inference, and the three approaches outlined in Sections \ref{sec:pModel}-\ref{sec:NPF}. Note that Figure \ref{fig:y1:SB} has substantially different axes to the other plots. The value of the parameters that created the linear data in the region $x<1$ were $a=10$, $b=50$, which are denoted by the white dashed lines.}
    \label{fig:y1}
\end{center}
\end{figure}
  
  Figure \ref{fig:y1} shows approximations of the posterior densities arising from the linear regression of this data using standard Bayesian inference, and the three approaches outlined in Sections \ref{sec:pModel}-\ref{sec:NPF}. The standard Bayesian linear regression's estimate of the parameters $a, b \in \mathbb{R}$ reflects the whole of the data. As such, the first thing to notice is that due to the mismatch between the data and the model in the region $x>1$, the estimates of $a$ and $b$ are very far away from the values that were used to create the data in the region $x<1$, so much so that we are forced to use different axes for Figure \ref{fig:y1:SB}. The three data selection approaches demonstrate some variability, but importantly the majority of the probability mass is located close to the value of the parameters that created the data in the region $x<1$. We note the eye-catching nature of the p-model posterior in Figure \ref{fig:y1:p}, which is a Gaussian mixture with $2^{20}$ components. In particular there are streaks at different angles in the posterior, representing different combinations of $a$ and $b$ which match well to certain subsets of the data. However, there is also a clear dominating mode in the density close to the value of the parameters that gave rise to the linear data in the region $x<1$. The non-product form posterior in Figure \ref{fig:y1:NPF} is slightly more diffuse than the data fidelity posterior in Figure \ref{fig:y1:DF}, which is due to the additional uncertainty in the data due to the inference of the value of the observational noise variance, and the differing influence of the fidelity priors in the product form and non-product form cases.
  
  \begin{table}[htp]
  \centering
\begin{tabular}{c|ccccc}
      & Standard & p-model & Data fidelity & Non-product form & ``True" value for $x<1$ \\ \hline
$a^*$ & 93.61    & 10.18   & 10.29         & 10.17            & 10                   \\
$b^*$ & -4.006   & 49.98   & 49.91         & 49.99            & 50                  
\end{tabular}
\caption{MAP estimates (to 4 s.f.) for the four Bayesian linear regression methodologies for data $\b{y}_1$ as shown in Figure \ref{fig:y1_data}}
\label{tab:y1:MAP}
\end{table}
  
  \begin{figure}[htp]
      \centering
      \begin{subfigure}[b]{0.48\textwidth}
      \includegraphics[width=\textwidth]{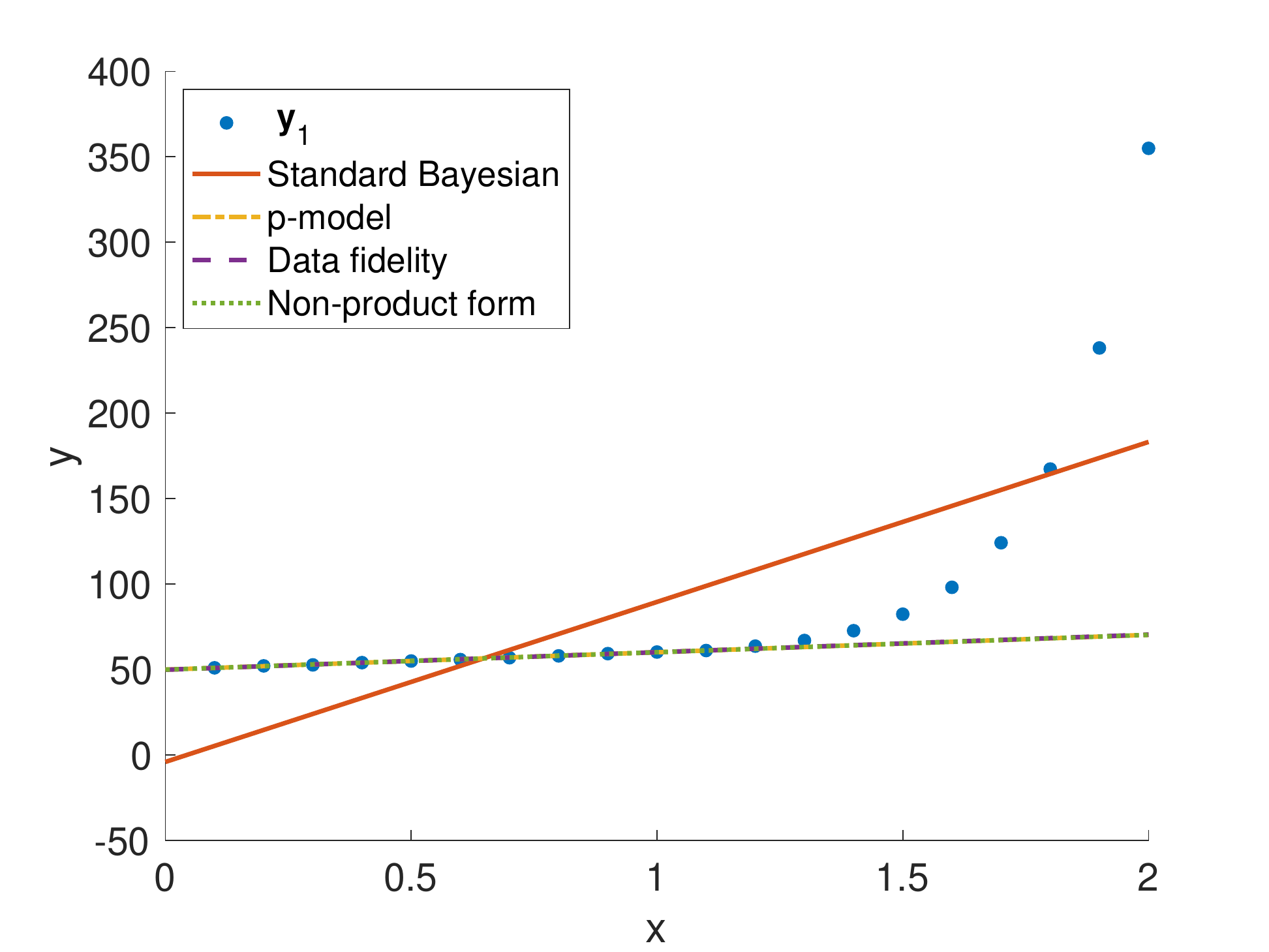}
      \caption{The MAP fits.}
      \label{fig:y1:MAPfits}
      \end{subfigure}
      \begin{subfigure}[b]{0.48\textwidth}
      \includegraphics[width=\textwidth]{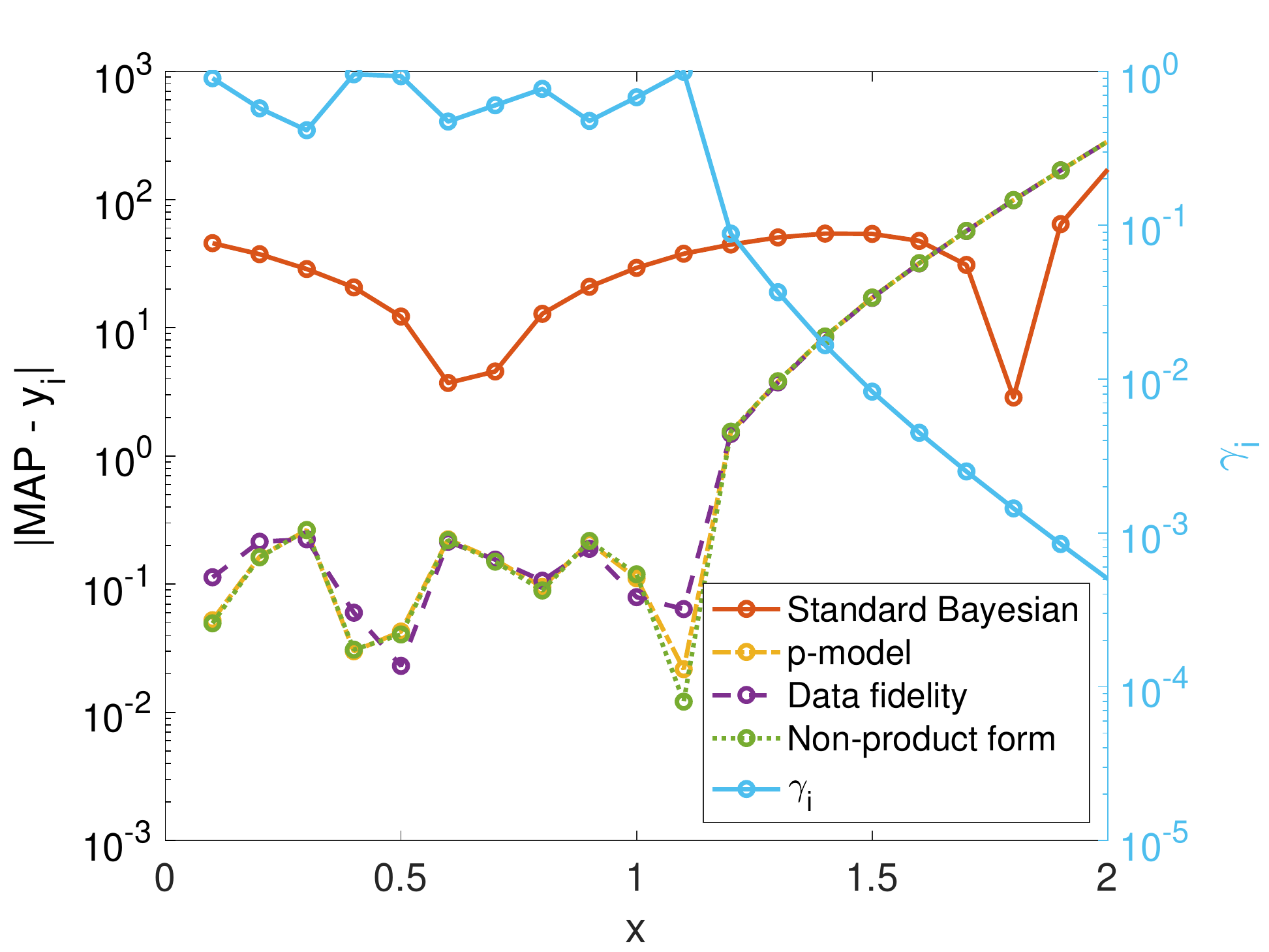}
      \caption{Data-model mismatches.}
      \label{fig:y1:MAPerrs}
      \end{subfigure}
      \caption{Plots to show the MAP fits of the four Bayesian linear regression approaches to the data $\b{y}_1$ as shown in Figure \ref{fig:y1_data}. (a) The MAP fits and the data. (b) The data-model mismatches of the MAP fits, and the MAP estimates of the $\gamma_i$ for the non-product form case.}
      \label{fig:y1:MAP}
  \end{figure}
  
  Figure \ref{fig:y1:MAP} shows the fits using MAP estimators of the four different Bayesian linear regression methodologies, as given in Table \ref{tab:y1:MAP}. These MAP estimates were calculated using a restarted BFGS implementation~\cite{fletcher2013practical}. Note that the non-product form example has posterior which is a joint density on the model parameters and the fidelity parameters $\{\gamma_i\}_{i=1}^{20}$; here the MAP shown is the values of the model parameters at the MAP of the joint density, rather than the MAP of the marginal density on the model parameters. The three data selection methods have MAP estimates which are very closely clustered, which fit to the data within $\mathcal{O}(\sigma)$ in the region $x<1$, where $\sigma = 0.1$ is the standard deviation of the observational noise. The fits to the exponential part of the data, as we would expect, are very poor. This is reflected in the MAP estimates of the fidelity parameters $\{\gamma_i\}_{i=1}^{20}$ in the non-product form case, which are plotted on the second axis (light blue) in Figure \ref{fig:y1:MAPerrs}, with the $\gamma_i$ decaying quickly for $i>10$. The standard Bayesian linear regression, in contrast, attempts to fit to all of the data with equal weight. This leads to a fit which is poor in all regions.
  
  This simple example clearly illustrates the advantages of Bayesian inference with data selection; the resulting distribution is concentrated close to parameter values which allow for very close matching of the model to the data in as big a subset of the data as possible. However, the impact on the inference of the data in the region which cannot be represented by the model is minimised.
  
  This example also demonstrates that the  approaches as given by the two data fidelity methodologies (for product and non-product form likelihoods) show relatively close agreement to the natural, but ultimately computationally intractable, $p$-model.
  
  \subsubsection{Example 2: Data corruption}
\begin{figure}[htp]
    \begin{center}
      \includegraphics[width=0.5\textwidth]{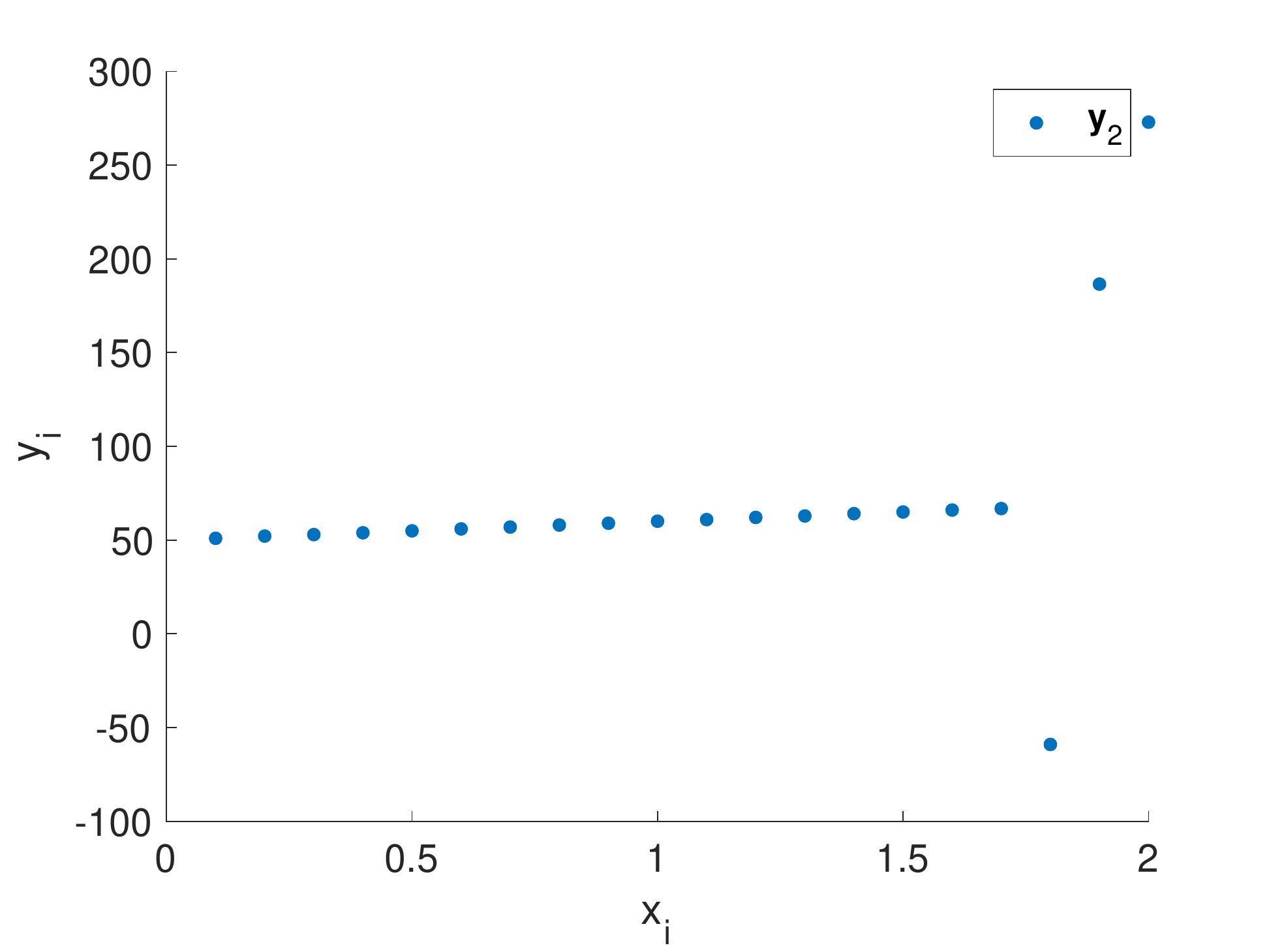}
      \caption{Plot to show data included in the vector $\b{y}_2$ on
        which we wish to conduct linear regression, displaying
        data-model mismatch. The parameters used were $a = 10$,
        $b=50$, $\sigma_1^2 = 0.1^2$, $\sigma_2^2 = 100^2$.}\label{fig:y3_data}
    \end{center}
  \end{figure}

  We consider a second example where a portion of the data is corrupted. We model this corruption by drastically increasing the observational noise variance for a subset of the data. Specifically, we consider the function
  \begin{equation}
      f_2(x) = ax + b \qquad x \in [0,1].
  \end{equation}
  We consider observations of this function given by $\b{y}_2 = [y_1, \ldots, y_{20}]^\top$ where
  \begin{equation}y_i = f_2(i/10) + \eta_i, \qquad i=1,\ldots, 20.\end{equation}
  The assumed statistical model in the inference is that $\eta_i \sim \mathcal{N}(0,\sigma_1^2)$, where $\sigma_1^2 = 0.1^2$ and the samples are i.i.d.. However, when creating the data $\b{y}_2$, as shown in Figure \ref{fig:y3_data}, we use the following approach:
  \begin{equation}
      \eta_i \sim \begin{cases} \mathcal{N}(0,\sigma_1^2 = 0.1^2), \qquad & i \leq 17,\\
      \mathcal{N}(0,\sigma_2^2 = 100^2), \qquad & i > 17.\end{cases}
  \end{equation}
Figure \ref{fig:y3_data} shows a realisation of this data on which we will
then attempt to conduct Bayesian linear regression.
  
\begin{figure}[htp]
\begin{center}
    \begin{subfigure}[b]{0.48\textwidth}
         \begin{center}
         \includegraphics[width=\textwidth]{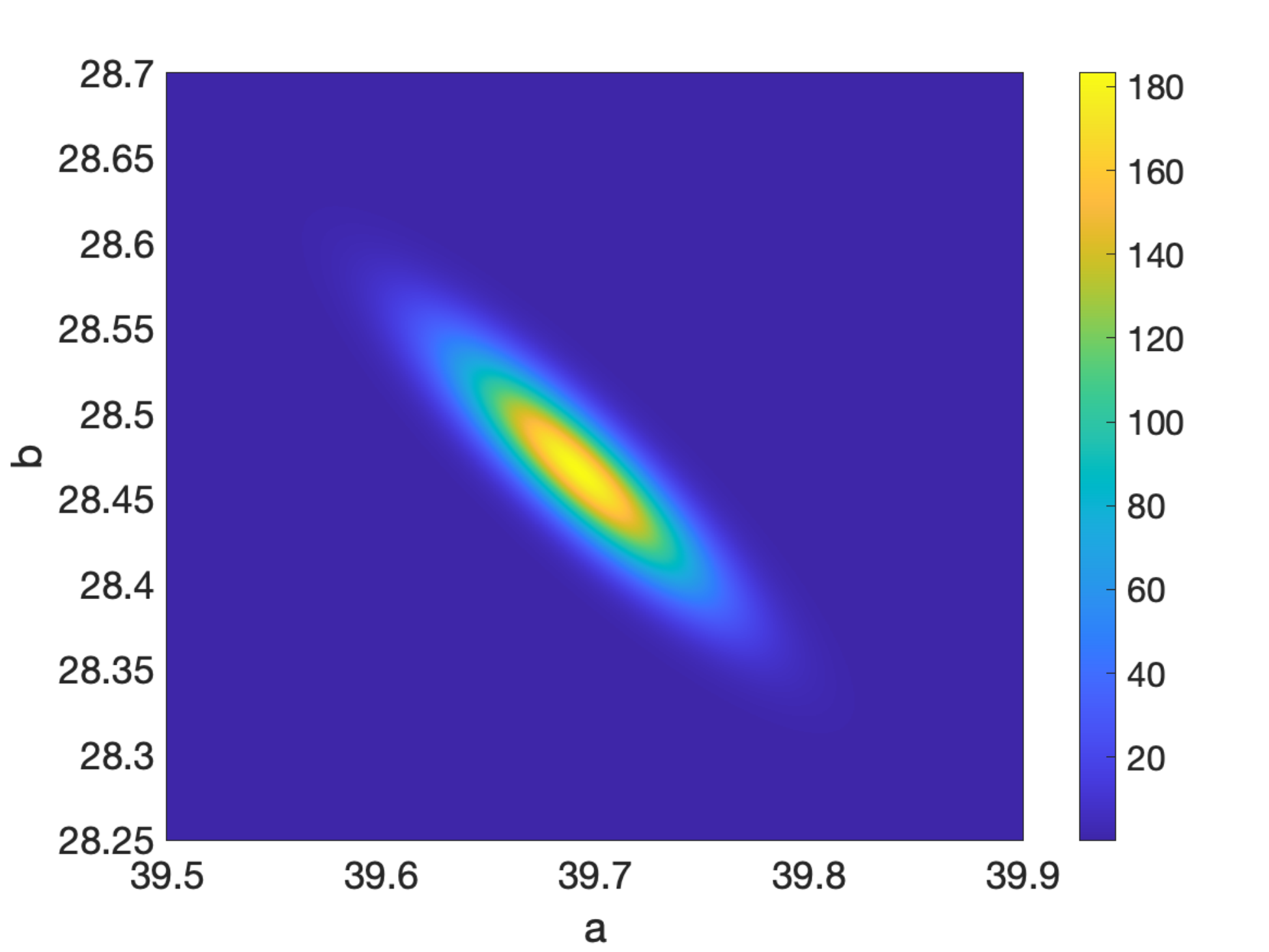}
         \caption{Standard Bayesian.}
         \label{fig:y3:SB}
         \end{center}
    \end{subfigure}
    \begin{subfigure}[b]{0.48\textwidth}
         \begin{center}
         \includegraphics[width=\textwidth]{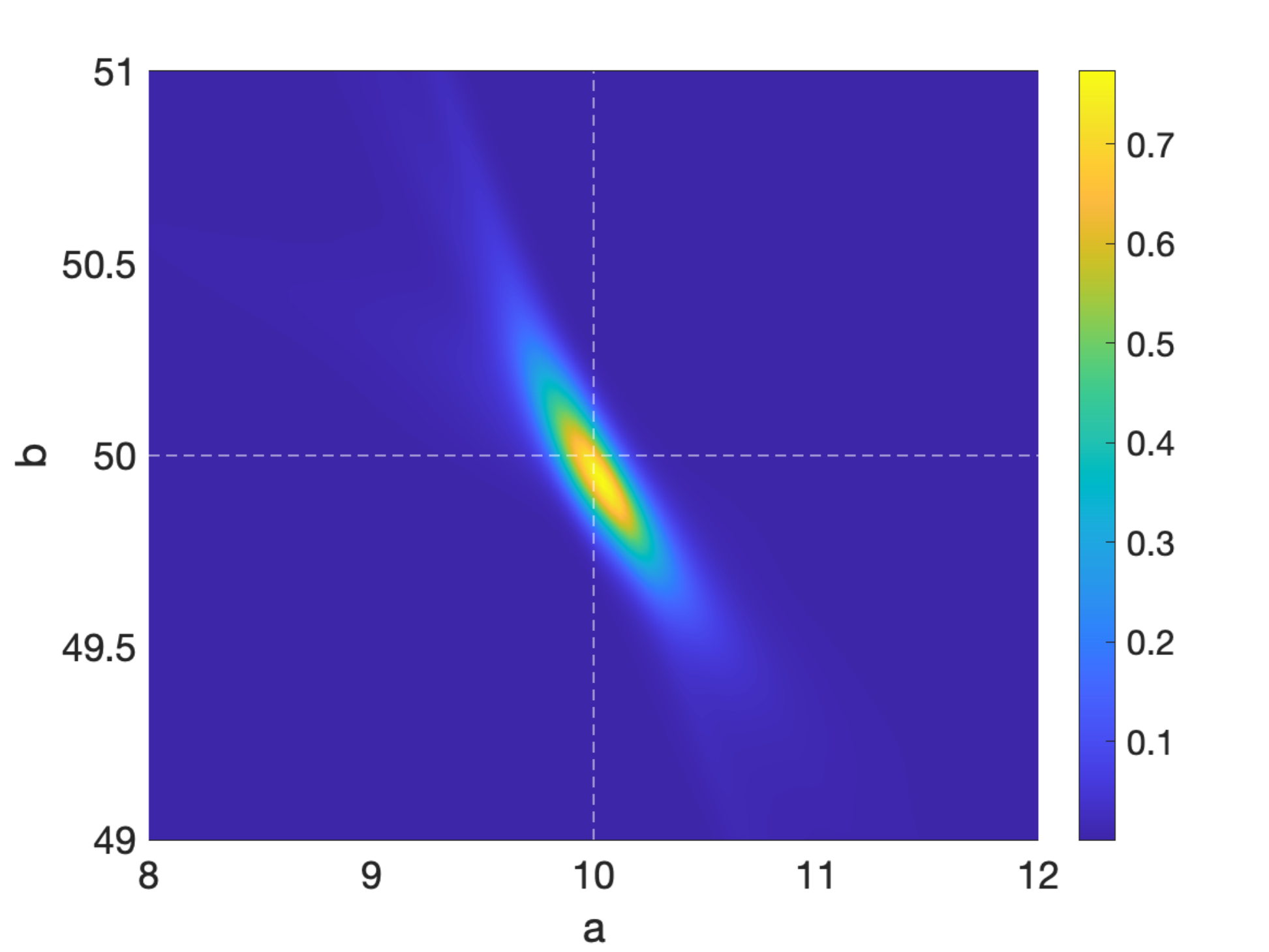}
         \caption{p-model.}
         \label{fig:y3:p}
         \end{center}
    \end{subfigure} \\
    \begin{subfigure}[b]{0.48\textwidth}
         \begin{center}
         \includegraphics[width=\textwidth]{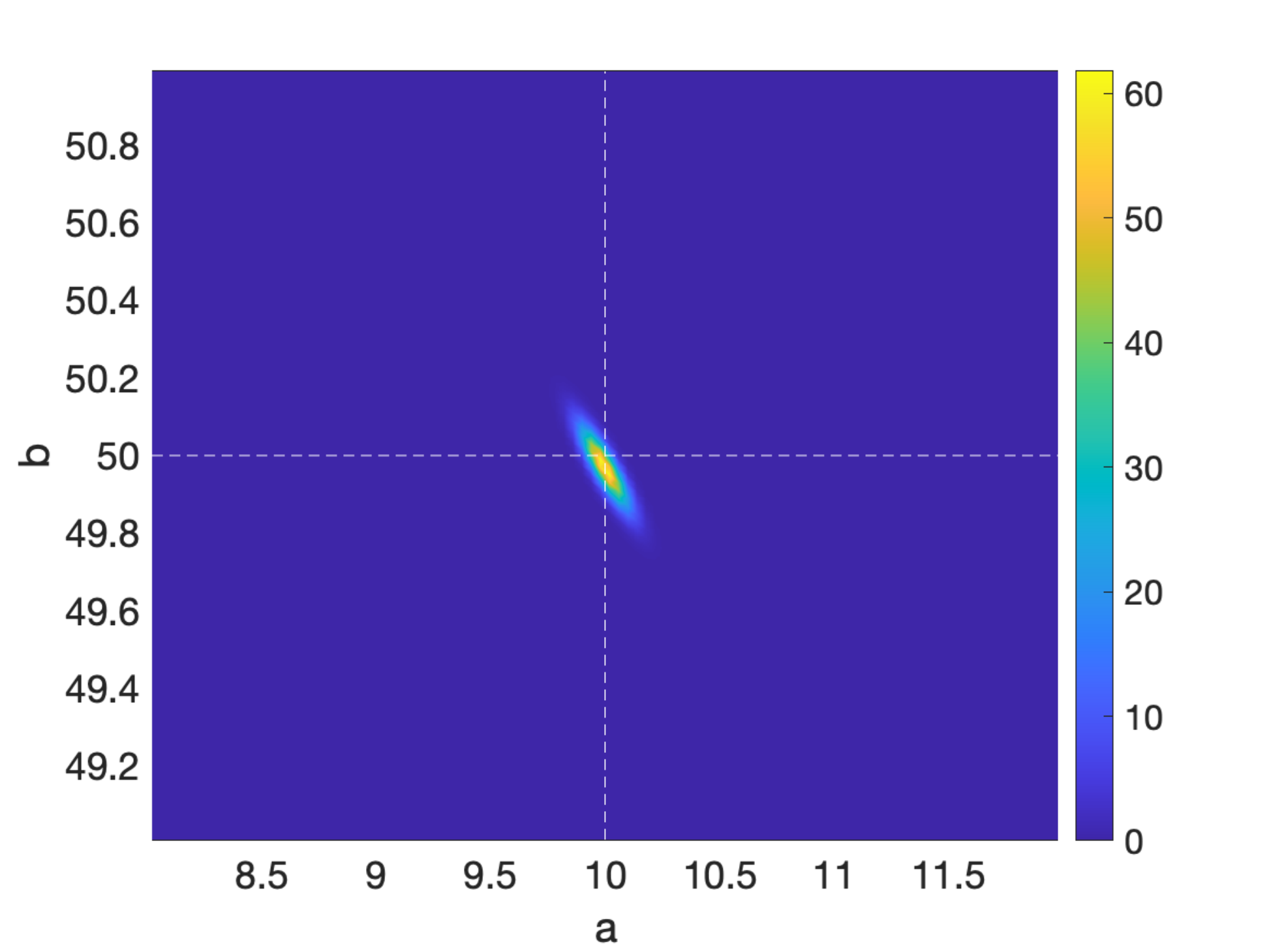}
         \caption{Data fidelity.}
         \label{fig:y3:DF}
         \end{center}
    \end{subfigure}
    \begin{subfigure}[b]{0.48\textwidth}
         \begin{center}
         \includegraphics[width=\textwidth]{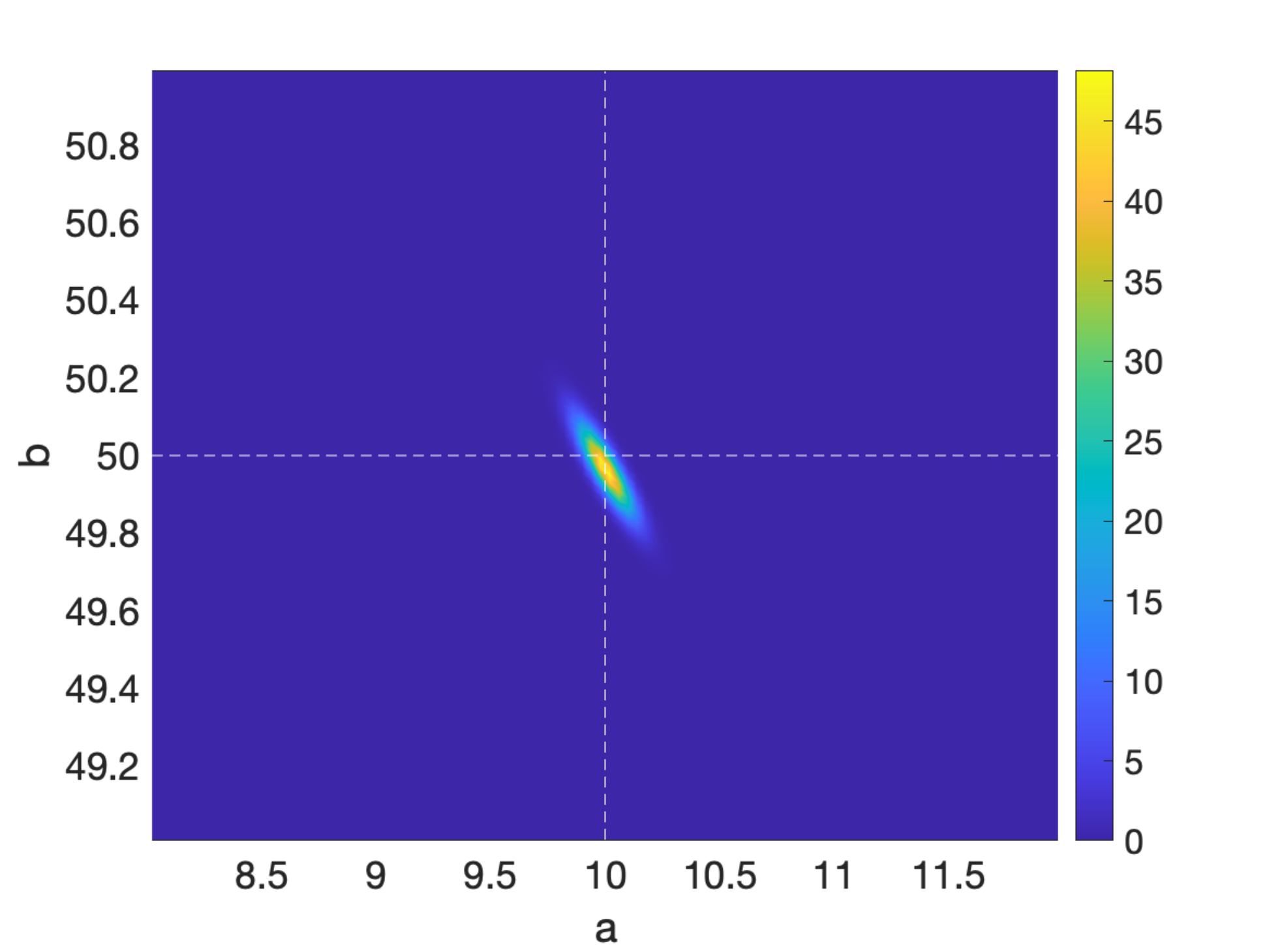}
         \caption{Non-product form.}
         \label{fig:y3:NPF}
         \end{center}
    \end{subfigure}
    \caption{The posterior densities arising from the data $\b{y}_2$, as given in Figure \ref{fig:y3_data}, using standard Bayesian inference, and the three approaches outlined in Sections \ref{sec:pModel}-\ref{sec:NPF}. Note that Figure \ref{fig:y3:SB} has substantially different axes to the other plots. The value of the parameters that created the data were $a=10$, $b=50$, which are denoted by the white dashed lines.}
    \label{fig:y3}
\end{center}
\end{figure}
  
  Figure \ref{fig:y3} shows approximations of the posterior densities arising from the linear regression of this data using standard Bayesian inference, and the three approaches outlined in Sections \ref{sec:pModel}-\ref{sec:NPF}. The standard Bayesian linear regression method is badly affected by the outliers in the data, whereas the three data selection methods identify these data points as outliers, and arrive at posterior distributions that are concentrated close to the value of the parameters that gave rise to the data.
  
  \begin{center}
  \begin{table}[htp]
\begin{tabular}{c|ccccc}
      & Standard & p-model & Data fidelity & Non-product form & ``True" value for $x<1$ \\ \hline
$a^*$ & 39.69    & 9.994   & 9.997         &  10.04         & 10                   \\
$b^*$ & 28.47   & 49.97   & 49.97        & 49.93         & 50                  
\end{tabular}
\caption{MAP estimates (to 4 s.f.) for the four Bayesian linear regression methodologies for data $\b{y}_2$ as shown in Figure \ref{fig:y3_data}.}
\label{tab:y3:MAP}
\end{table}
\end{center}
  
  \begin{figure}[htp]
      \centering
      \begin{subfigure}[b]{0.48\textwidth}
      \includegraphics[width=\textwidth]{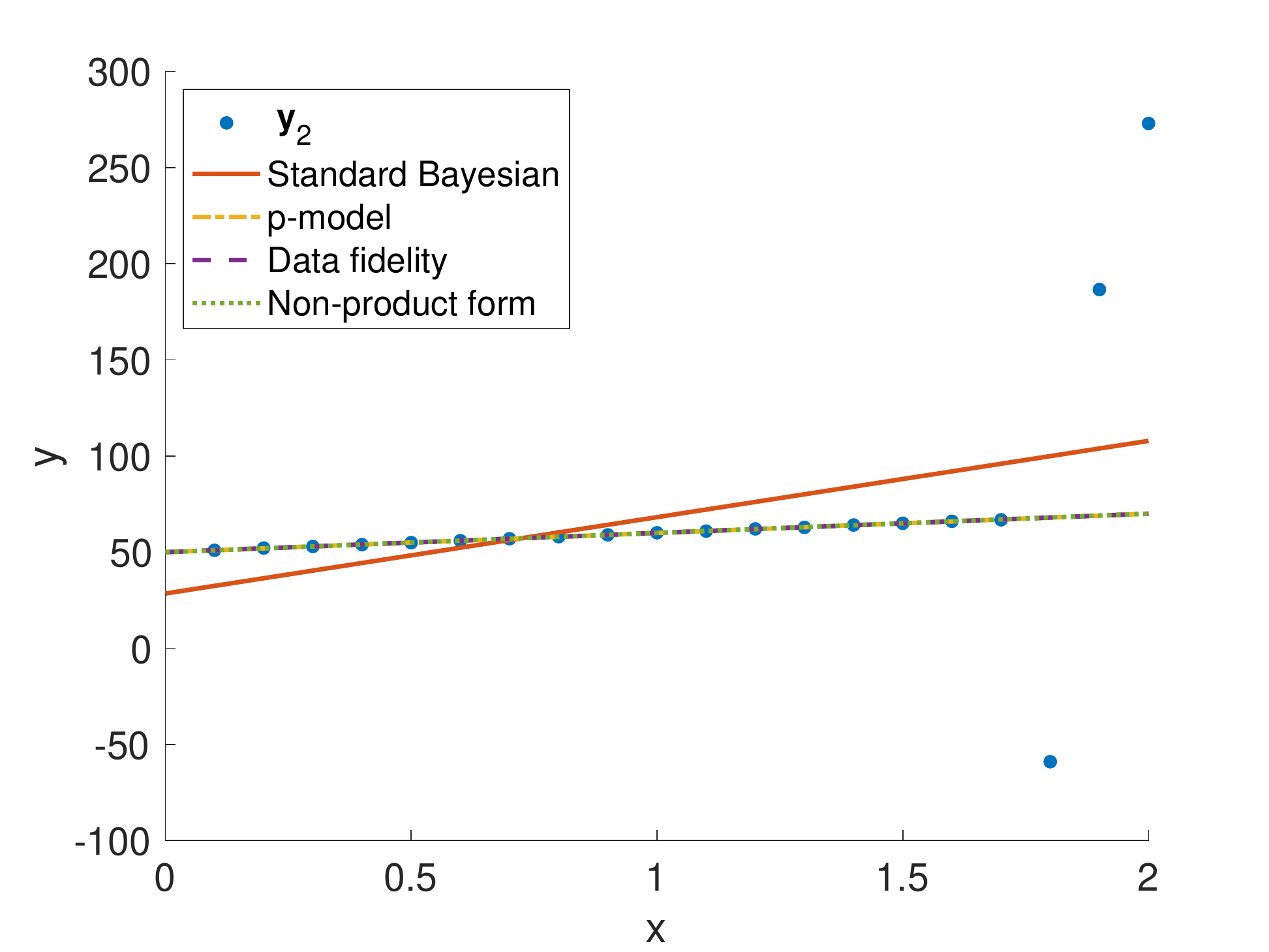}
      \caption{The MAP fits.}
      \label{fig:y3:MAPfits}
      \end{subfigure}
      \begin{subfigure}[b]{0.48\textwidth}
      \includegraphics[width=\textwidth]{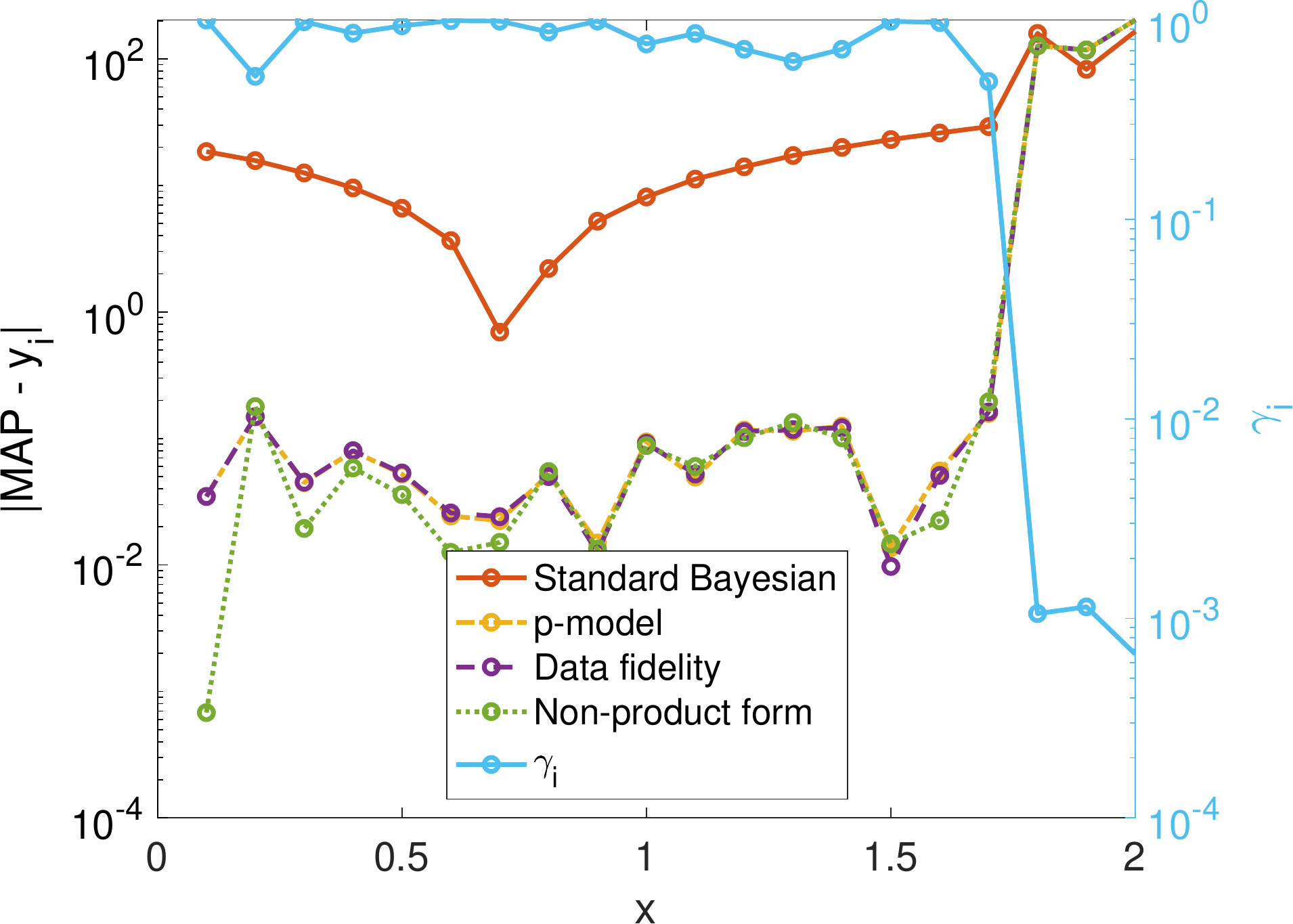}
      \caption{Data-model mismatches.}
      \label{fig:y3:MAPerrs}
      \end{subfigure}
      \caption{Plots to show the MAP fits of the four Bayesian linear regression approaches to the data $\b{y}_2$ as shown in Figure \ref{fig:y3_data}. (a) The MAP fits and the data. (b) The data-model mismatches of the MAP fits, and the MAP estimates of the $\gamma_i$ for the non-product form case.}
      \label{fig:y3:MAP}
  \end{figure}
  
  Figure \ref{fig:y3:MAP} shows the fits using MAP estimators of the four different Bayesian linear regression methodologies, as given in Table \ref{tab:y3:MAP}. The standard Bayesian linear regression gives us a poor fit for the whole data range, due to the influence of the corrupted data points. However, all three Bayesian data selection approaches give MAP estimates which fit the first 17 data points extremely well. We note that the remaining data points have much lower values for $\gamma_i$ for the non-product form case (which would also be reflected in the marginal posteriors of the fidelity terms in the data fidelity method of Section \ref{sec:fid}). This example demonstrates the power of Bayesian data selection, and its ability to minimise the effect of outliers in the inference of model parameters.
  
  \subsubsection{Example 3: Multiple feasible parameter values}\label{sec:MFPV}
\begin{figure}[htp]
    \begin{center}
      \includegraphics[width=0.5\textwidth]{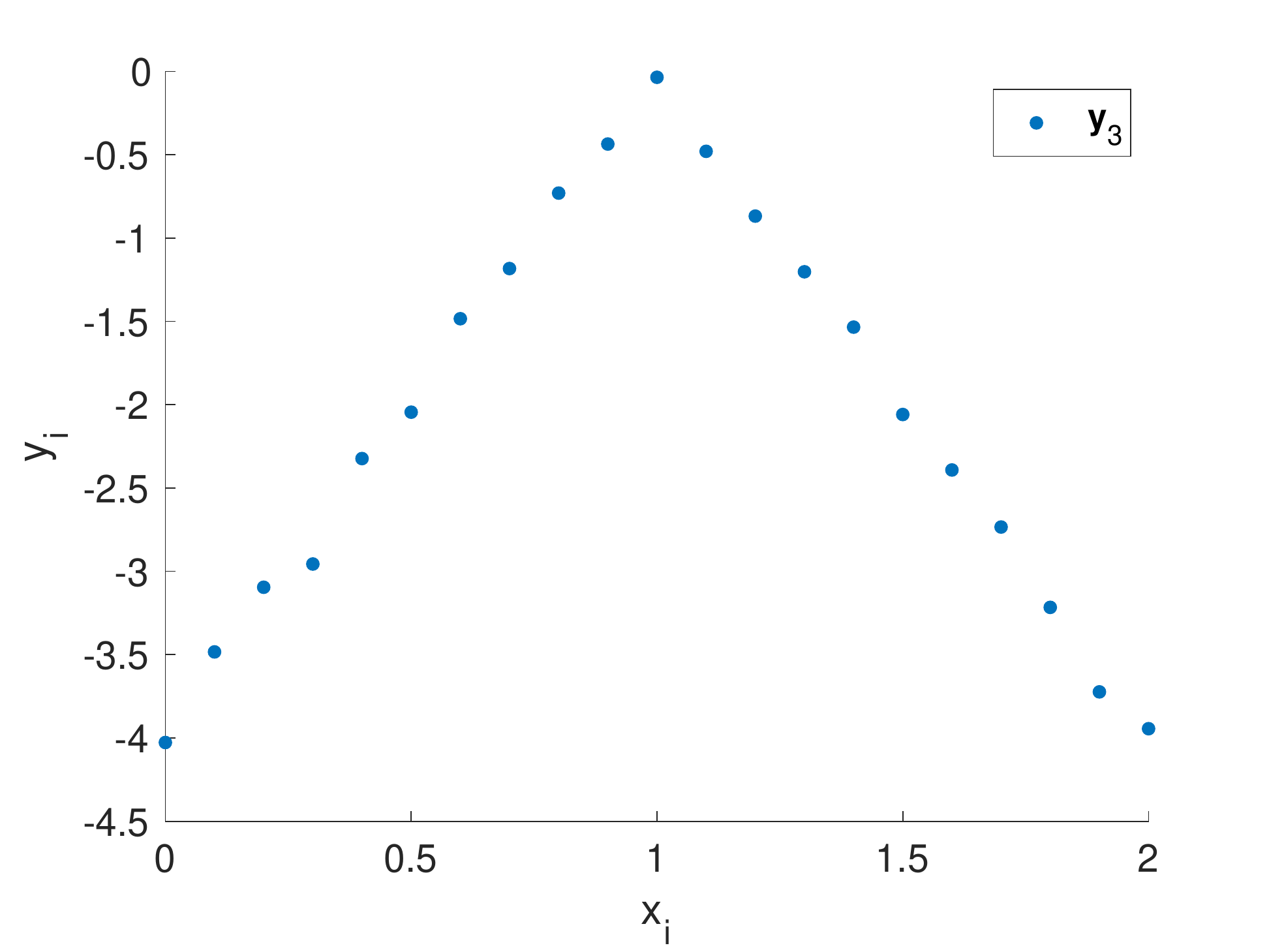}
      \caption{Plot to show data included in the vector $\b{y}_3$ on
        which we wish to conduct linear regression, displaying
        piecewise fit to the model. The parameters used were $a_1 = 4$,
        $b_1=-4$, $a_2 = -4$, $b_2 = 4$, $\sigma^2 = 0.1^2$.}\label{fig:y2_data}
    \end{center}
  \end{figure}

  We consider a third example where differing values of the model parameters can be fit to different subsets of the data. In this case, we consider noisy observations of a piecewise linear function:
  \begin{equation}
      f_3(x) = \begin{cases} a_1 x + b_1 \qquad & x \in [0,1] \\ a_2 x + b_2 \qquad & x \in (1,2].\end{cases}
  \end{equation}
  In particular we consider the case that the function $f_3 \in C([0,2])$, so that $a_1 + b_1 = a_2 + b_2$. This time we revert to  i.i.d. observational noise, such that the observations $\b{y}_3 = [y_0, \ldots, y_{20}]$ are given by
  \begin{equation}
      y_i = f(i/10) + \eta_i, \qquad \eta_i \sim \mathcal{N}(0,\sigma^2), \qquad i=1,\ldots,20.
  \end{equation}
  Figure \ref{fig:y2_data} shows a realisation of this data on which we will
then attempt to conduct Bayesian linear regression.

\begin{figure}[htp]
\begin{center}
    \begin{subfigure}[b]{0.48\textwidth}
         \begin{center}
         \includegraphics[width=\textwidth]{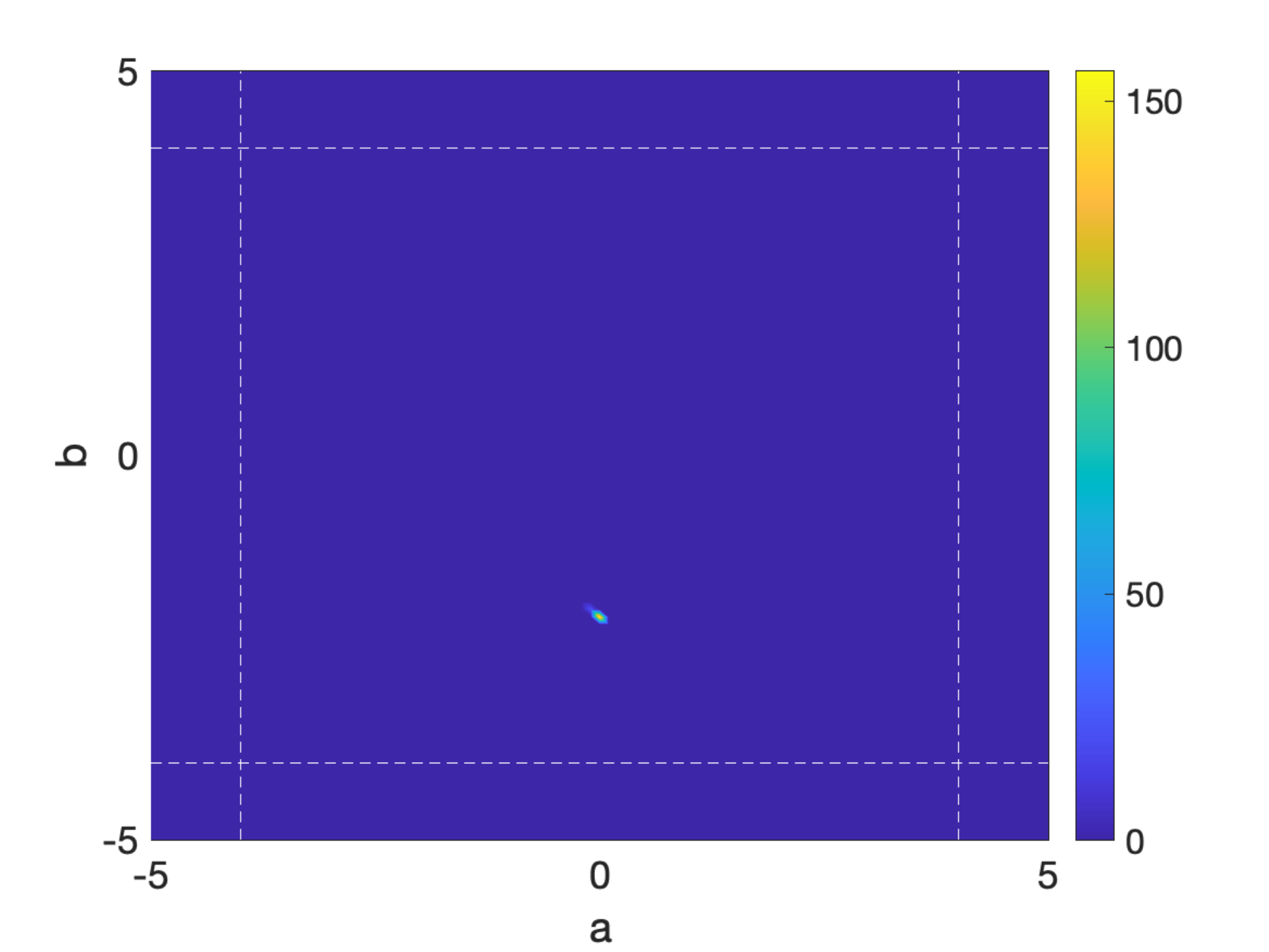}
         \caption{Standard Bayesian.}
         \label{fig:y2:SB}
         \end{center}
    \end{subfigure}
    \begin{subfigure}[b]{0.48\textwidth}
         \begin{center}
         \includegraphics[width=\textwidth]{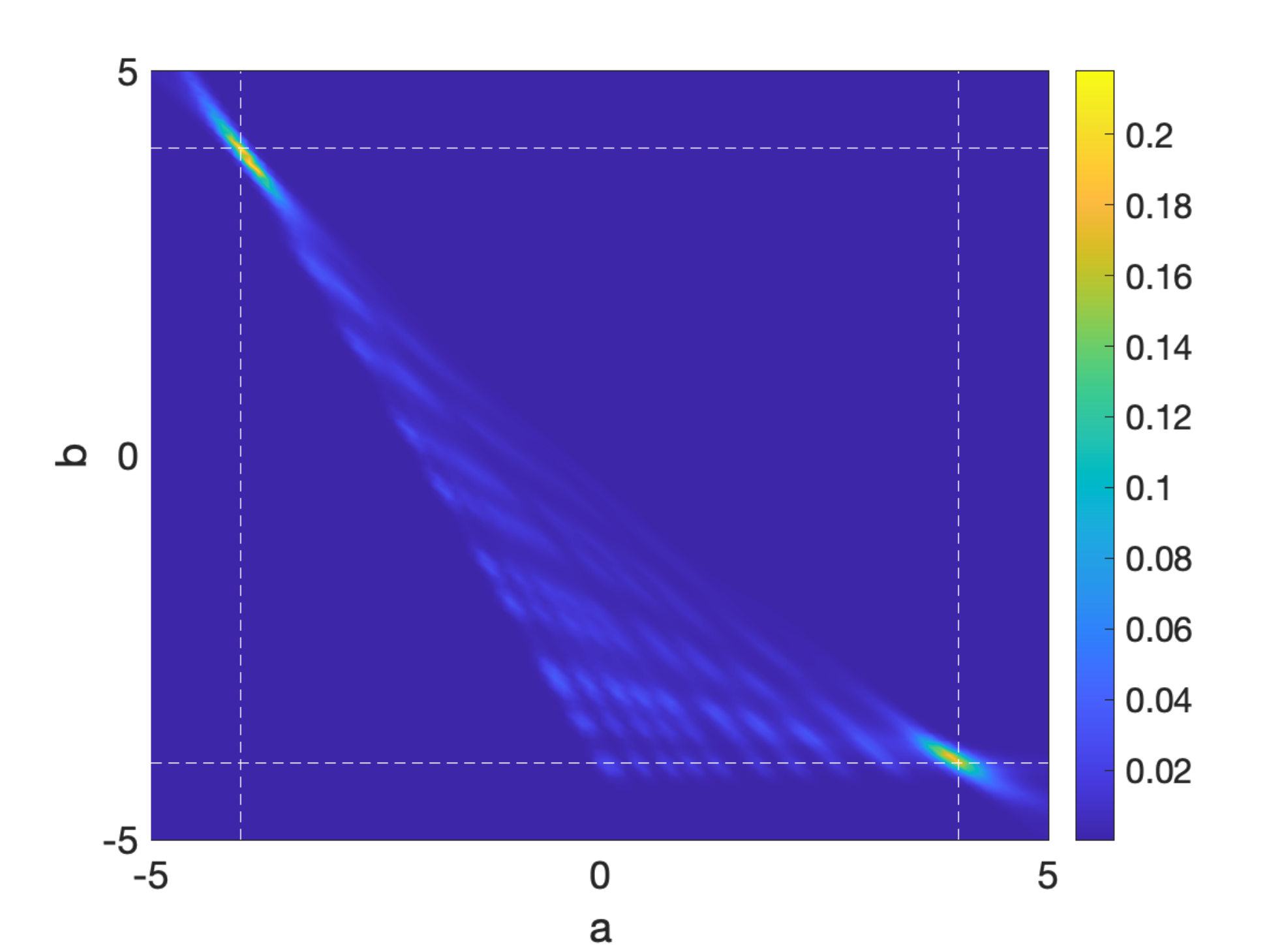}
         \caption{p-model.}
         \label{fig:y2:p}
         \end{center}
    \end{subfigure} \\
    \begin{subfigure}[b]{0.48\textwidth}
         \begin{center}
         \includegraphics[width=\textwidth]{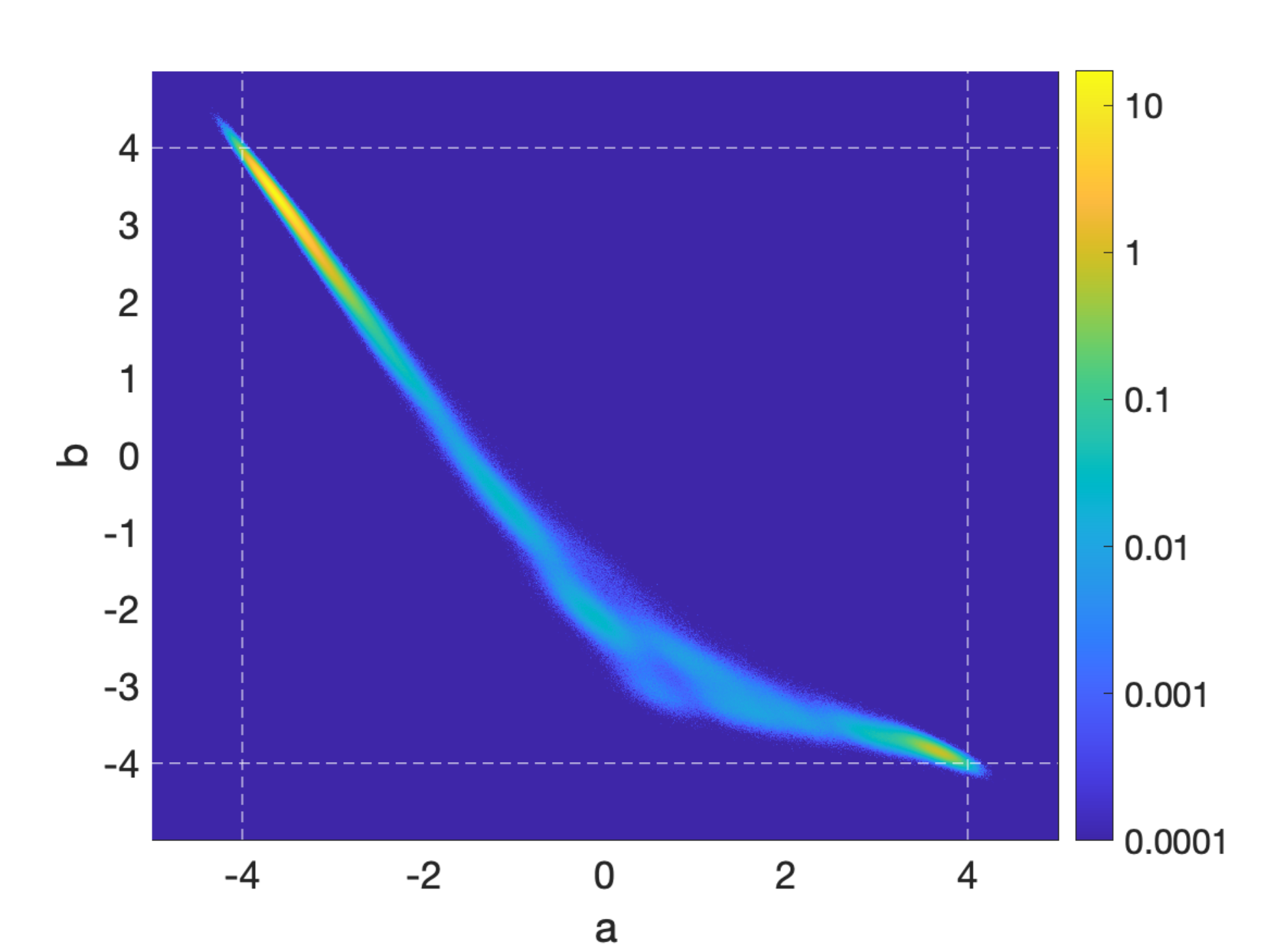}
         \caption{Data fidelity (log scale).}
         \label{fig:y2:DF}
         \end{center}
    \end{subfigure}
    \begin{subfigure}[b]{0.48\textwidth}
         \begin{center}
         \includegraphics[width=\textwidth]{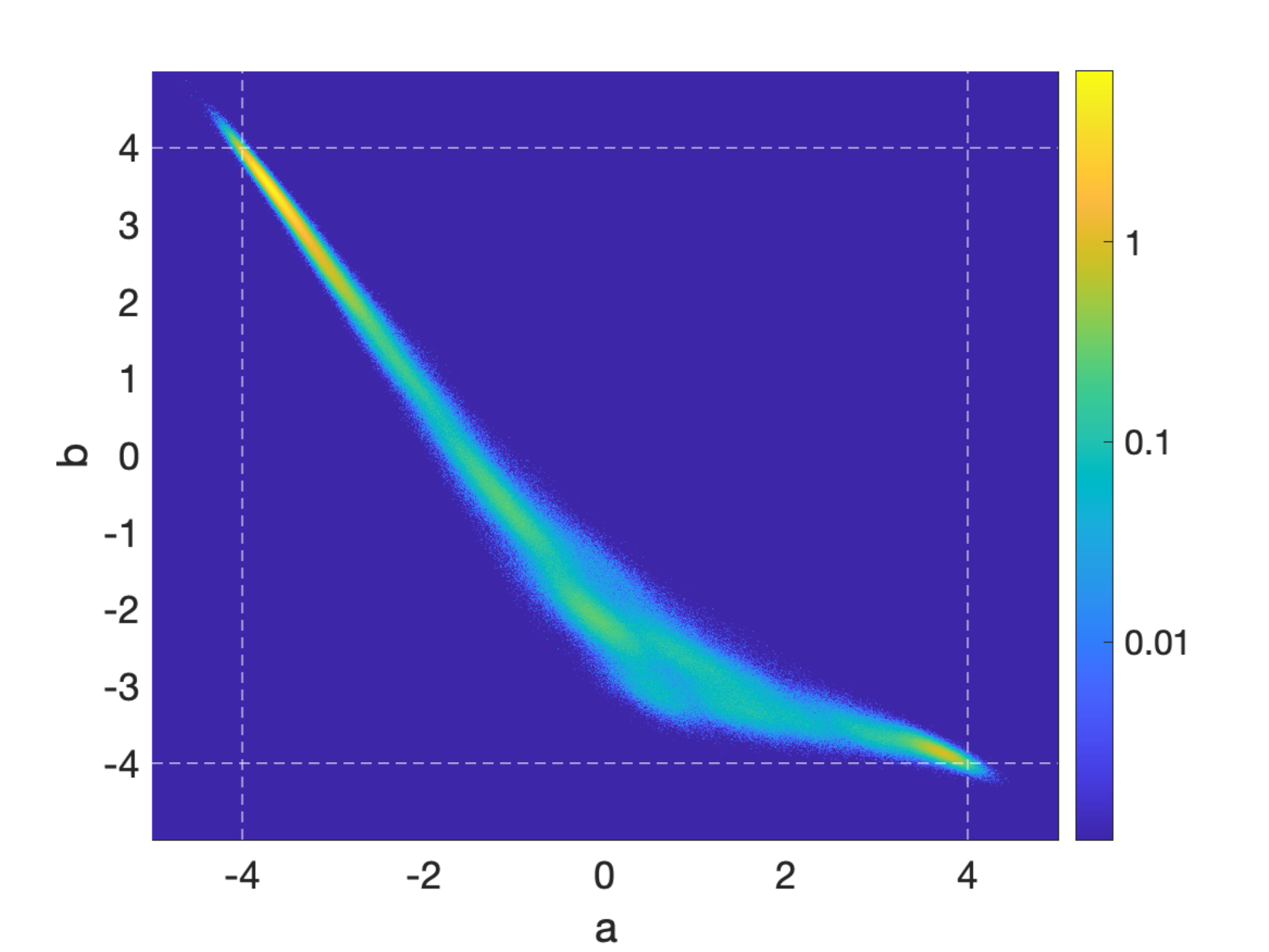}
         \caption{Non-product form (log scale).}
         \label{fig:y2:NPF}
         \end{center}
    \end{subfigure}
    \caption{The posterior densities arising from the data $\b{y}_3$, as given in Figure \ref{fig:y2_data}, using standard Bayesian inference, and the three approaches outlined in Sections \ref{sec:pModel}-\ref{sec:NPF}. The value of the parameters that created the data were $a_1 = 4$,
        $b_1=-4$, $a_2 = -4$ and $b_2 = 4$, which are denoted by the white dashed lines.}
    \label{fig:y2}
\end{center}
\end{figure}
  
  Figure \ref{fig:y2} shows approximations of the posterior densities arising from the linear regression of this data using standard Bayesian inference, and the three approaches outlined in Sections \ref{sec:pModel}-\ref{sec:NPF}. The standard Bayesian linear regression finds the values of $a$ and $b$ that can be best fit to the whole data set, and so in this scenario it completely fails to find either of the sets of parameter values that created the data. In contrast, the data selection methods lead to posterior distributions with two significant modes concentrated close to the two different sets of parameter values that were used to create the data. The second of these modes is not so visible in the case of the data fidelity approaches, and as such we present the logarithms of the densities. This is in contrast to the $p$-model, where both modes are clearly visible. We purposely picked a data scenario such that there was approximately equal evidence (modulo the impact of observational noise) for one set of parameter values as the other, with 11 of the 21 data points being consistent either with $a_1$ and $b_1$, or $a_2$ and $b_2$. However, the heights of these modes seems to be very sensitive to the realisation of the noise in the data, since if we use a different random number generator seed for the data synthesis, in some cases the relative heights of the modes switched. The $p$-model seems to be more robust to these issues. However, in this particular instance we are looking at a small data scenario, specifically so that we can employ the $p$-model, and this sensitivity to the observational noise realisation is likely to dissipate as the size of the data increases.
  
  \begin{center}
  \begin{table}[htp]
\begin{tabular}{c|ccccc}
      & Standard & p-model & Data fidelity & Non-product form & ``True" values \\ \hline
$a_1^*$ & -0.02341    & 3.849   & 3.681        &  3.826         & 4            \\
$b_1^*$ & -2.070   & -3.874   &  -3.852        &  -3.884        & -4 \\        
\hline
$a_2^*$ &  -    & -3.925   &  -3.644        &  -3.844         & -4                   \\
$b_2^*$ &  -   &  3.874   &  3.415        &  3.745        &  4 \\           
\hline 
$\frac{\pi(a^*_1,b^*_1)}{\pi(a^*_2,b^*_2)}$ & -  &   1.274      &     1.017          &       1.016     &  -   \\ 
\end{tabular}
\caption{Estimates from the dominant peaks (to 4 s.f.) for the four Bayesian linear regression methodologies for data $\b{y}_3$ as shown in Figure \ref{fig:y2_data}. The negative log (unnormalised) density at these modes is given in order to compare the relative heights of the peaks.}
\label{tab:y2:MAP}
\end{table}
\end{center}

Table \ref{tab:y2:MAP} shows the estimates of the model parameters at the dominant modes (1 in the case of the standard Bayesian regression, 2 in the case of data selection methods). Also given are the ratios of the densities at the peaks for the data selection methods, in order to show the relative heights of the modes. Surprisingly, despite the second mode being less visible in the plots of the two data fidelity methods, the actual peak heights of the two modes are extremely similar, reflecting the very similar amount of evidence for each of the sets of parameter values.

  \begin{figure}[htp]
      \centering
      \begin{subfigure}[b]{0.48\textwidth}
      \includegraphics[width=\textwidth]{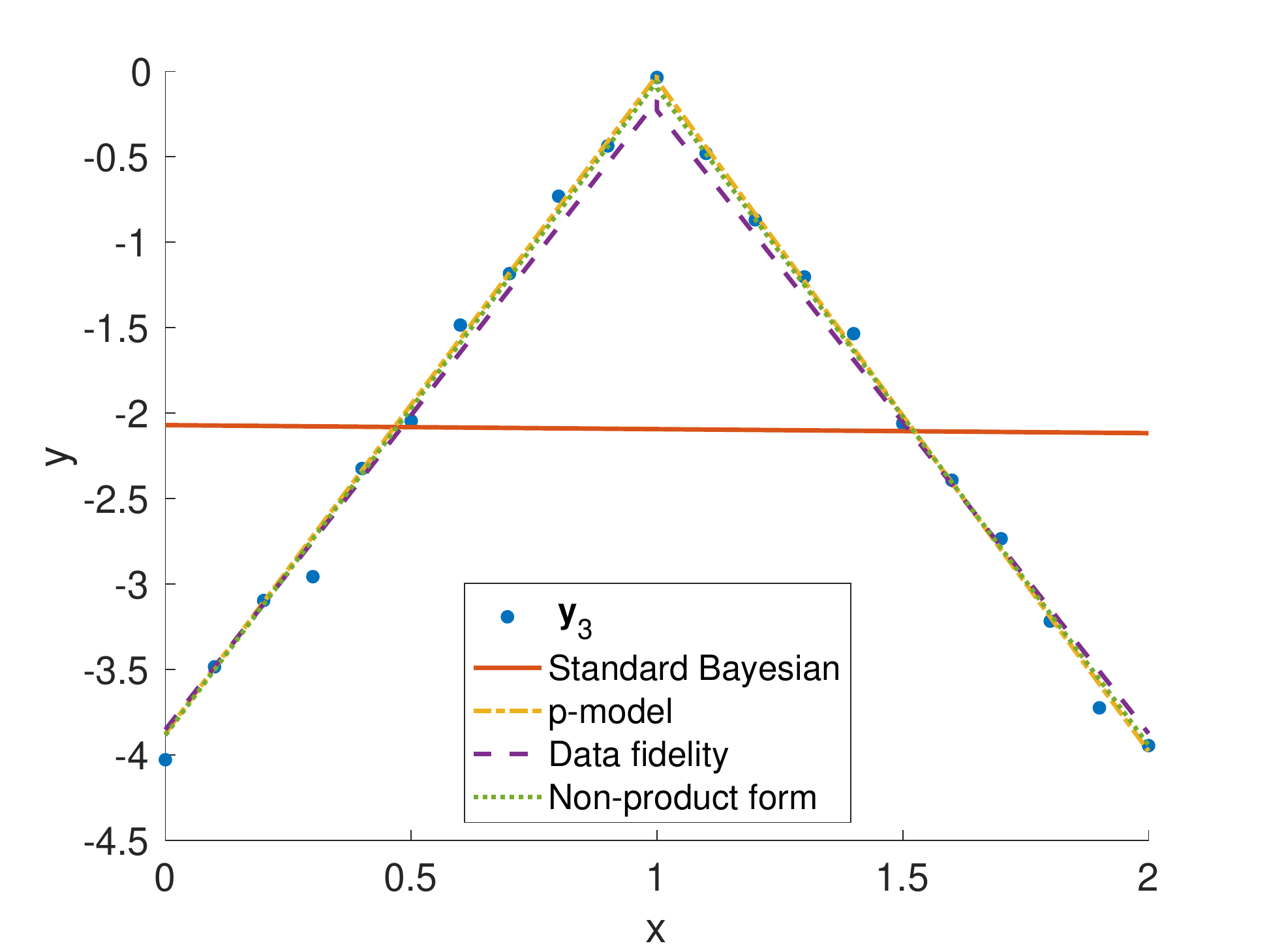}
      \caption{The mode fits.}
      \label{fig:y2:MAPfits}
      \end{subfigure}
      \begin{subfigure}[b]{0.48\textwidth}
      \includegraphics[width=\textwidth]{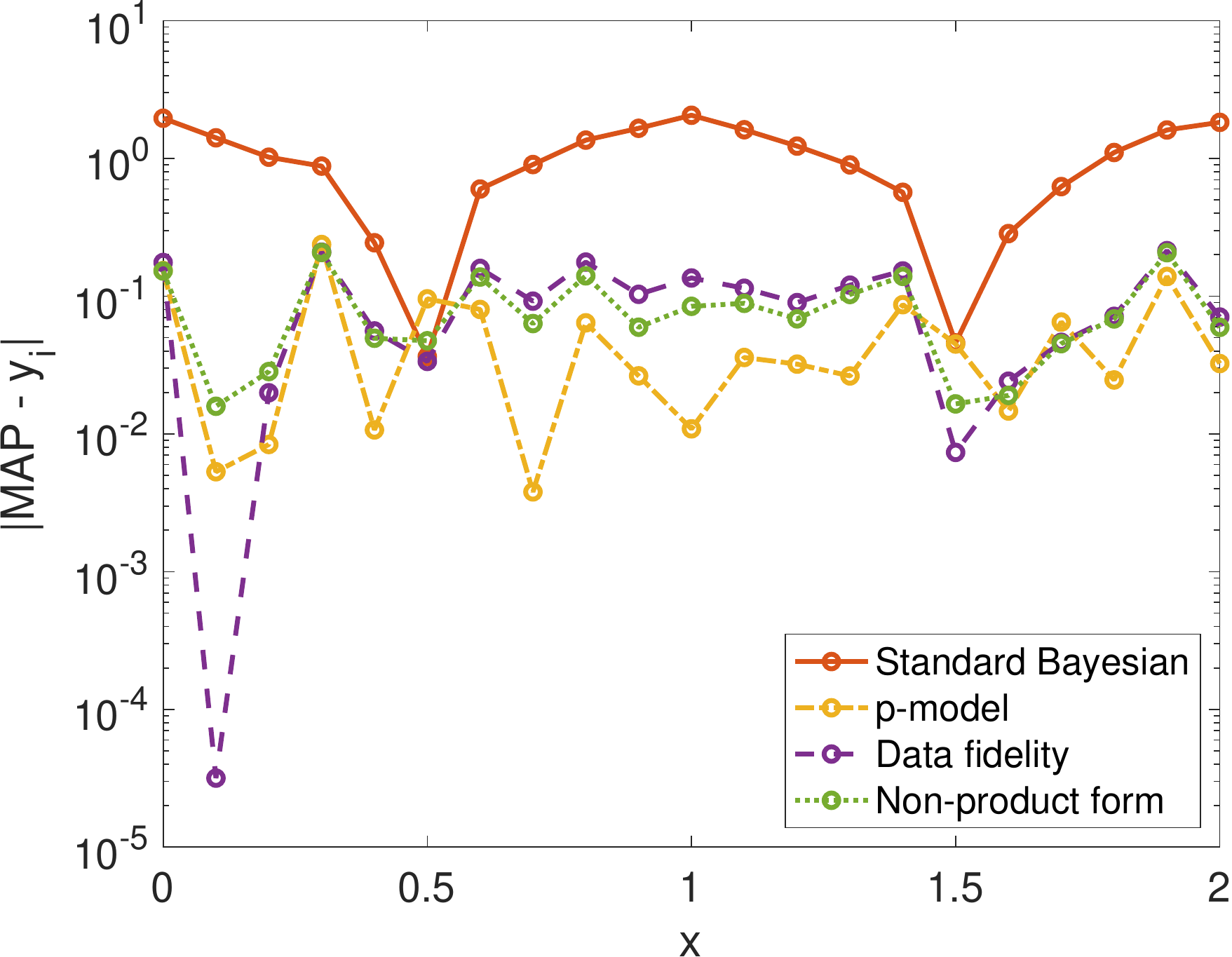}
      \caption{Data-model mismatches.}
      \label{fig:y2:MAPerrs}
      \end{subfigure}
      \begin{subfigure}[b]{0.48\textwidth}
      \includegraphics[width=\textwidth]{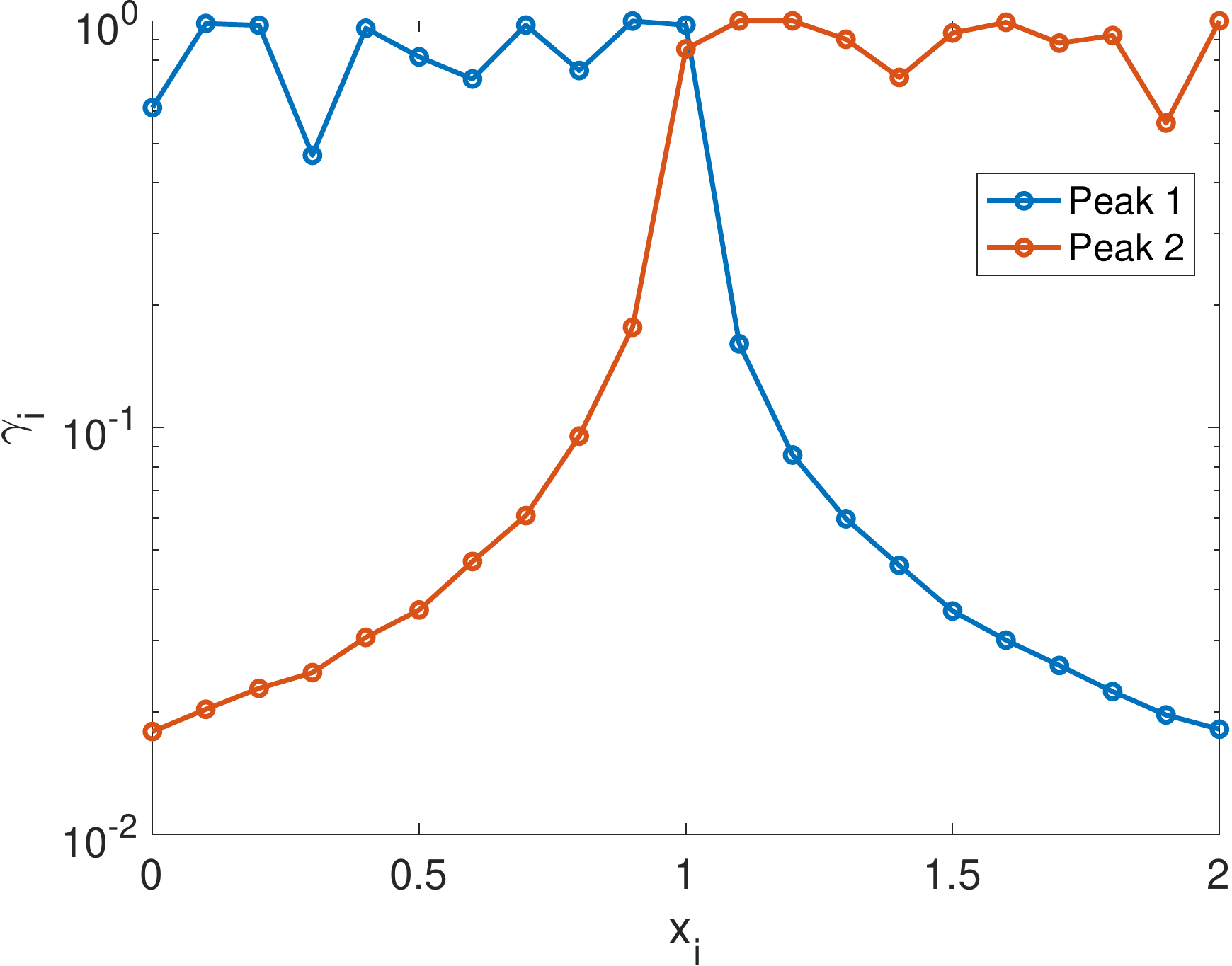}
      \caption{Mode fits of the $\gamma_i$.}
      \label{fig:y2:MAPgams}
      \end{subfigure}
      \caption{Plots to show the mode fits of the four Bayesian linear regression approaches to the data $\b{y}_3$ as shown in Figure \ref{fig:y2_data}. (a) The mode fits and the data. (b) The data-model mismatches of the mode fits. (c) The mode estimates of the $\gamma_i$ for the non-product form case.}
      \label{fig:y2:MAP}
  \end{figure}
  
  Figure \ref{fig:y2:MAP} shows the fits to data at the dominant modes in the densities for each of the four linear regression methods, estimated using restarted BFGS. In the case of the data selection methods, the two peaks indicate a piecewise linear fit is preferable to a linear fit. What is more, the regions in which each parameter set is the best fit can be identified by looking at the value of the $p$, $\tau$ or $\gamma$ parameters. This can be seen for example in  Figure \ref{fig:y2:MAPgams}, where we plot the estimates of the $\gamma_i$ at the two peaks in the joint density. It is clear from this plot that the first peak indicates parameter values which fit to data well in the region $[x_0, x_{10}] = [0,1]$, and the second peak indicates parameter values which fit to data well in the region $[x_{10},x_{20}] = [1,2]$. Given this information, we present the piecewise linear functions indicated by the parameter values at the two peaks in Figure \ref{fig:y2:MAPfits}. As can also be seen in Figure \ref{fig:y2:MAPerrs}, the standard Bayesian linear regression  demonstrates the old adage that a stopped clock is right twice a day. All of the Bayesian data selection approaches demonstrate a good fit to all of the data.
  
  \subsubsection{Conclusions}
  In these three examples, we have demonstrated a step change in the quality of inference in linear regression in certain cases where the standard Bayesian linear regression does not work well. In all of these settings, there is a different reason that a single linear model cannot be fit well to the whole data set: non-linearity of the model in certain regions, corruption of data, and data which can only be fit piecewise to the model. In particular this means that practitioners are no longer required to make subjective choices about data ``cleaning", or the practice of removing data outliers, or identifying regions which we believe our model can be fit. The resulting inference can be highly sensitive to these choices, but by automating these choices through Bayesian data selection, we arrive at a set of methods which are robust to a range of challenging issues with data and models, the likes of which occur frequently in real life applications.

  \section{Application to an inverse problem for an ordinary differential equation}\label{sec:ODE}
  In this section we demonstrate Bayesian data selection for an inverse problem for a simple ODE. In particular, we consider the scenario where the posterior density will be approximated using a numerical approximation of the solution of the differential equation, where the quality of the approximation decays in time. This is a common scenario in many applications where, for example, the system of differential equations is chaotic, and so any numerical approximation of the solution will diverge from the true solution in finite time. For simplicity, we will not consider a chaotic system of equations, but instead we will employ a numerical approximation whose error we know will grow in time.
  
  We consider the following initial value problem (IVP) with unknown initial condition,
  \begin{equation}\label{eq:ODE}
  \frac{dX}{dt} = \lambda X, \qquad X(0) = x_0.
  \end{equation}
  Our aim is to infer, given the value of $\lambda$ and noisy observations of the solution to the ODE $X(t)$ at a sequence of times $t_1 < \ldots < t_N$, the value of the initial condition $x_0$. We assume mean zero Gaussian additive observational noise, giving us observations
  \begin{equation}\label{eq:obs}
      Y_i = \mathcal{G}_i(x_0) + \eta_i, \qquad \eta_i \sim \mathcal{N}(0,\Sigma),
  \end{equation}
  where $\mathcal{G}_i(x_0) = x_0\exp(\lambda t_i)$. Naturally this ODE can be solved without numerical approximation, but we simulate a scenario where the exact solution is not available, as is the case in many applications of interest. We employ a forward Euler approximation of the solution, and for ease assume that the observation times $t_i = i \Delta_{\rm{Obs}}$ are uniformly spaced with time step $\Delta_{\rm{Obs}}$. For simplicity we also pick  the timestep in the Euler approximation $\Delta t$ so that $\Delta_{\rm{Obs}} = n \Delta t$ is an integer multiple of $\Delta t$. Then the $i$th element of the approximate observation operator $\tilde{\mathcal{G}}$ is given by
  \begin{equation}\label{eq:approx}
      \tilde{\mathcal{G}}_i(x_0) = (1 + \lambda \Delta t)^{i\times n}.
  \end{equation}
  For small $\Delta t$ and small times, this is a good approximation. However, if $\lambda > 0$, then no matter how small $\Delta t$ is, for larger times the absolute error in the approximation will grow very large. Naturally, this causes issues when trying to use this approximation in order to sample from a posterior distribution on the initial condition, where the true value may have low density with respect to the approximate posterior density.
  
  \begin{figure}[htp]
      \centering
      \begin{subfigure}[b]{0.48\textwidth}
      \includegraphics[width=\textwidth]{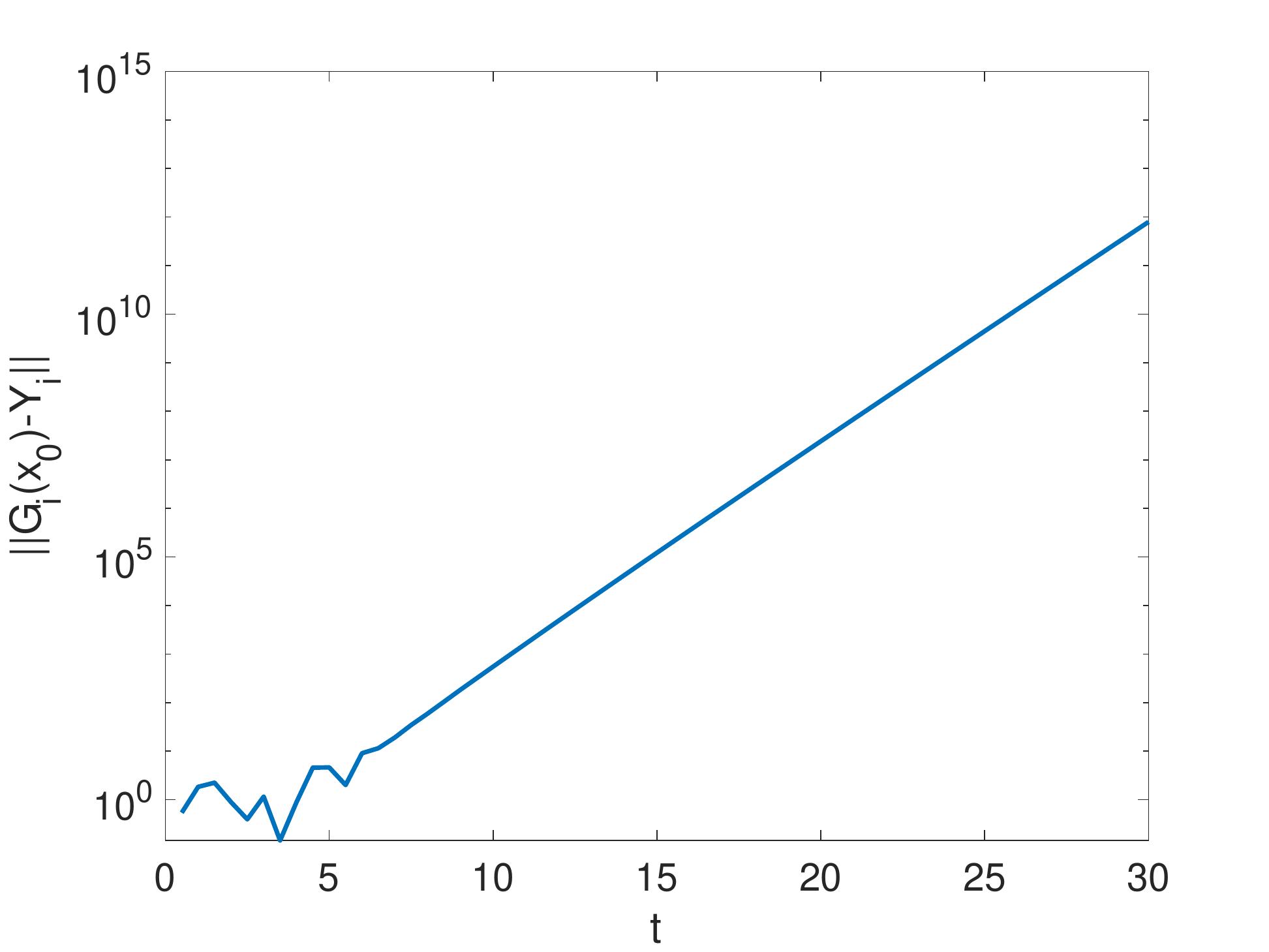}
      \caption{Absolute data-model mismatches.}
      \label{fig:error}
      \end{subfigure}
      \begin{subfigure}[b]{0.48\textwidth}
      \includegraphics[width=\textwidth]{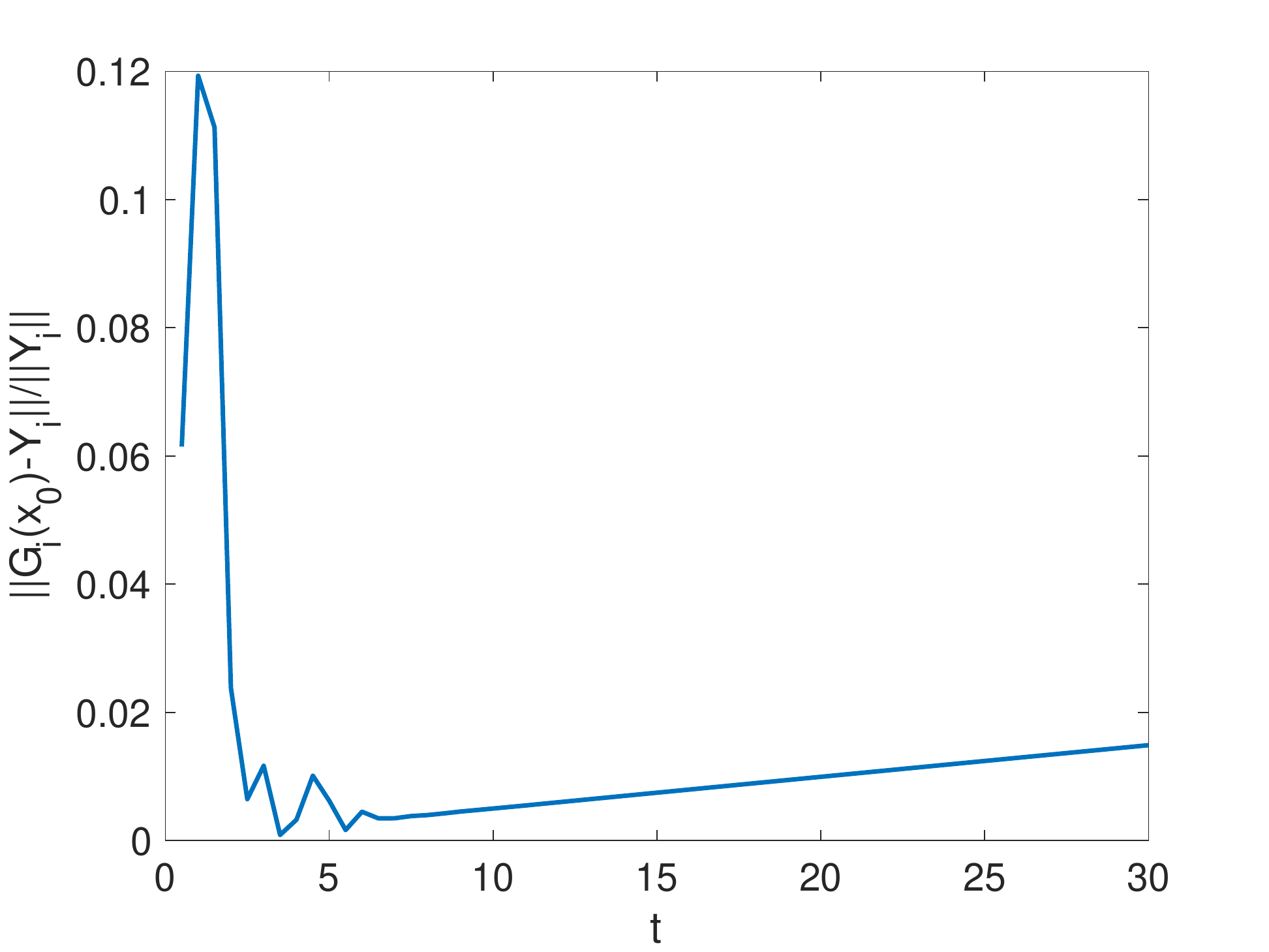}
      \caption{Relative data-model mismatches.}
      \label{fig:rel_error}
      \end{subfigure}
      \caption{Plots to show the (a) absolute and (b) relative error between the observations $Y$ given by \eqref{eq:obs} and the approximate observation operator \eqref{eq:approx} with observation times $t_i = i/2$, $i\in \{1,\ldots,60\}$, $\lambda = 1$, $\Sigma=1$, and $\Delta t = 10^{-3}$, evaluated at the ``true" value of the initial condition $x_0 = 5$.}
      \label{fig:errors}
\end{figure}

Figure \ref{fig:errors} demonstrates how the data $Y$ and the approximate observational operator $\tilde{\mathcal{G}}(x_0)$ diverge even when evaluated at the value of the initial condition $x_0=5$ which was used to create the data. The larger initial relative error fluctuations are caused by the relative size of the solution and the observational noise, but then we see a modest steady growth in relative error as $t$ increases. Importantly, the absolute error grows enormously in a very short time - these errors simply do not comply with the statistical model that we have assumed in \eqref{eq:obs}, and as such the approximate posterior distribution may not be concentrated close to the ground truth. However, for early times $t<5$ the absolute errors are dominated by the noise of the observations, which indicates that these observations could be used to accurately infer the value of the initial condition. This makes this problem a prime candidate for Bayesian data selection.

  \begin{figure}[htp]
      \centering
      \begin{subfigure}[b]{0.48\textwidth}
      \includegraphics[width=\textwidth]{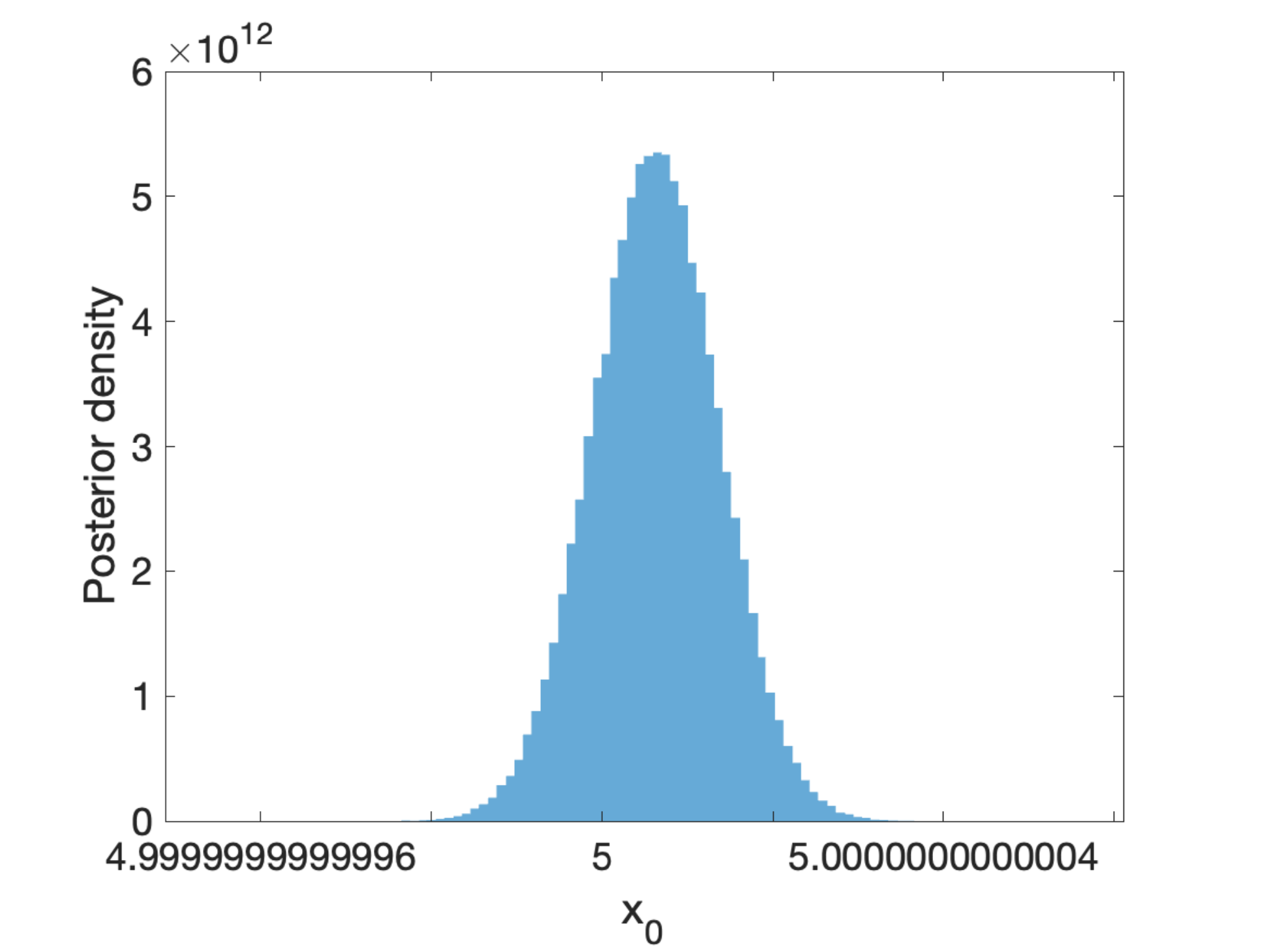}
      \caption{Exact $\mathcal{G}$, no data selection.}
      \label{fig:AnoBDS}
      \end{subfigure}
      \begin{subfigure}[b]{0.48\textwidth}
      \includegraphics[width=\textwidth]{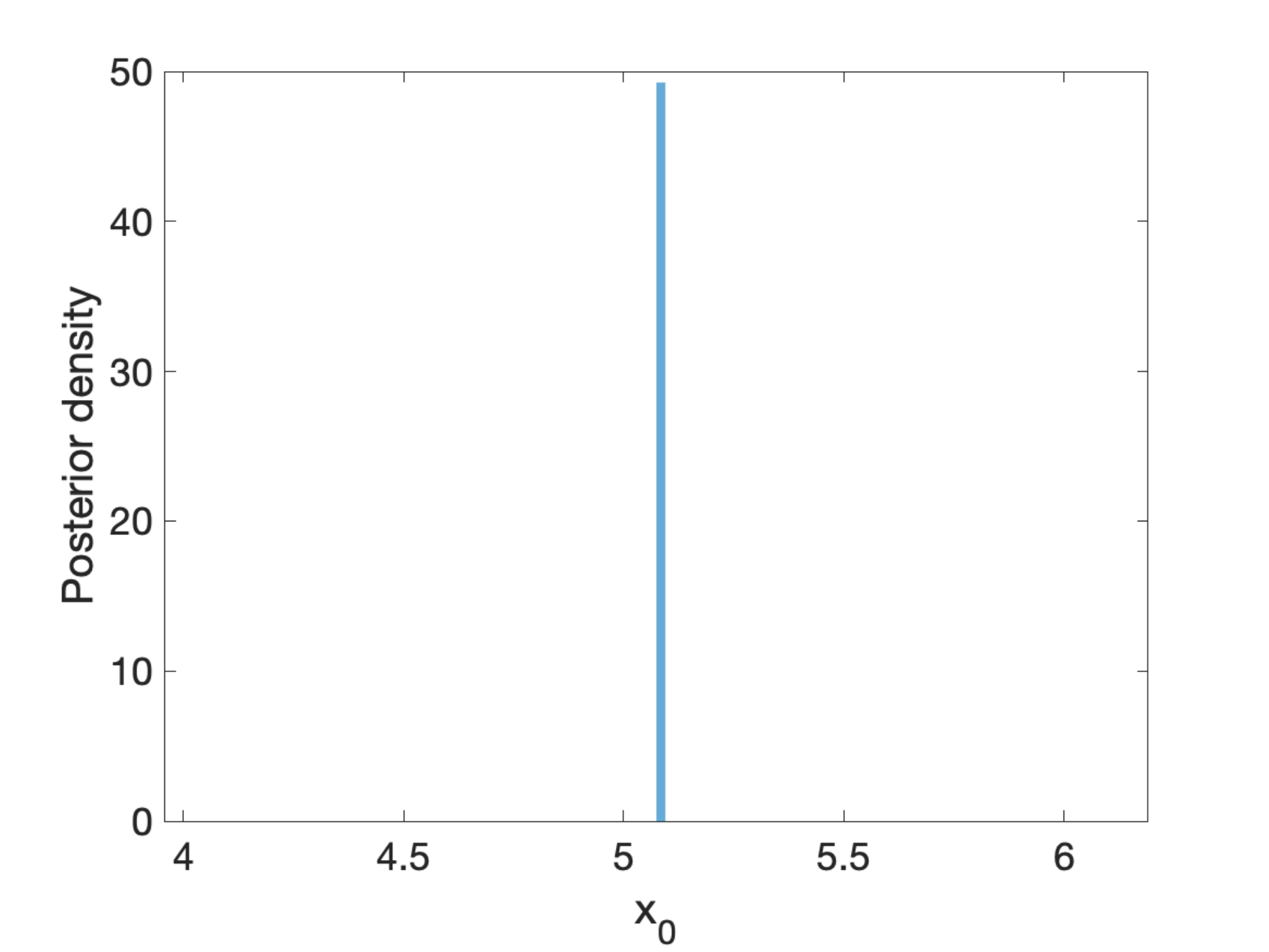}
      \caption{Approx. $\mathcal{G}$, no data selection.}
      \label{fig:anoBDS}
      \end{subfigure}\\
      \begin{subfigure}[b]{0.48\textwidth}
      \includegraphics[width=\textwidth]{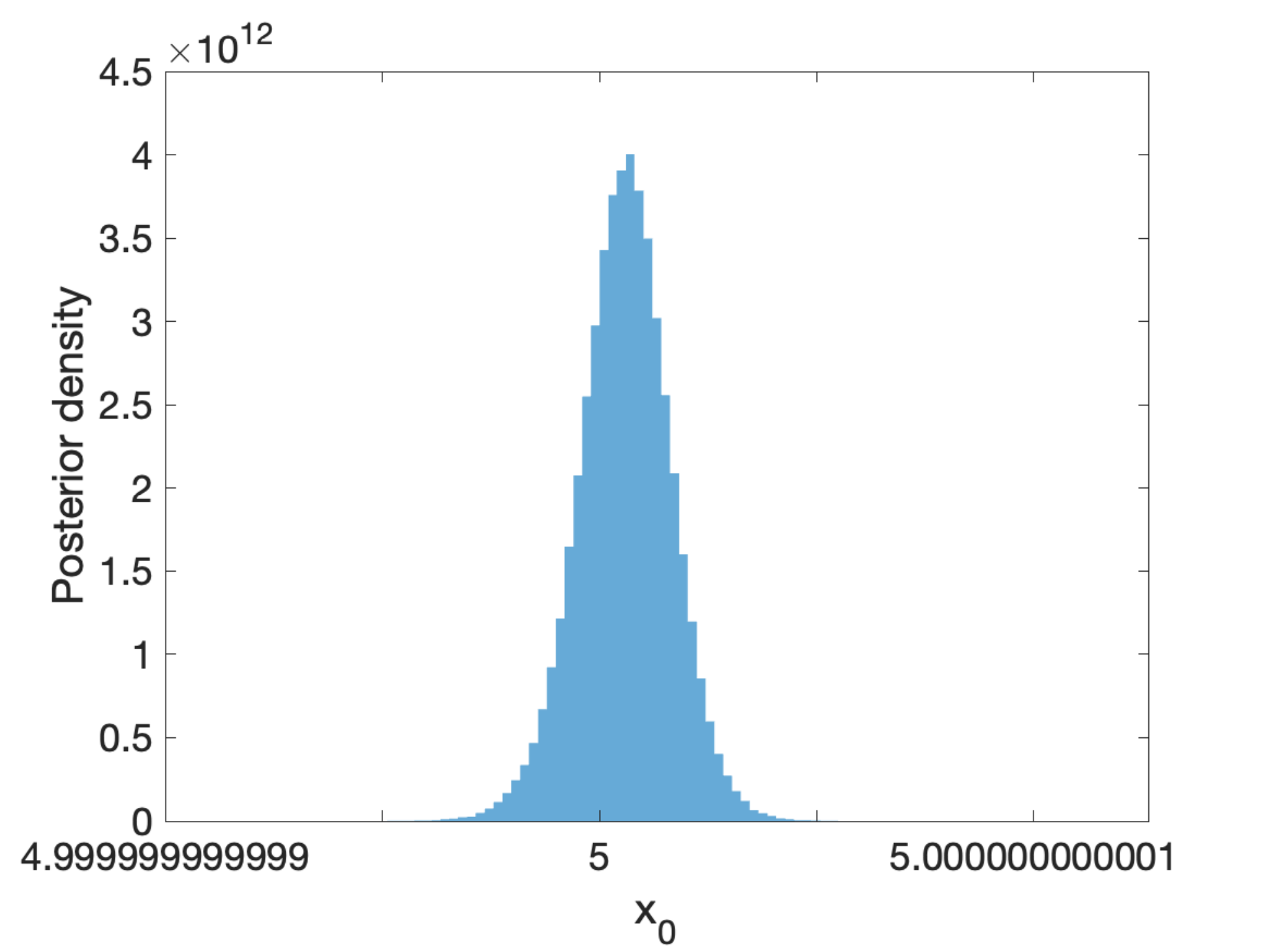}
      \caption{Exact $\mathcal{G}$, data selection.}
      \label{fig:ABDS}
      \end{subfigure}
      \begin{subfigure}[b]{0.48\textwidth}
      \includegraphics[width=\textwidth]{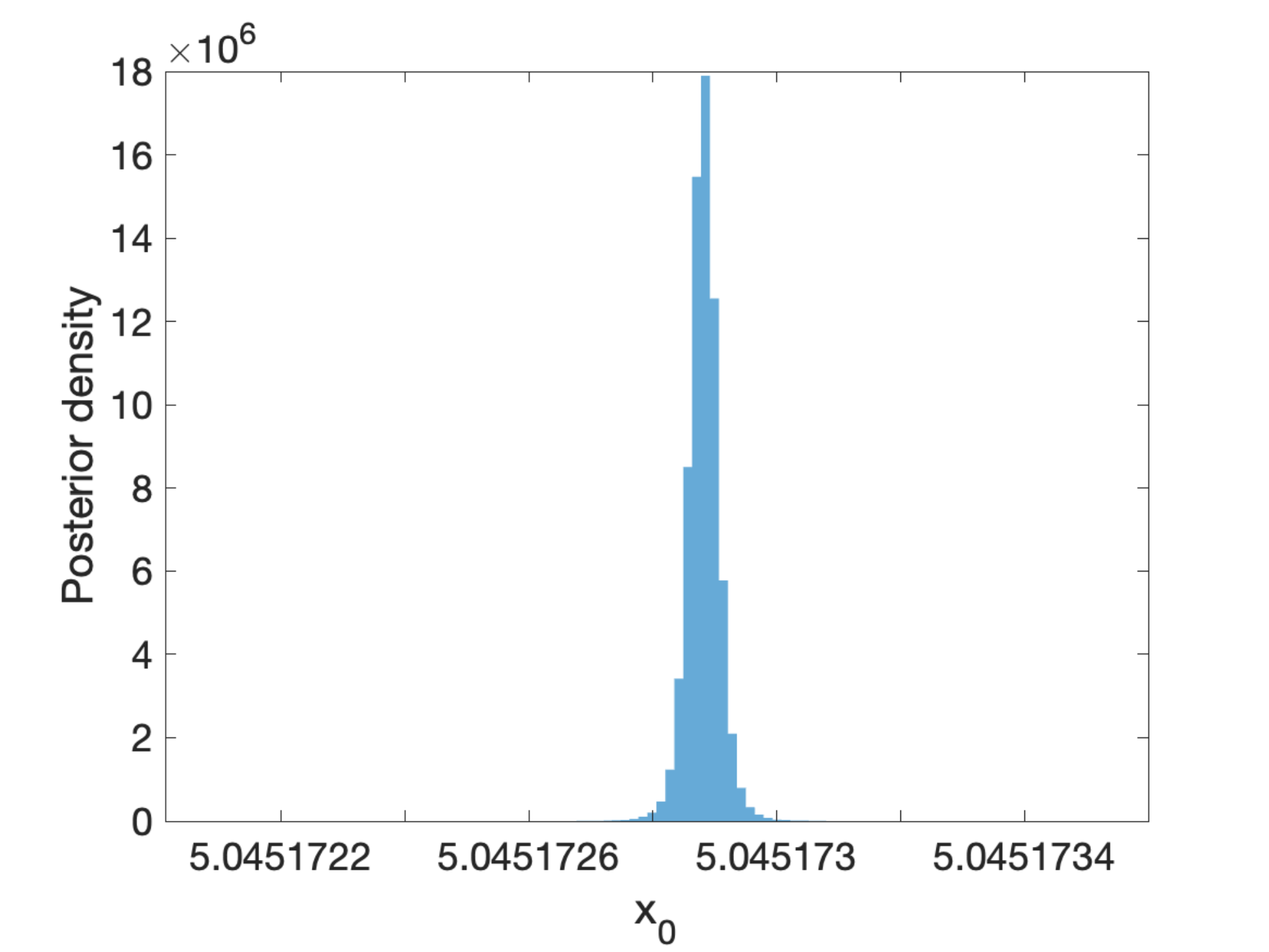}
      \caption{Approx. $\mathcal{G}$, data selection.}
      \label{fig:aBDS}
      \end{subfigure}\\
      \begin{subfigure}[b]{0.48\textwidth}
      \includegraphics[width=\textwidth]{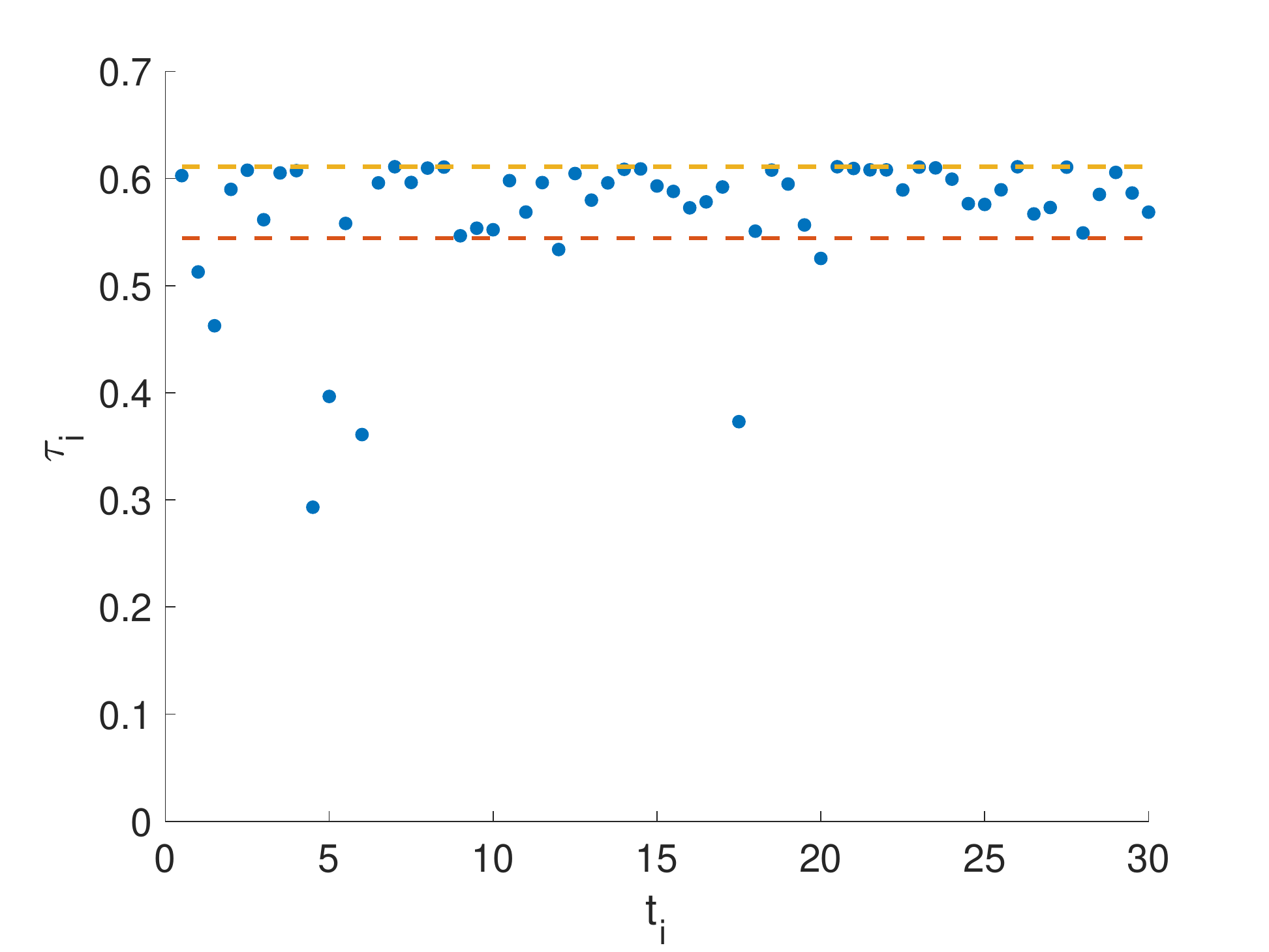}
      \caption{Mean values of $\tau_i$, exact $\mathcal{G}$.}
      \label{fig:Ataus}
      \end{subfigure}
      \begin{subfigure}[b]{0.48\textwidth}
      \includegraphics[width=\textwidth]{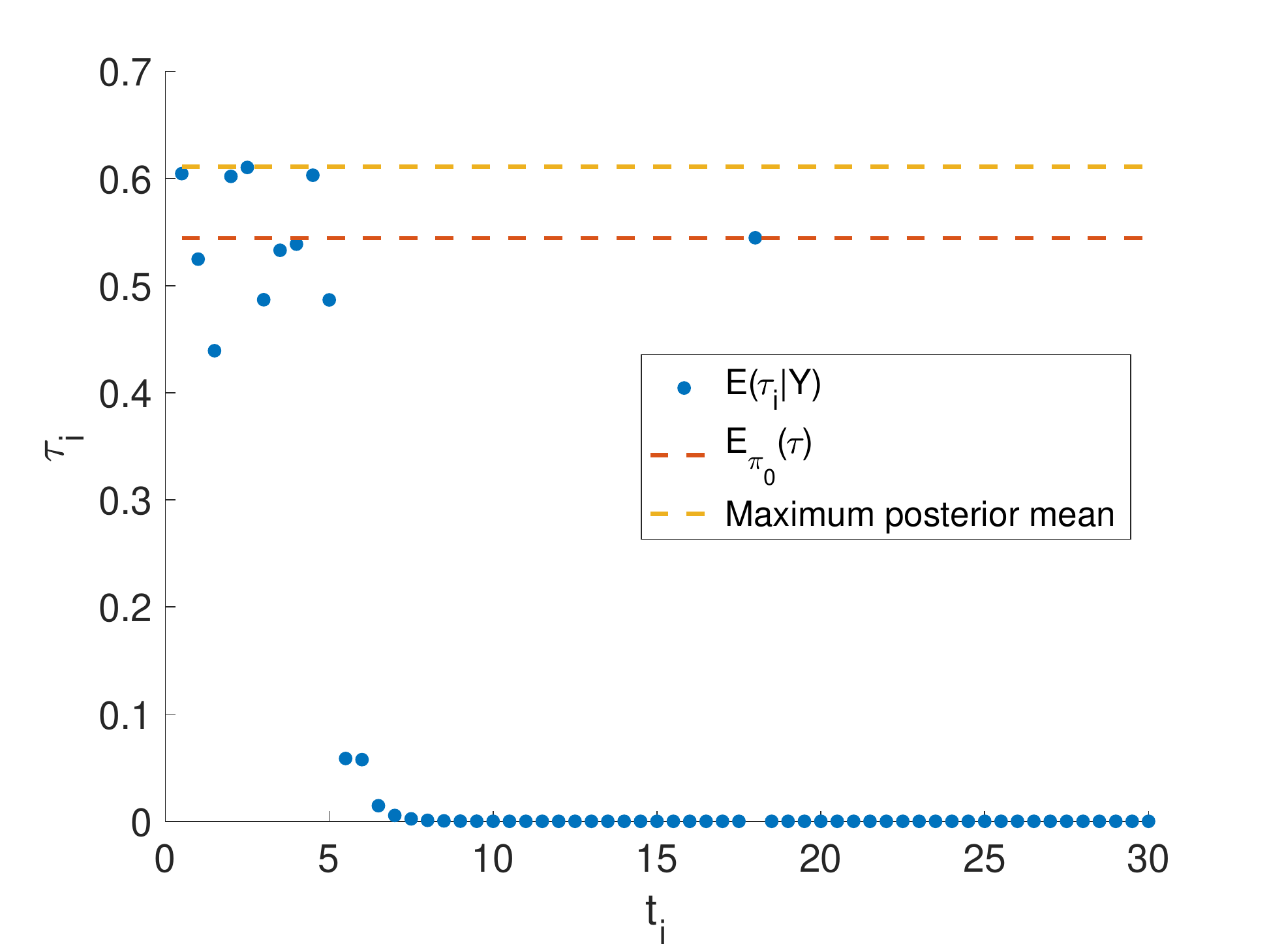}
      \caption{Mean values of $\tau_i$, approx. $\mathcal{G}$.}
      \label{fig:ataus}
      \end{subfigure}\\ 
      \caption{Posterior distributions on the initial condition $x_0$ (ground truth is $x_0 = 5$) (a) with exact observation operator and no data selection, (b) with approximate observation operator and no data selection, (c) with exact observation operator and data selection, (d) with approximate observation operator and data selection. Mean values of the $\tau_i$ (e) with exact observation operator, (f) with approximate observation operator, plotted alongside the prior mean on $\tau_i$ (red) and the maximum possible posterior mean (orange). }\label{fig:ODE}
\end{figure}

We conduct inference for this problem, both with and without Bayesian data selection, and with either the approximate observation operator \eqref{eq:approx} or the exact observation operator, in order to highlight the benefits of our approach. As in the example in Figure \ref{fig:errors}, we let the ground truth $x_0 = 5$, $\lambda=1$, $\Sigma = 1$ and $t_i = i/2$, $i\in \{1,\ldots,60\}$. The approximate observation operator \eqref{eq:approx} was used with $\Delta t = 10^{-3}$. 

Figure \ref{fig:ODE} shows the posterior distributions arising from these problems on the initial condition $x_0$ and the fidelity parameters $\tau_i$. Figures \ref{fig:AnoBDS} and \ref{fig:ABDS} show posteriors which arise from using the exact observation operator. Since the data and the observation operator are consistent, the posterior distributions are extremely tightly concentrated around a region encompassing the ground truth $x_0 = 5$, leading to a successful inference. There are small differences between these distributions due to the inclusion of Bayesian data selection; the inclusion of uncertainty about whether to include each observation leads to a slightly more diffuse posterior, with an increase in variance from $9.37\times 10^{-27}$ to $1.45 \times 10^{-26}$. The consistency of the data and model can be seen in the posterior means of the $\tau_i$, as shown in Figure \ref{fig:Ataus}, which are all significantly above zero. The vast majority of the means lie between the prior mean and the maximum possible posterior mean, which indicates that we have good evidence of a good fit between model and data for these points in observation space.

Figures \ref{fig:anoBDS} and \ref{fig:aBDS} show the posterior distributions which arise from using the approximate observation operator with $\Delta t = 10^{-3}$, with and without Bayesian data selection respectively. Without Bayesian data selection, as shown in Figure \ref{fig:anoBDS}, we arrive at a numerical Dirac; random walk proposals are accepted but only for posterior variances which are so small that the floating point representations of all of the posterior samples are identical. This is because of the huge discrepancies between the model and data, which cause the problem to become extremely stiff, in the sense that there are extremely strong restoring forces in both directions. There is a significant difference between the mean of this distribution, $\mathbb{E}(x_0|Y) = 5.0748$ and the ground truth $x_0 = 5$, despite the absolute sureity of the posterior.

With Bayesian data selection, we also have a significant difference in the mean of the posterior $\mathbb{E}(x_0|Y) = 5.0452$ and the ground truth. However, we have learned a great deal more about the problem from conducting the inference. The majority of the data have fidelity posterior means very close to zero. The best approach in this scenario is to use the fidelity posteriors to identify regions in observation space on which the model is able to well-represent the data. We note that due to the independence of the priors on the fidelity parameters, we are able to see singularities in the underlying parameter field, at around $t=5$ and $t=18$ in Figure \ref{fig:ataus}. The high value at $t=18$ is problematic, because this a coincidental good fit at this point caused by an overestimation of the unknown parameter, and an approximation of the model which is significantly lower than the true model. The singularity at $t=5$ also represents problematic behaviour, since we do not in reality have one region $0<t<5$ where the model approximation is exact, and another where it is bad, it is more of a gradual decay in the quality. Low fidelity parameters indicate that there is a significant difference between model and data which cannot be explained through the modelled observational noise, but it is likely that points close to this in observation space are also subject to significant reduction in the quality of the model approximation which may be detrimental to the accuracy of the resulting inference.

\subsection{Bayesian data selection with logit-GP prior} \label{sec:usingLGP}
Due to these issues, we employ the logit-GP prior for the fidelity field as described in Section \ref{sec:learn}. The logit of the field (evaluated element-wise) is modelled as a GP with constant mean function $m(x) = m \in \mathbb{R}$ and squared exponential kernel $K(x,y) = \sigma_\tau^2 \exp \left ( -\frac{1}{2l^2}\|x - y\|_2^2\right ).$ The values of the parameters $m = 1, \sigma_\tau = 1$ were chosen so that the resulting prior evaluated at any given point has a distribution reasonably similar to that employed above. The lengthscale of the kernel $l= 0.5$ was chosen to be the time between observations of the ODE.

In this case, since we are not using a conjugate prior, it is necessary to sample from the whole augmented state space, including all $N=60$ fidelity values. Due to large differences in scales of the posteriors on the model parameter $X_0$ and the fidelity parameters, we used adaptive Metropolis-within-Gibbs. Within each iteration, moves were proposed first on $X_0$, and then on the fidelity parameters. In each case, a Gaussian proposal centred on the current state was used, with covariances $\beta_X^2$ and $\beta_\gamma^2 \Sigma^\tau$ respectively. The scaling parameters $\beta_X, \beta_\gamma>0$ were tuned adaptively during a burn-in phase to give average acceptance probabilities of around 23.4\% (which although may not be optimal for low dimensions, is a reasonable rule of thumb). This implementation demonstrated good mixing leading to good estimates within a short time.

  \begin{figure}[htp]
      \centering
      \begin{subfigure}[b]{0.48\textwidth}
      \includegraphics[width=\textwidth]{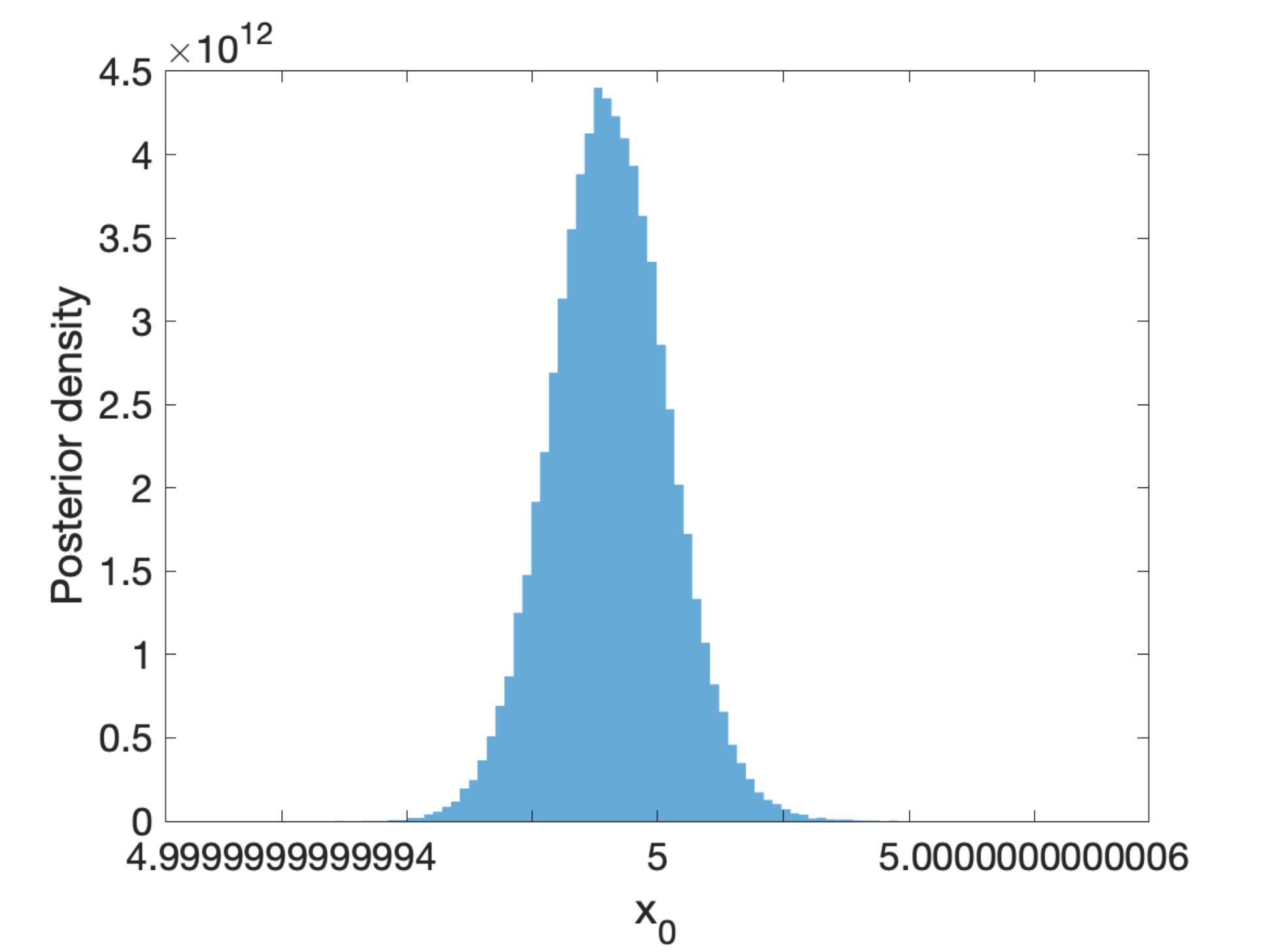}
      \caption{Exact $\mathcal{G}$, data selection, logit-GP prior.}
      \label{fig:acc_GP}
      \end{subfigure}
      \begin{subfigure}[b]{0.48\textwidth}
      \includegraphics[width=\textwidth]{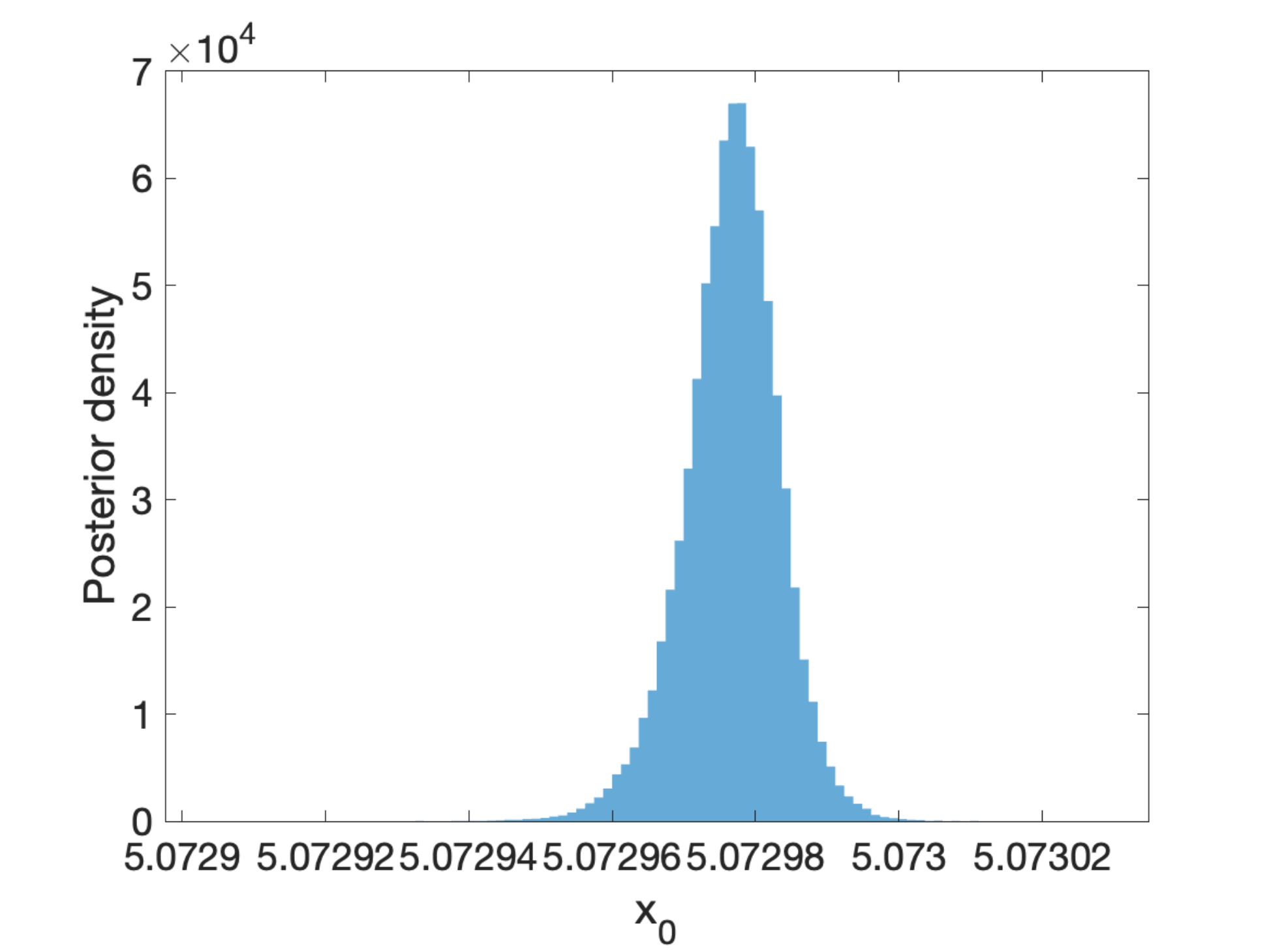}
      \caption{Approx. $\mathcal{G}$, data selection, logit-GP prior.}
      \label{fig:inacc_GP}
      \end{subfigure}\\
      \begin{subfigure}[b]{0.48\textwidth}
\includegraphics[width=\textwidth]{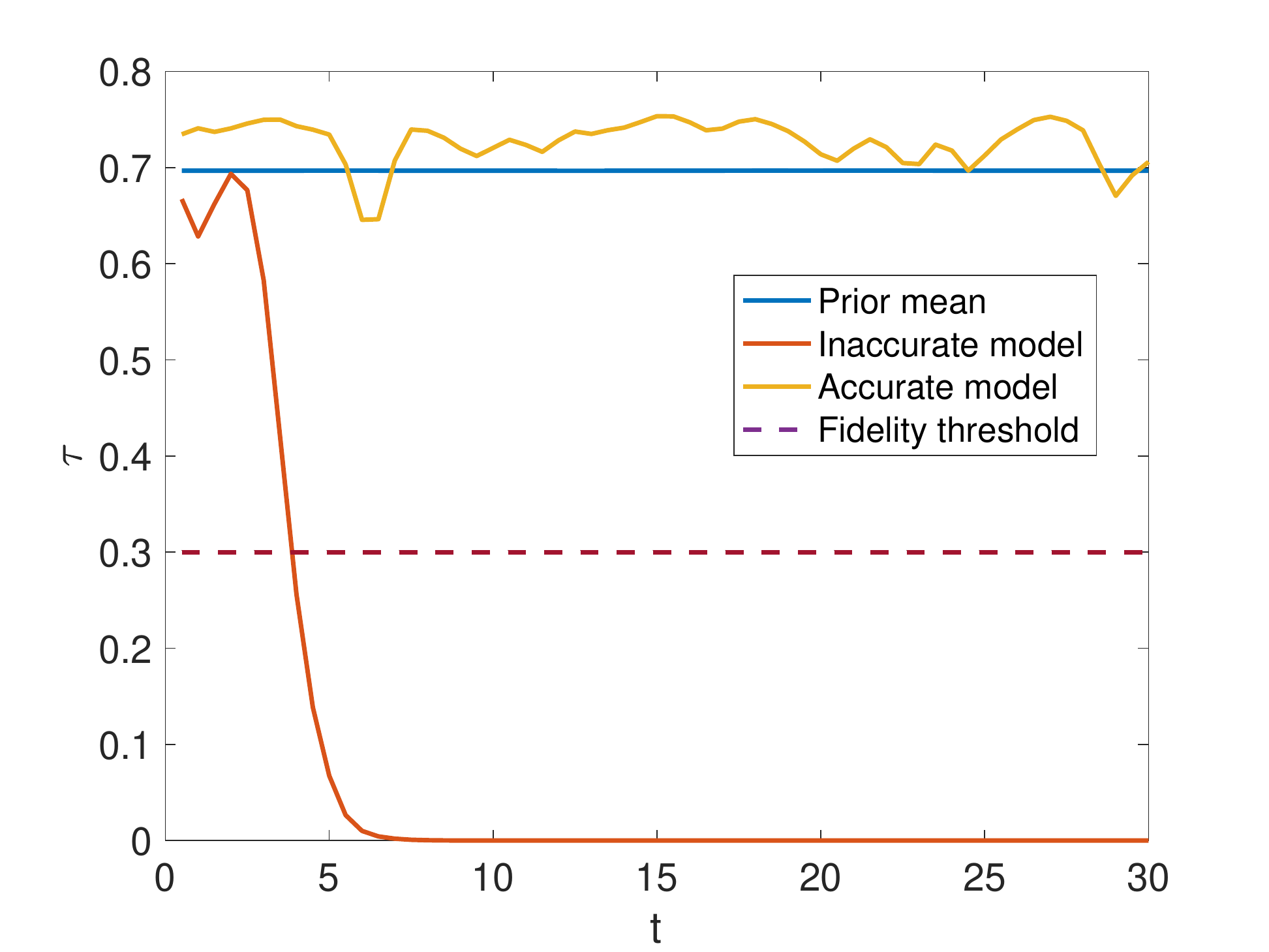}
      \caption{Posterior means on the fidelity field.}
      \label{fig:fidelity_av}
      \end{subfigure} 
      \caption{(a)-(b) Posterior distributions using Bayesian data selection with the logit-GP prior on the initial condition $x_0$ (ground truth is $x_0 = 5$) (a) with exact observation operator, (b) with approximate observation operator. (c) The posterior means of the fidelity field for both the inaccurate and accurate model approximations, using the logit-GP prior, along with the prior mean value and the proposed fidelity threshold. }\label{fig:ODE_GP}
\end{figure}

Figure \ref{fig:ODE_GP} presents the results of this numerical experiment. In (a) and (b) we present the arising posterior distributions on the initial condition $x_0$. We note that the plot with the accurate model is extremely similar to that seen in the equivalent plot Figure \ref{fig:ODE}(c). However there seems to be an even worse estimate of the initial condition when the model is inaccurate, peaking at around 5.07298 instead of around 5.04517 as in Figure \ref{fig:ODE}(d). However this is somewhat missing the point, as in this scenario we are aiming to find the regions in observation space for which our model is an accurate representation of reality. The pertinent information is contained within Figure \ref{fig:fidelity_av}, which reveals this region. This is contrast with Figure \ref{fig:ataus} where a single datapoint was incorrectly identified as being in a region with good data/model match.

One option for interpreting this information is to decide upon a threshold of the fidelity means above which we identify a point in observation space as being in this region. How this threshold is chosen is up for debate, but can be estimated by identifying properties of the region that we would like to hold. For data in this region, we would hope that the posterior distribution on the model parameters would be largely unchanged (other than a small diffusive effect) with or without data selection. This property can be assessed without knowledge of the ground truth or an accurate model, which makes it widely applicable. It is also likely that the appropriate threshold level found here, namely $\tau = 0.3$, is likely to be close to the correct value in other scenarios, as it represents a certain distance in probability between the model and the data above which we are likely to have large discrepancies that will corrupt the inference.

  \begin{figure}[htp]
      \centering
      \begin{subfigure}[b]{0.48\textwidth}
      \includegraphics[width=\textwidth]{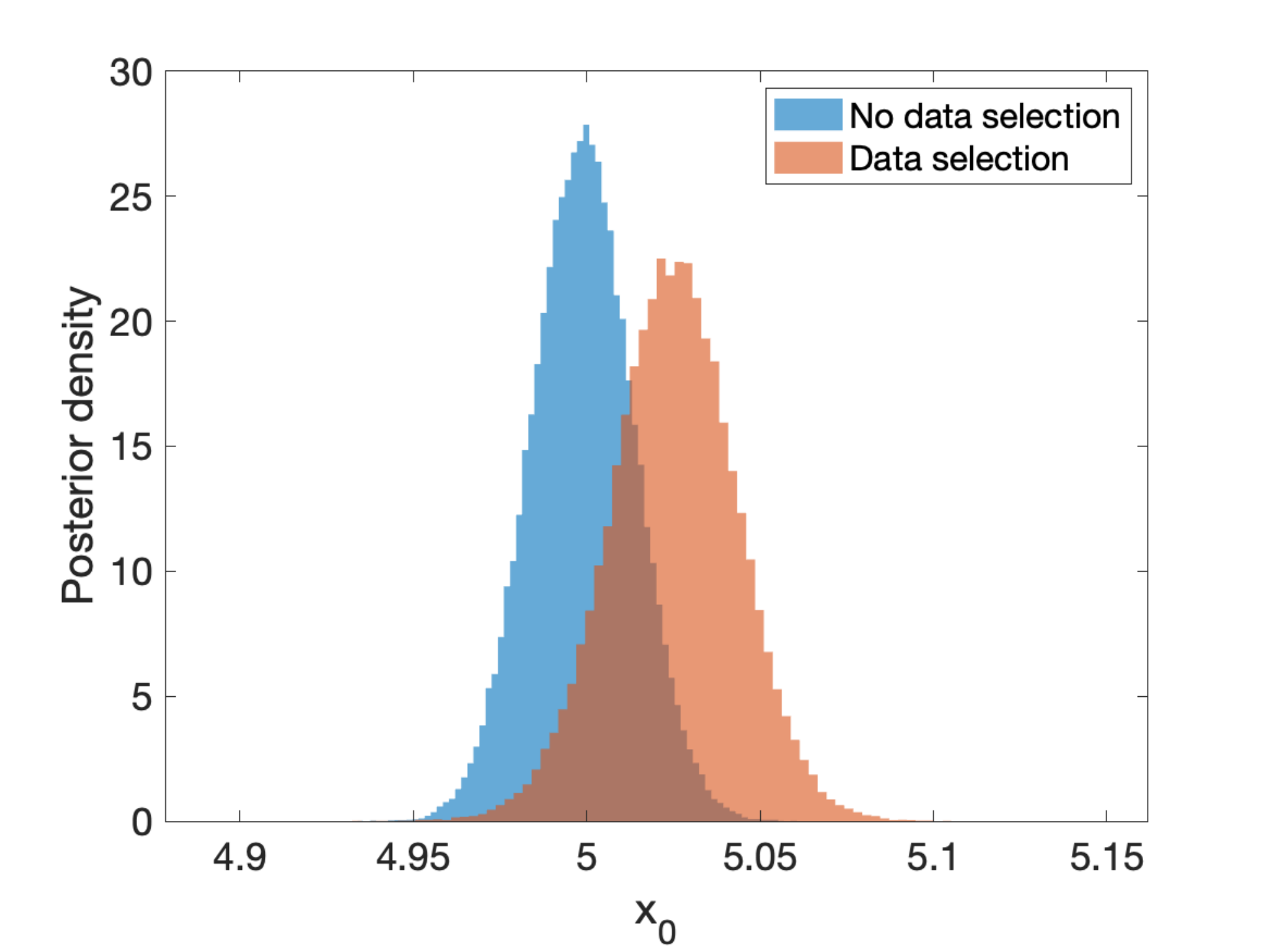}
      \caption{Posterior distributions on $x_0$ with threshold $\tau = 0.25$.}
      \label{fig:threshp25}
      \end{subfigure}
      \begin{subfigure}[b]{0.48\textwidth}
      \includegraphics[width=\textwidth]{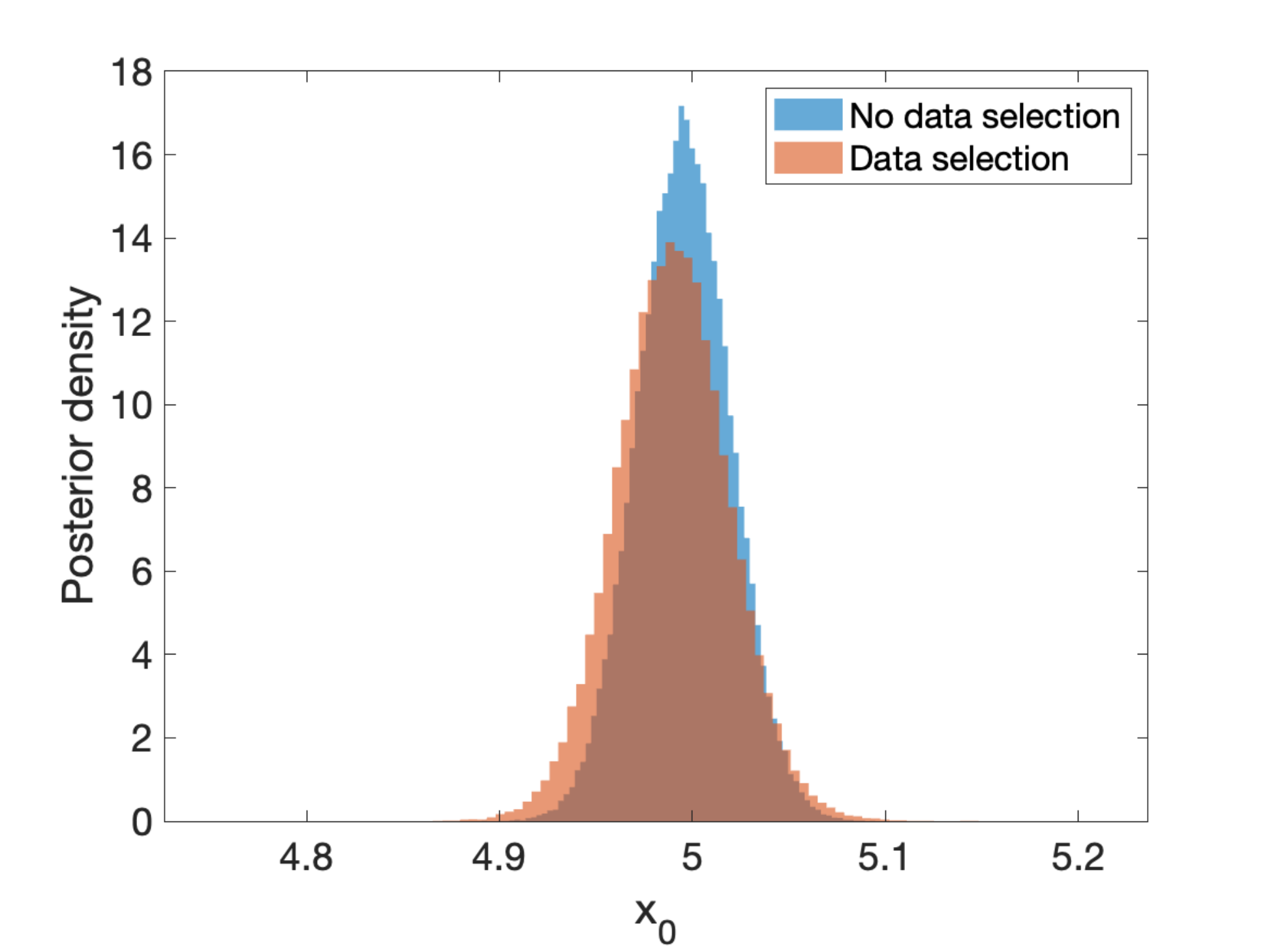}
      \caption{Posterior distributions on $x_0$ with threshold $\tau = 0.30$.}
      \label{fig:threshp3}
      \end{subfigure}
      \caption{Posterior distributions on $x_0$ (ground truth is $x_0 = 5$) using data truncated to region for which $\tau>0.3$ in the experiment given in Figure \ref{fig:ODE_GP}, with and without data selection (also using the logit-GP prior) with threshold (a)$\tau = 0.25$, (b)$\tau = 0.30$}\label{fig:thresh}
\end{figure}

Figure \ref{fig:thresh} shows the the posterior distributions on $x_0$ when using two different values of the threshold, $\tau = 0.25$ and $\tau = 0.30$. With the lower level, there is a clear difference to the posteriors with and without data selection. With the higher level, the data selection posterior is a little more diffuse than the posterior without data selection, but they are extremely similar, indicating that we have a good agreement between data and model in this region. Although the data selection posterior is sensitive to this threshold, it is worth noting that the posterior without data selection on the truncated data is not, which is reassuring.

\section{Discussion} \label{sec:discussion}
In this work we have presented several hierarchical approaches to Bayesian data selection, where the suitability of data for model parameter inference is assessed alongside inference of the model parameters, within a Bayesian framework. Our approach is to acknowledge that in general, data, models, and observational models are all imperfect. There are many possible strategies to deal with this, for example existing methodologies for inference of model error. With Bayesian data selection we take a different route, in that we aim to find regions in observation space for which the chosen model is a good approximation of the data. This approach is the result of a philosophy which acknowledges the imperfect nature of both data and models, and provides a wealth of information not accessible by standard Bayesian inference. As we demonstrated with examples of linear regression, the method is robust to a range of challenging situations in which standard inference fails; models with a partial fit to data, corrupted data, and problems for which the model is a good fit to the data with different parameter values in different regions in observation space. Moreover we were able to choose conjugate priors for the fidelity values in many cases, leading to a method with little additional computational cost, over and above standard Bayesian inference.

In the example regarding a Bayesian inverse problem for the initial value of an ODE, we demonstrated the effectiveness of aiming to find the regions of the data for which a neighbourhood of observations have good agreement between the model and data. With a coarse approximation of the ODE, only observations made with small times were able to be well approximated by the model. By employing inference on a scalar field, with a logit-GP prior, which describes our belief in the model's ability to represent the data, we were able to identify at what point the approximate model is insufficiently accurate to conduct inference. Using the region discovered using our methods for inference enabled us to accurately estimate the true value of the initial condition. This approach could be very useful in a number of applications where we expect our model approximation to deteriorate in quality over time, for example in chaotic models, which are commonplace in oceanography, meteorology, and astronomy.

Moreover, the changes in the likelihood due to data selection smooth the potential landscape, leading to potential reduction in the overall computational burden of characterising the posterior distribution using methods such as MCMC, as compared with standard Bayesian inference. All of these factors are indicative of an approach and family of methods that are very broadly applicable to a wide range of applications in statistics, data science, engineering and the physical sciences.



\bibliographystyle{siam}
\bibliography{refs.bib}

\end{document}